\documentclass[12pt]{article}
\usepackage[utf8]{inputenc}
\usepackage{graphicx}
\usepackage{color}
\usepackage[T1]{fontenc}
\usepackage[top=2.5cm, bottom=2.5cm, left=1in, right=1in]{geometry}
\usepackage{textcomp}
\usepackage{latexsym}
\usepackage[mathcal]{eucal}
\usepackage{amssymb}
\usepackage{amsthm}
\usepackage{amsmath,amsfonts}
\usepackage{eufrak}
\usepackage{multicol}
\usepackage{float}
\usepackage{amscd}
\usepackage{verbatim}
\usepackage{subfigure} 
\usepackage[authoryear]{natbib}
\usepackage[colorlinks,citecolor=blue,urlcolor=blue]{hyperref}
\usepackage[font=footnotesize,labelfont=bf]{caption} 
\usepackage{setspace}
\newcommand\hide[1]{#1}

\title{Coherence-based Time Series Clustering for Brain Connectivity Visualization}
\author{\hide{Carolina Euan}\footnote{\hide{King Abdullah University of Science and Technology 
(KAUST), CEMSE Division, Thuwal 23955, Saudi Arabia}}~, 
Ying Sun\addtocounter{footnote}{-1}\footnotemark  ~ ~and Hernando Ombao\addtocounter{footnote}{-1}\footnotemark~ \addtocounter{footnote}{1}\footnote{\hide{Department of Statistics, University of California at Irvine (UCI), Irvine, CA 92697, USA}}
}
\begin{document}
\bibliographystyle{plainnat}

\maketitle

\begin{abstract}
We develop the hierarchical cluster coherence (HCC) method for brain signals, 
a procedure for characterizing connectivity in a network by clustering nodes or groups of
channels that display high level of coordination as measured by "cluster-coherence". 
While the most common approach to measuring dependence between clusters is 
through pairs of single time series, our method proposes cluster coherence which 
measures dependence between whole clusters rather than between single elements. 
Thus it takes into account both the dependence between clusters and within channels in a cluster.  
Using our method, the identified clusters contain time series that exhibit 
high cross-dependence in the spectral domain. That is, these clusters correspond 
to connected brain regions with synchronized oscillatory activity. In the 
simulation studies, we show that the proposed HCC outperforms commonly 
used clustering algorithms, such as average coherence and minimum coherence based methods.
To study clustering in a network of multichannel electroencephalograms 
(EEG) during an epileptic seizure, we applied the HCC method  
and identified connectivity in alpha $(8-12)$ Hertz and 
beta $(16-30)$ Hertz bands at different phases of the recording: before 
an epileptic seizure, during the early and middle phases of the seizure episode. 
To increase the potential impact of this work in neuroscience, we 
also developed the HCC-Vis, an R-Shiny app (RStudio), which can be downloaded from this
\url{https://carolinaeuan.shinyapps.io/hcc-vis/}.
\end{abstract}

\noindent%
\vfill

  \section{Introduction}
The study of electroencephalograms (EEGs) times series data is naturally related 
to features in the frequency domain, since EEGs capture brain signals as  
superposed waveforms with varying amplitudes on the scalp. 
There exists a direct relationship between energy distribution among different 
frequency bands and brain activity. The frequency bands of interest to neuroscience 
are in the range of $(0,50)$ Hertz (Hz), where the frequency bands are named as 
follows: delta $(0,4)$ Hz, theta $(4,8)$ Hz, alpha $(8,12)$ Hz, beta $(12,30)$ Hz and 
gamma $(30,50)$ Hz. Energy in the lower frequency bands, delta and theta, 
are typically related to resting state of the brain, while activities in the higher 
frequency bands, alpha and beta, are more likely related to an activity or 
disease. The dependency between brain signals observed in different 
brain regions reflects brain connectivity. A first approach to study brain connectivity is via covariance, 
correlation or precision matrices, which is common used on fMRI (functional magnetic 
resonance images) data. In this line, methods have been developed to identify changes in brain networks structure 
\citep{Cribben2017}. Among dependency measures, 
coherence is the most commonly used in the study of EEG signals \citep{Fiecas10,Bowyer2016}.

From a statistical point of view, EEG data can be treated as multivariate time series collected 
from electrodes placed on the scalp, and brain activity can be characterized by 
a set of spatially correlated time series.
Thus, to quantify the connectivity between brain regions, we need to 
measure dependency or coherence between sets of time series. Moreover,
many applications in neuroscience require identifying highly correlated brain 
regions. One approach to explore the connectivity between signals is to 
set a threshold for the observed coherence values; then, two signals are 
connected if the observed coherence exceeds the threshold.
This procedure is simple and computationally fast. However, 
it could result in groups where some time series are low correlated. 
A better option is to study brain connectivity via brain signals clustering.  
Compared to the first approach, clustering methods will produce groups of 
time series that that show high correlation between channels within a 
cluster and low correlation between different clusters. 
With this approach, clusters of time series with 
high dependence in the frequency domain correspond to connected brain regions.  
To identify these groups, the clustering method must 
take into account the properties of the EEG data.
In particular, a clustering method based on the
spectral features of the EEG data will produce groups of brain signals that are easily interpreted.
In this paper, we propose a new time series clustering method based 
on coherence to describe brain connectivity. 

The primary contributions of this paper are the following: (1.) We propose the use of a measure for divergence between clusters that is based on coherence between clusters taken as a whole, rather than by pairs of individual signals in each cluster. Thus, the measure automatically takes into account the dependence between channels within each cluster. (2.) The resulting clustering algorithm identifies those channels that 
are highly correlated interpreted as connected brain regions. (3.) We propose new visualization plots of the 
clustering results and we developed the HCC-Vis, an interactive Shiny app, to visualize the data clustering results, which is useful for the neuroscience research.     
       
The use of spectral features of time series for classification purposes has attracted the attention 
of many researchers. 
The general goal is to identify clusters of time series with 
similar spectral densities or similar oscillatory patterns.  
\citet{Krafty2016} has proposed a model to explain the between group and the within group spectra variability when the classes are known. In case of an unsupervised learning or unknown classes, clustering methods are used. Among the existing clustering methods for time series, \citet{Caiado06} were the first who proposed 
using periodograms of time series to define similarity in a hierarchical clustering algorithm. 
A more recent study has proposed the hierarchical spectral clustering (HSM) method 
\citep{Euan18}, which also considers the spectral density for the classification of time series. 
Unlike classic hierarchical methods, the HSM method does not need a linkage function 
between clusters, since when merging clusters all time series contained in the cluster are used to update the spectral density estimation
and the dissimilarity between clusters is computed by using the updated spectral densities.
Another classification method  
is based on the discrete wavelet transform of the time series \citep[see ][]{Maharaj07}, for a more 
robust clustering result under non-stationary variances. 
Although wavelet decomposition is different from Fourier analysis, the wavelet 
variances provide information similar to the spectral density. Therefore, both methods will produce
clusters of time series with similar spectra.
However, the similarity in spectra does not necessarily 
imply correlation between the time series; it is possible to have two independent 
time series generated from the same model with the same spectral density
but zero correlations.
 
Our goal is to develop a clustering method that builds highly correlated clusters 
by using the spectral features of the time series. In frequency domain, 
coherence is a measure of the correlation between random processes (see Section 
\ref{SMethod1.1}). Therefore, coherence plays a central role in this paper. 
\citet{Maharaj2010} proposed 
coherence-based clustering procedures for classifying time series. 
They considered hierarchical and non-hierarchical procedures;
among the hierarchical clustering, they proposed to measure similarity 
between clusters as average coherence or minimum coherence.
They showed that  coherence can identify linearly related groupings of time 
series, even in cases with non-stationary variances. However, 
their measures based on average and minimum coherence
are pairwise measures, and do not take into account the within cluster structure. 
To address the within-cluster structure, we need to generalize coherence 
to measure the correlation between two clusters of time series. 

In the literature, we can find two proposed extensions of coherence: 
block  coherence \citep{Nedungadi11} and canonical coherence \citep{Takahashi14}. Block coherence at frequency $\omega$ 
between two sets of time series, $\mathbf{X}$ and $\mathbf{Y}$, 
is defined as 
$\displaystyle 1- \frac{\det (S_{\mathbf{X,Y}}(\omega))}{\det (S_{\mathbf{X}}(\omega) )\det (S_{\mathbf{Y}}(\omega))}$, where $S(\omega)$ denotes the spectral matrix. 
However, block coherence overestimates the correlation between two blocks of 
time series, even when there is only correlation between few members of the blocks. 
To overcome this limitation, canonical coherence was proposed. Canonical  
coherence projects a block of time series into canonical variables, then computes the block coherence 
between the canonical variables from different blocks. This procedure improves the block coherence, but
the identified clusters may lack interpretability.
In this paper, we develop a new notion of coherence, cluster coherence, that measures the 
correlation between groups of time series. Since cluster coherence is defined by 
the pairwise coherence matrix, without projecting the time series, 
the observed dissimilarity values can be interpreted in terms of the original signals.
Then, we apply the cluster coherence in a hierarchical algorithm.
Our proposal, the hierarchical cluster coherence (HCC) method can be applied to any stationary 
(or locally stationary) set of time series to find highly correlated groups. 

The rest of this paper is organized as follows. Section 2 introduces the frequency analysis of multivariate time series. Section 3 presents the cluster coherence measure with comparisons to average coherence, and introduces the HCC clustering algorithm. In Section 4,
we conduct simulation studies to compare the performance of the HCC method with two other hierarchical clustering methods that uses average or minimum coherence as similarity measures. Simulation designs are inspired by the EEG data application. Finally, in Section 5, 
we study an EEG recorded during an epileptic seizure to identify brain
regions connected in the alpha and beta bands.

\section{Spectral Analysis for Multivariate Time Series}\label{SMethod1.1}
In this section, we introduce the basic concepts of frequency domain analysis for multivariate time series. 
As previously stated, the EEG signal from an electrode located on the scalp can be considered
 as realizations of a brain process. In this paper, we will refer to the electrode located at the 
 scalp as a channel and in the case of multiple electrodes as a multi-channel.
 
 Let $\{\mathbf{X}(t)\}=\{(X_1(t),\ldots,X_N(t))^T\} $ be a $N$-multivariate stationary time series with mean $\mathbf{0}$ and covariance matrix function
 $\Gamma(\cdot)$ with absolutely summable elements denoted by $\gamma_{j,k}(\cdot)$. 
 Then the function
$
  f_{j,k}(\omega)=\sum_{-\infty}^{\infty}\gamma_{j,k}(h) e^{-i 2 \pi \omega h}$, $-\frac{1}{2}\leq \omega \leq \frac{1}{2}, 
$
is called the cross spectrum or cross spectral density of $\{X_j(t)\}$ and $\{X_k(t)\}$ for $j,k=1,\ldots,N$ \citep[see][]{Brockwell2006, Shumway2011}. The matrix
$$S(\omega)=\begin{pmatrix}
              f_{1,1}(\omega) &  f_{1,2}(\omega) & \cdots &  f_{1,N}(\omega)\\
               f_{2,1}(\omega) &  f_{2,2}(\omega) & \cdots &  f_{2,N}(\omega)\\
               \vdots& \ddots& \ddots& \vdots \\
               f_{N,1}(\omega) &  f_{N,2}(\omega) & \cdots &  f_{N,N}(\omega)
            \end{pmatrix}
,$$
is called the spectral density matrix of $\{\mathbf{X}(t)\}$ where $f_{j,j}(\omega)$ corresponds to the univariate spectral densities of  $\{X_j(t)\}$ for $j=1,\dots, N$.

Statistical inference in frequency analysis of multivariate time series is based on the periodogram. Let 
$\{\mathbf{X}(1),\ldots,\mathbf{X}(T)\}$ be an observed $N$-multivariate time series; then the periodogram matrix 
at the Fourier frequencies $w_j= j/T$ is defined to be  
$$\mathbf{I}(\omega_j)=\frac{1}{T}\left(\sum_{t=1}^T \mathbf{X}(t) e^{-i t 2\pi \omega_j }\right)\left(\sum_{t=1}^T \mathbf{X}(t) e^{-i t 2\pi \omega_j }\right)^*= \sum_{|k|<T} \hat{\Gamma}(k) e^{-i t 2\pi \omega_j }, $$
where $*$ denotes complex conjugate transpose, 
$\hat{\Gamma}(k)= T^{-1} \sum_{t=1}^{T-k} ((\mathbf{X}(t+k)-\overline{\mathbf{X}}) (\mathbf{X}(t)-\overline{\mathbf{X}})$ is the sample covariance matrix and 
$\overline{\mathbf{X}}=\frac{1}{T}\sum_{t=1}^T \mathbf{X}(t)$. Though, the periodogram is not a consistent estimator of the spectral density, consistent estimators can be constructed by smoothing the periodogram (\citealp[see ][Chap 11]{Brockwell2006}, \citealp[ ][Chap 4]{Shumway2011} ). 

Notice that contrary the univariate case, the cross spectrum is not constrained to be positive. It is in fact complex-valued. However, the spectral matrix needs to be a positive definite spectral density matrix. To preserve positive-definiteness, of the spectral matrix estimator, the classical procedure is to use the same degree of smoothness in all the spectral estimates. If it is not reasonable to assume the same degree of smoothness for all spectral functions, there exist nonparametric methods based on the Whittle likelihood to estimate the spectral matrix with different degree of smoothness \citep[see][]{Pawitan96, Rosen07}.

Let $\{\mbox{d}Z_1(\omega),\ldots, \mbox{d}Z_N(\omega)\}$  be the zero mean orthogonal increment processes in the univariate spectral representation, then the cross spectrum has a similar property to the spectral density, 
$$f_{j,k}(\omega)\mbox{d}\omega= \mathbb{E}[\mbox{d}Z_j(\omega)\overline{\mbox{d}Z_k(\omega)}]= cov(\mbox{d}Z_j(\omega),\mbox{d}Z_k(\omega)),$$
for $j,k=1,\ldots,N$.
If $\omega_1\neq \omega_2$, then $\mathbb{E}[\mbox{d}Z_j(\omega_1)\overline{\mbox{d}Z_k(\omega_2)}]=0$. Thus, the squared coherence, at frequency $\omega$, is defined as
\begin{equation}
 \kappa_{j,k}(\omega)= \frac{|f_{j,k}(\omega)|^2}{f_{j,j}(\omega)f_{k,k}(\omega)} ,\qquad -\frac{1}{2}\leq \omega \leq \frac{1}{2},\label{Coh}
\end{equation}
which measures the correlation between 
$\mbox{d}Z_j(\omega)$ and $\mbox{d}Z_k(\omega)$.

Coherence is the analog of cross-correlation in the frequency domain and is a widely used measure for analyzing EEG signals. Coherence could vary across frequencies but remains constant across time. The concept of evolutionary coherence has been studied by \citet{Ombao08}. Among dependency
measures, coherence is the most commonly used in the study of EEG signals. An alternative to coherence for studying functional connectivity is partial coherence, which measures the direct linear association between any pair components of a multivariate time series after removing the linear effects of the other components \citep{Fiecas10,Fiecas11}. In this paper, we will focus on coherence and how to extend this concept to measure dependency between clusters of EEG signals (or multivariate time series).

\section{Hierarchical Cluster Coherence Method} 
We propose the Hierarchical Cluster Coherence (HCC) method to describe brain connectivity.
The HCC method uses a hierarchical algorithm to identify clusters containing highly correlated time series, 
based on our coherence-based measure of similarity.
In \ref{SubHCC1} we define the cluster coherence between two groups of time series.   
In \ref{SubHCC2} we present the HCC clustering algorithm.  

\subsection{Cluster Coherence}\label{SubHCC1}
Consider the $N$-multivariate stationary time series $\{\mathbf{X}(t)\}$, where $\mathbf{X}(t)=[\mathbf{X}_1(t), \mathbf{X}_2(t)]$ and $\mathbf{X}_j(t)$ are stationary multivariate time series of dimension $n_1$ and $n_2=N-n_1$, respectively. We can think of $\mathbf{X}_1(t)$ as a collection of time series in one cluster and $\mathbf{X}_2(t)$ as a second collection of time series in a different cluster. 
Consider the coherence matrix of  $\{\mathbf{X}(t)\}$, as a block matrix for each $\omega$, i. e., 
\begin{eqnarray}
\mathbf{C}(\omega)=\begin{pmatrix}
              \mathbf{C}_{1,1}(\omega) &  \mathbf{C}_{1,2}(\omega)\\
               \mathbf{C}_{2,1}(\omega) &  \mathbf{C}_{2,2}(\omega)
            \end{pmatrix}, \label{SpecM}
\end{eqnarray}
where $\mathbf{C}_{j,k}(\omega)$ is the squared coherence between the entire first cluster of time series $\mathbf{X}_j(t)$ and the entire second cluster of time series $\mathbf{X}_k(t)$. Let $\{\lambda_1^1(\omega),\ldots,\lambda_{n_1}^1(\omega)\}$ and $\{\lambda_1^2(\omega),\ldots,\lambda_{n_2}^2(\omega)\}$ be the normalized eigenvalues of the within cluster coherence matrices $ \mathbf{C}_{1,1}(\omega)$
 and $  \mathbf{C}_{2,2}(\omega)$, and  let $\{\lambda_1(\omega),\ldots,\lambda_N(\omega)\}$ be the normalized eigenvalues of $\mathbf{C}(\omega)$. We define the cluster coherence between $\mathbf{X}_1(t)$ and $\mathbf{X_2}(t)$ as
\begin{eqnarray}
 CCo(\omega)&=&\left(\sum_{j=1}^N \left|\lambda_{[j]}(\omega)-\lambda_{[j]}^*(\omega)\right|^p  \right)^{(1/p)}
  \label{MCoh}
\end{eqnarray}
where $\lambda_{[j]}(\omega)$ is the $j$-th largest eigenvalue of  $\{\lambda_1(\omega),\ldots,\lambda_N(\omega)\}$ and $\lambda_{[j]}^*(\omega)$ is the $j$-th largest eigenvalue of  $\{\lambda_1^1(\omega),\ldots,\lambda_{n_1}^1(\omega), \lambda_1^2(\omega),\ldots,\lambda_{n_2}^2(\omega)\}$.

\noindent \textit{Remark} 1. Cluster coherence compares the eigenvalues of the coherence matrix between clusters with the independent block case, i.e., when $ \mathbf{C}_{1,2}(\omega)=\mathbf{C}_{2,1}(\omega)=\mathbf{0}$. In this sense, it is clear that larger values of the between cluster coherence indicate highly correlated time series in the frequency domain. We define cluster coherence in terms of the $L^p$ norm, but we consider only the cases where $p=1$ or $2$. 

\noindent \textit{Remark} 2. Notice that when $\mathbf{X}_1(t)$ and $\mathbf{X_2}(t)$ are uncorrelated, then the set of eigenvalues of the whole spectra matrix $\mathbf{C}(\omega)$ is equal to merge the eigenvalues of 
$  \mathbf{C}_{1,1}(\omega)$ and $\mathbf{C}_{2,2}(\omega)$ for all $\omega$. This fact produces a value of cluster coherence equal to zero. 
\noindent \textit{Remark} 3. In contrast, if $\mathbf{X}_1(t)$ and $\mathbf{X_2}(t)$ are perfectly correlated, we can consider that there exists a latent variable $Z_1(\omega)$ in the frequency domain that is shared by components of $\mathbf{X}_1(t)$ and $\mathbf{X_2}(t)$ at frequency $\omega$. This will produce that the largest eigenvalue of $\mathbf{C}(\omega)$ is close to one and the rest of the eigenvalues will be close to zero. Consequently, cluster coherence will be close to 1. 

\vspace{.5cm}

\noindent \textit{Properties of $CCo$}  $(\omega)$:
\begin{enumerate}
 \item[P1.] The cluster coherence between $\mathbf{X}_1(t)$ and $\mathbf{X}_2(t)$ is bounded between $0$ and $1$, i.e, 
$0 \leq CCo(\omega) \leq 1$. This can be deduced from the definition of cluster coherence 
between $\mathbf{X}_1(t)$ and $\mathbf{X}_2(t)$, where the norm of all of the eigenvalues is equal to 1 by construction. 
 \item[P2.] If $\mathbf{X}_1(t)$ and $\mathbf{X}_2(t)$ are uncorrelated then
 $CCo(\omega)=0$ for all $\omega$. When $\mathbf{X}_1(t)$and $\mathbf{X}_2(t)$ 
 are uncorrelated at frequency $\omega$, $\lambda_{[j]}(\omega)=\lambda_{[j]}^*(\omega)$ for $j=1,\ldots,N$. 
 \item[P3.] If $\mathbf{X}_1(t)$ and $\mathbf{X}_2(t)$ are perfectly correlated then
 $CCo(\omega)=1$ for all $\omega$. When  $\mathbf{X}_1(t)$ and $\mathbf{X}_2(t)$ are perfectly correlated, $\mathbf{C}_{j,k}(\omega)= \mathbf{1}_{n_j \times n_k}$ (all-ones matrix of dimension $n_j \times n_k$) for all $\omega$ and $j,k=1, 2$. Then,
 $\lambda_{[1]}(\omega)=1$ and $\lambda_{[j]}(\omega)=0$ for $j=2,\dots,N$ and $\lambda_{[1]}^*(\omega)=\lambda_{[2]}^*(\omega)=1/2$ and $\lambda_{[j]}^*(\omega)=0$ for $j=3,\dots,N$.
 \item[P4.] If $n_1=n_2=1$ and $p=1$, then $CCo(\omega)=\kappa_{1,2}(\omega)$ (Equation \eqref{Coh}). This is because 
$n_1=n_2=1$ and $p=1$, hence $\{\lambda_{[1]}(\omega),\lambda_{[2]}(\omega)\}=\{\frac{1+\kappa_{1,2}(\omega)}{2},\frac{1-\kappa_{1,2}(\omega)}{2}\}$ and $\{\lambda_{[1]}^*(\omega),\lambda_{[2]}^*(\omega)\}=\{\frac{1}{2},\frac{1}{2}\}$. Finally, by computing \ref{MCoh},  $CCo(\omega)=\kappa_{1,2}(\omega)$.
 \end{enumerate}

\noindent \textit{Comparison of the CCo with average coherence and block coherence.}

The most commonly used measures for cluster dependency are average 
coherence (AC) and minimum coherence (MC). AC and MC between  $\mathbf{X}_1(t)$ and $\mathbf{X_2}(t)$ 
are defined, respectively, to be 
For example, in \eqref{SpecM} mean($\mathbf{C}_{1,2}(\omega)$) and 
$\min (\mathbf{C}_{1,2}(\omega))$, where mean($A$) is the average over all elements of $A$ and 
min($A$) is the minimum of all elements of $A$. 
Therefore, AC and MC consider only the pairwise coherence 
in $\mathbf{C}_{1,2}(\omega)$, neglecting within coherences $\mathbf{C}_{1,1}(\omega)$ and $\mathbf{C}_{2,2}(\omega)$. In contrast, cluster coherence measures dependency using the complete information of the coherence matrix represented by 
eigenvalues. This measure reflects an obvious difference between the within-cluster 
dependency and the between-cluster dependency, enhancing the clustering performance.

Cluster coherence has advantages over average 
coherence, especially in situations where the within clustering dependency is strong. 
We illustrate an example based on a mixture of autoregressive processes. 
Mixture models are useful in studying multi-channels EEG data, since these models can 
represent signals generated by different amplitudes \citep{Wang16}. Our goal is to obtain results 
with high connectivity between EEG channels in the presence of noisy signals, which would be 
unobtainable using average coherence. In addition we will compute the block coherence (defined in the introduction) to compare with cluster coherence.

\begin{figure}
 \centering
 \subfigure[\label{SZ}]{\includegraphics[scale=.3]{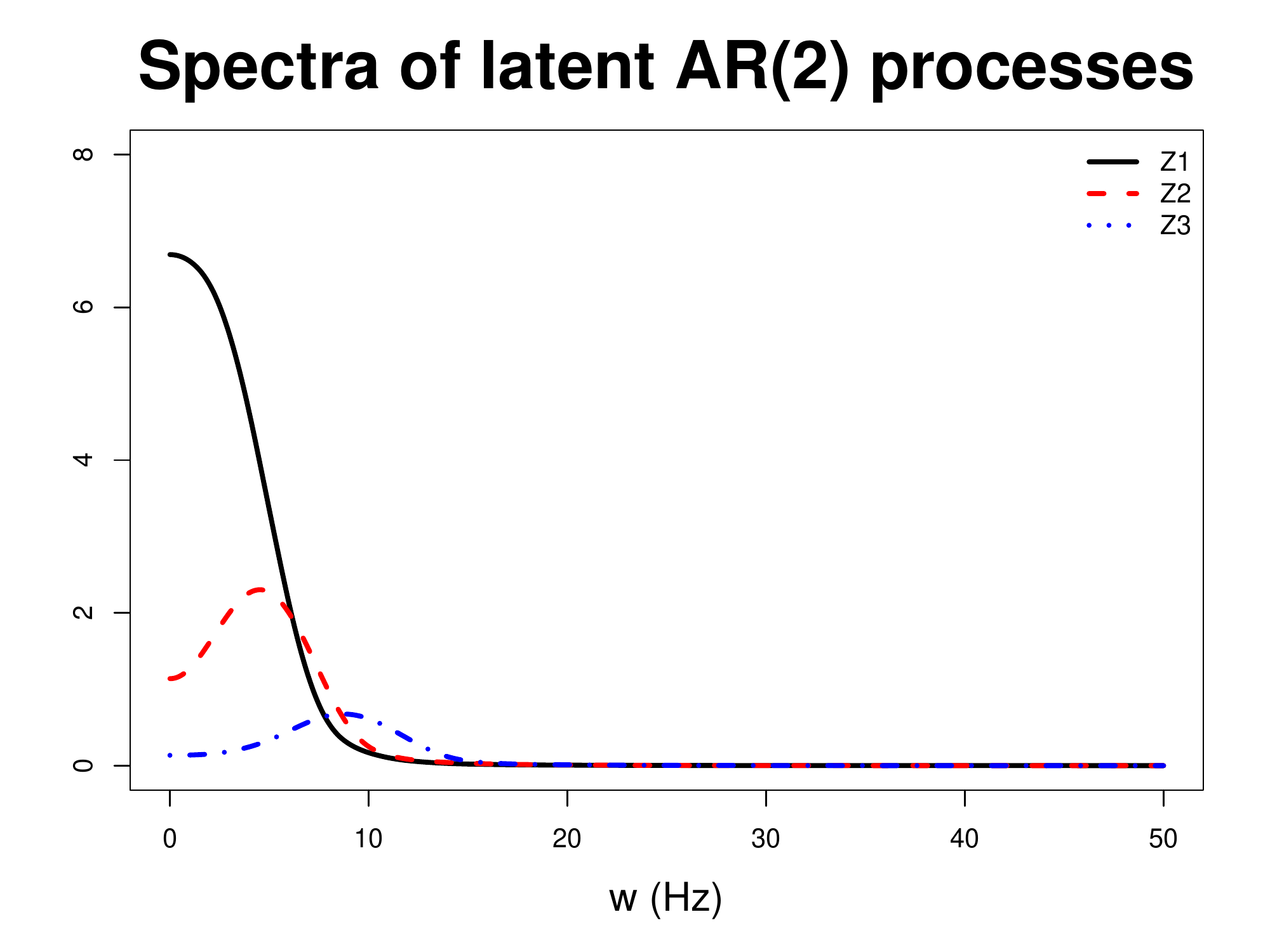}}
 \subfigure[\label{SX}]{\includegraphics[scale=.3]{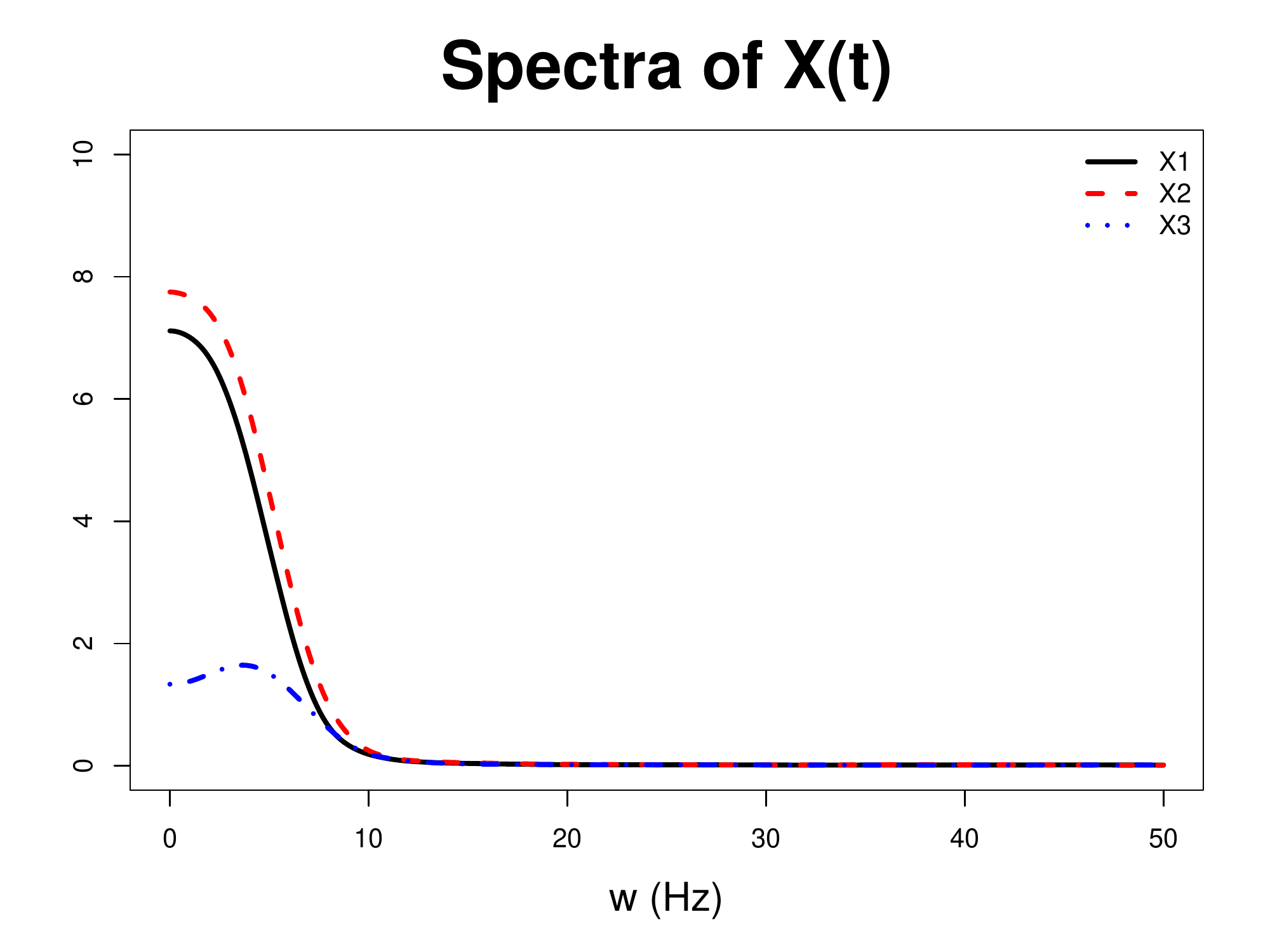}}\\
 \subfigure[\label{Exam1Coha}]{\includegraphics[scale=.3]{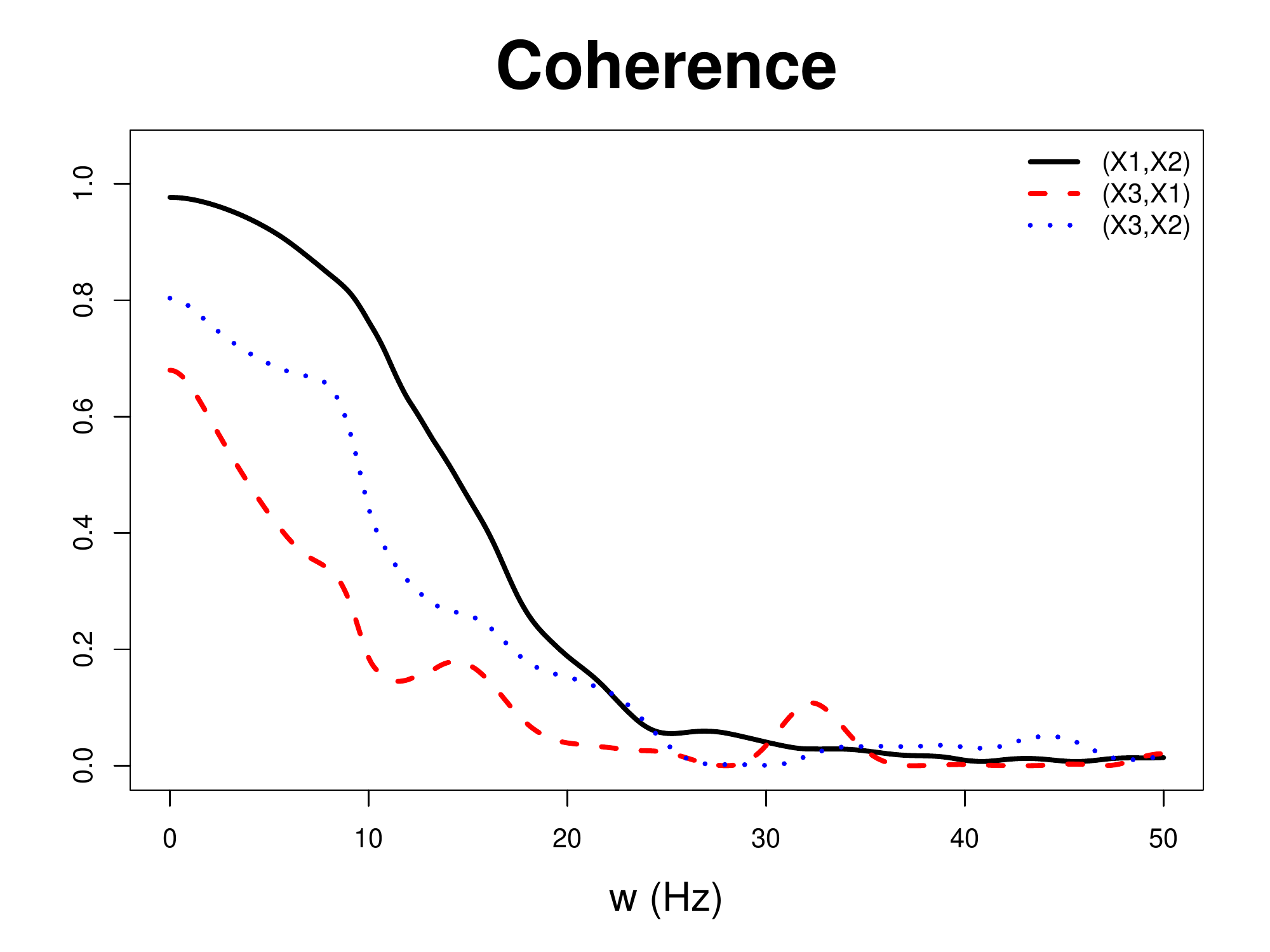}}
\subfigure[\label{Exam1Cohb}]{\includegraphics[scale=.3]{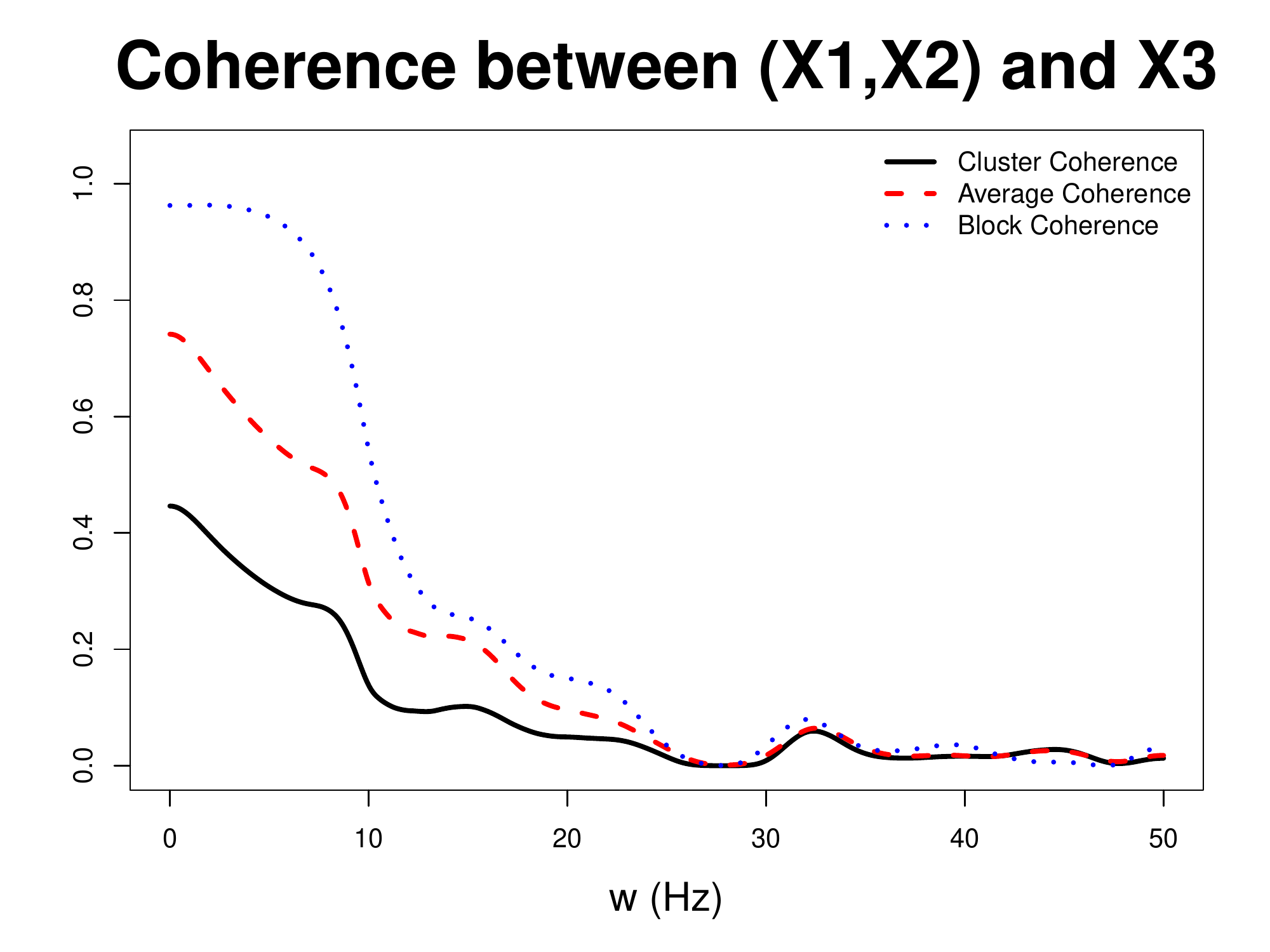}}
\caption{Estimated spectral densities of: (a) latent AR(2) processes and (b) observed signals. Estimated coherence between signals: (c) pairwise coherence and (d) coherence measures between clusters
$\{X_1(t), X_2(t)\}$ and $\{X_3(t)\}$.}
\end{figure}

\noindent \textit{IIustration 1.} Let $Z_1(t)$, $Z_2(t)$ and $Z_3(t)$ independent AR(2) latent processes with the unimodal spectral density peaks at $3,5$ and $9$ Hz, respectively. $\mathbf{X}(t)$ is a three-variate time series generated by a mixture of these latent processes,  
\begin{eqnarray*}
X_1(t)&=& Z_1(t) + 0.2 Z_2(t)  + \varepsilon_1(t)\\
X_2(t)&=& Z_1(t) + 0.6 Z_2(t)  + \varepsilon_2(t)\\
X_3(t)&=& 0.3 Z_1(t) + 0.7 Z_2(t) + 0.3 Z_3(t) + \varepsilon_3(t),
\end{eqnarray*}
where $\varepsilon_i(t)$ are white noise sequences. We simulate one realization with $1000$ time points and compute the cluster coherence using $p=1$. Figure \ref{SZ} shows the estimated spectra of the latent variables, these laten variables represents activity on delta, theta and alpha bands. Figure \ref{SX} shows the estimated spectral densities for univariate time series $X_j(t)$. $X_1(t)$ and $X_2(t)$ have more similar spectral densities, which is a consequence of being highly correlated, than $X_3(t)$. 
Figure \ref{Exam1Coha} shows the estimated coherence on frequency bands delta, theta and alpha. We observe that $X_1(t)$ and $X_2(t)$ are highly correlated signals. However, 
$X_3(t)$ is highly correlated with $X_2(t)$ but less correlated with $X_1(t)$. In this sense, it is reasonable to have two clusters, $\{X_1(t),X_2(t)\}$ and $\{X_3(t)\}$.

Figure \ref{Exam1Cohb} shows the estimated functions of cluster coherence, average coherence and block coherence, respectively. 
Using the average coherence or block coherence, it is very likely that $\{X_1(t),X_2(t)\}$ and $\{X_3(t)\}$ will be merged as one cluster, since these values are bigger than $0.5$ on delta and theta bands. This will produce a bigger cluster than expected. 
However, when using cluster coherence, $\{X_1(t),X_2(t)\}$ and $\{X_3(t)\}$ may not be merged, since 
cluster coherence is lower than $0.4$.

\noindent \textit{Remark 4.} This example exhibits the case when  
block coherence overestimates the correlation between two clusters of time series due to the 
correlation between few members of the cluster ($X_2(t)$ and $X_3(t)$). 

\subsection{Clustering Algorithm}\label{SubHCC2}
The HCC method uses cluster coherence as a measure of similarity between clusters in each iteration of the clustering algorithm. 
\bigskip

\noindent \textbf{\textit{Hierarchical cluster coherence (HCC) algorithm.}}
Let $\mathbf{X}(t)=[X_1(t) X_2(t) \cdots X_N (t)]$ be a set of time series and let $\Omega_{12}=(\omega_1,\omega_2)$ be the frequency band of interest. The procedure starts with $i=1$ and $k_0=N$ clusters, with each cluster containing a single signal.\\
\noindent
 \textbf{Step 1.} Estimate the coherence matrix $\mathbf{C}(\omega)$ at frequency $\omega$.\vspace{.1cm}\\
 \textbf{Step 2.} Compute the initial dissimilarity matrix at band $\Omega_{12}$, $D(\Omega_{12}) = 1-\frac{1}{\omega_2-\omega_1}\int_{\omega_1}^{\omega_2}\mathbf{C}(\omega)$d$\omega$.\vspace{.1cm}\\
 \textbf{Step 3.} Find the two clusters with the lowest dissimilarity and save this 
value as a characteristic.\vspace{.1cm}\\
 \textbf{Step 4.} Merge the signals of the two most similar clusters, reduce the number of clusters by one, i.e., $k_i=k_{i-1}-1$, and increase  $i$ accordingly , i.e., $i=i+1$.\vspace{.1cm}\\
 \textbf{Step 5.} Compute the dissimilarity between the new cluster and the existing ones as $D(\Omega_{12})=1-\frac{1}{\omega_2-\omega_1}\int_{\omega_1}^{\omega_2}CCo(\omega)$d$\omega$.\vspace{.1cm}\\
 \textbf{Step 6.} Repeat Steps 3-5 until $k_i=1$.

In real applications, we need to choose the number of clusters. We propose to use the scree plot of the 
minimum dissimilarity value to decide the number of clusters. In this sense, we chose the minimum number of clusters $k$ such that the dissimilarity value of $k+1$ will not decrease significantly.   

\subsection{Estimation of the Spectra}
The HCC method requieres estimating the spectral matrix. 
As mentioned in Section 2, the most common estimator for the spectral matrix is the periodogram 
matrix whose elements are the periodogram and the cross periodogram. 
However, the periodogram matrix is not consistent estimators for the spectral matrix. A common method to construct a consistent estimator is to smooth the periodogram. 

\noindent \textit{Smoothing the periodogram.}
A class of estimators for the spectral matrix is the following \citep[see ][]{Brockwell2006},

$$\hat{S}(\omega_j)=\sum_{|k|\leq m_T} W_T(k) \mathbf{I}(\omega_{j+k}),$$
\noindent
where $W_n$ is a kernel function, $m_T$ is the smoothing window size and $T$ denotes the time series length. To ensure the mean square consistency of the estimator the kernel function must satisfied that $\sum_{|k|\leq m_T} W_T^2(k) \rightarrow 0$ when $T \rightarrow \infty$. And to preserve positive-definiteness, $m_T$ is the same for all the spectral estimates.

In practice, we need to choose the size of the smoothing window, $m_T$. \citet{Ombao2001} proposed a criterion to select the size of the smoothing window based on the generalized crossvalidation (GCV) function derived from the gamma deviance. In our simulation studies and real data analysis, we used a smoothed version of the periodogram to estimated the spectral matrix. We choose a Fejer kernel that can be obtained by using the \textit{kernel} function in R. And we applied the GCV criteria to select the size of the smoothing window. 
The election of the spectral density estimator can be modified.
In simulation studies we consider the same estimator for all methods, then comparisons are based on clustering results rather than spectral density estimators. 

Finally, the auto-spectral estimates are the elements of the diagonal of $\hat{S}(\omega_j)$ and 
the cross-spectral estimates are the elements of the off diagonals of $\hat{S}(\omega_j)$. Let us denote by $\hat{f}_{k,l}(\omega_j)$, $k,l=1,\ldots,N$, the auto-spectral and cross-spectral estimates, then the estimator for pairwise coherence at frequency $\omega_j$ is 
$$ \hat{\kappa}_{k,l}(\omega_j)= \frac{|\hat{f}_{k,l}(\omega_j)|^2}{\hat{f}_{k,k}(\omega_j)\hat{f}_{l,l}(\omega_j)}.$$

An alternative nonparametric estimator of the spectral density is the multitaper spectral estimator \citep{Walden00} which has also important optimality properties. A review on multitaper spectral analysis can be found in \citet{Babadi2014}.

\section{Simulation Study}
In this section, we test the performance of the proposed clustering method by using simulated examples. 
Firstly, we show the differences between HCC method with the HSM clustering method. Secondly, we explore a simulation study 
to compare the HCC method with other two hierarchical clustering methods, by using average coherence (HAC)
or by using minimum coherence (HMC), which are the most common used among hierarchical methods.

Simulated examples are based on a mixture of autoregressive  (AR) models of order 2.
The reason for choosing AR(2) models is that 
these models can accurately represent oscillations and have been useful to analyze EEG data. 
We consider two examples of a mixture of AR(2) models with fixed 
coefficients and with spatially-varying coefficients.

To improve comparison of the clustering results, we develop a visualization tool. This tool contains 
two plots; a scree plot and a cluster merging plot. 
In the scree plot, we consider the dissimilarity value obtained when merging two clusters.
For a fixed number of clusters $k$ we plot this dissimilarity value; this graph is located in the displayed upper plot. 
The clustering merging plot shows the dynamic of the clustering method;
on the x-axis we shall have the number of clusters and 
on the y-axis colors denote time series belonging 
to the same cluster. From $k$ to $k-1$ clusters, 
one color should disappear, which means that the time series in that cluster were merged with another cluster.
Finally, this visualization tool could be used to help scientists choose the number of clusters.  

\subsection{Comparison Between HCC and HSM}
\begin{figure}
 \centering
 \subfigure[HCC with $p=1$]{\includegraphics[scale=.4]{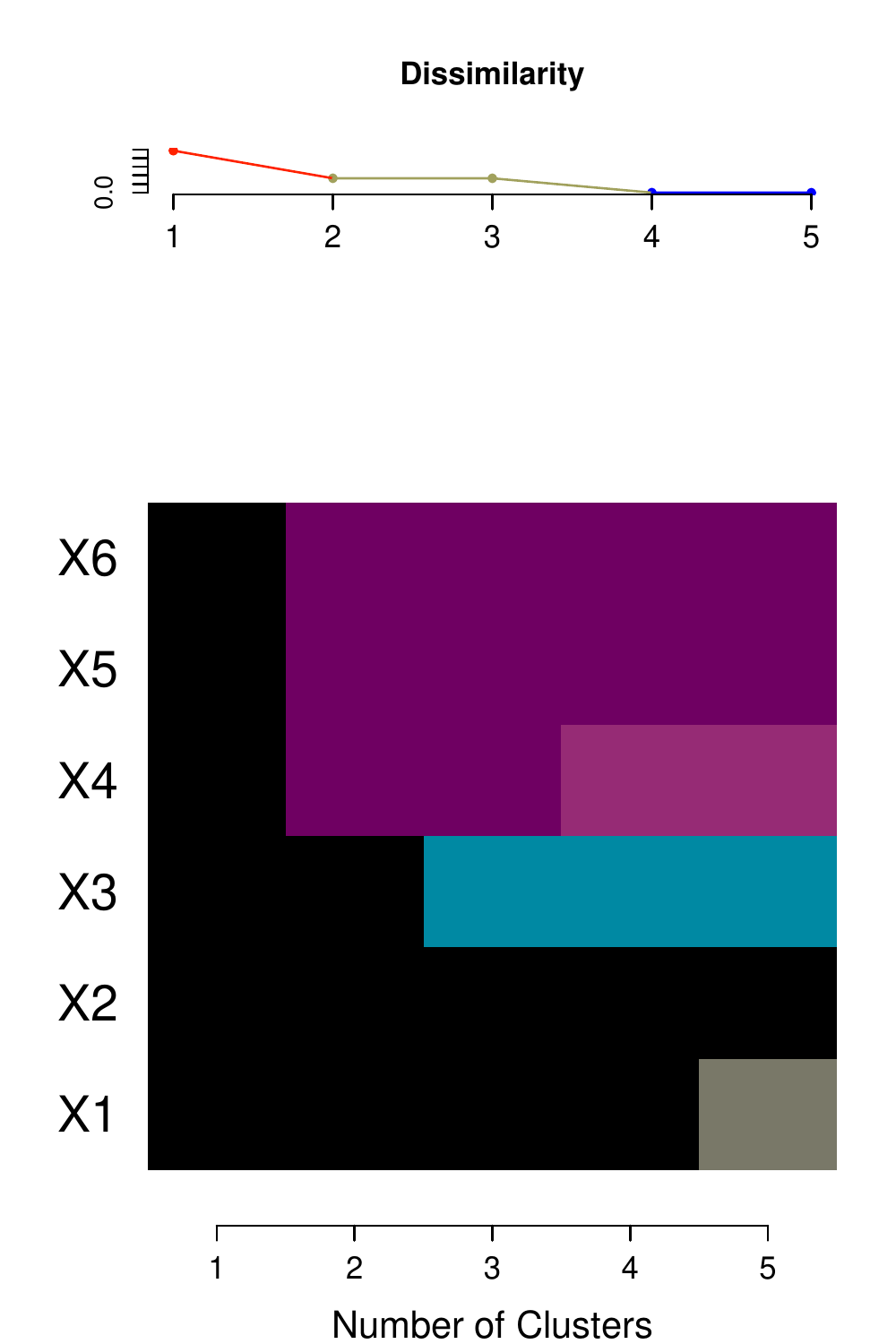}}
 \subfigure[HCC with $p=2$]{\includegraphics[scale=.4]{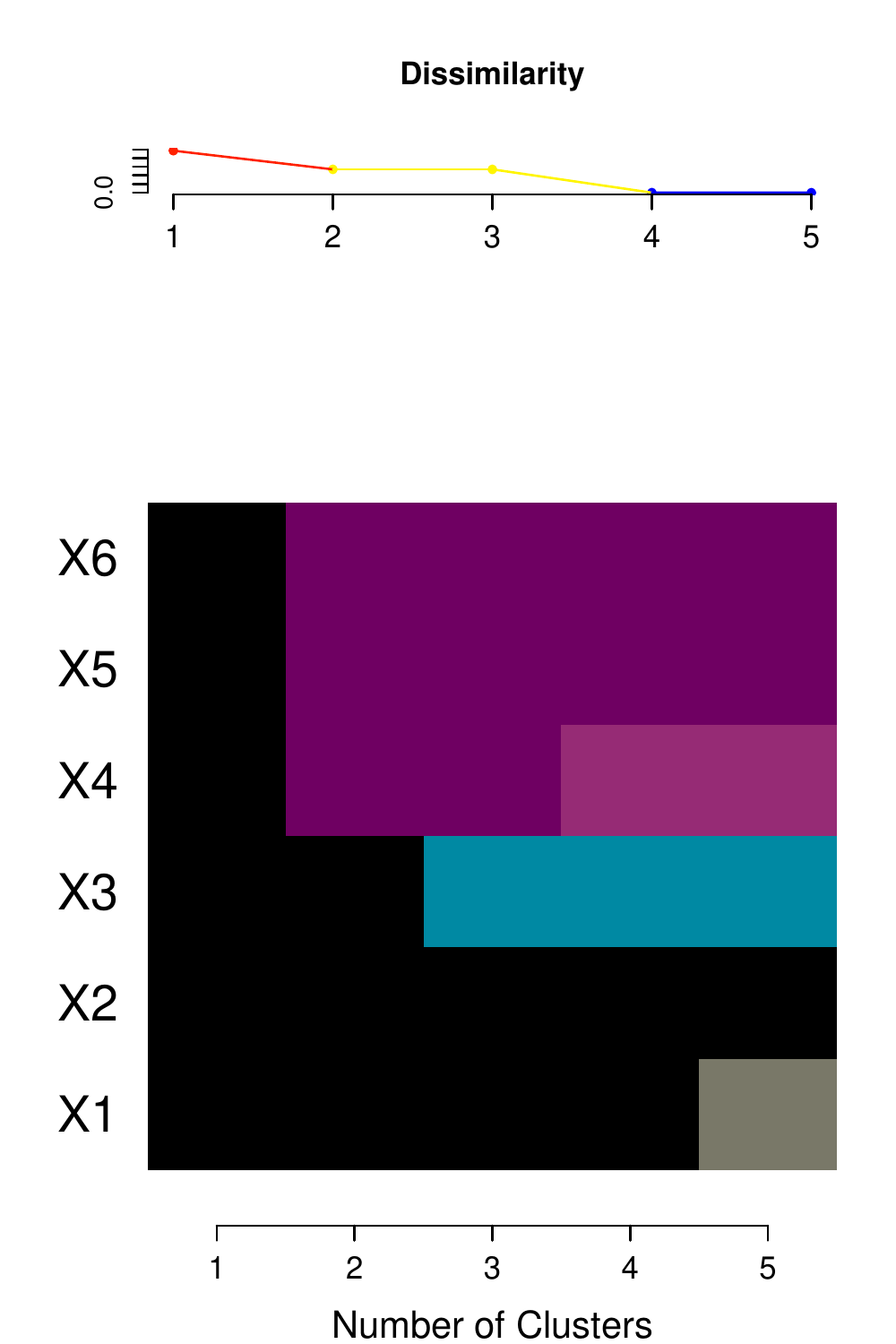}}
\subfigure[HSM Clustering]{\includegraphics[scale=.4]{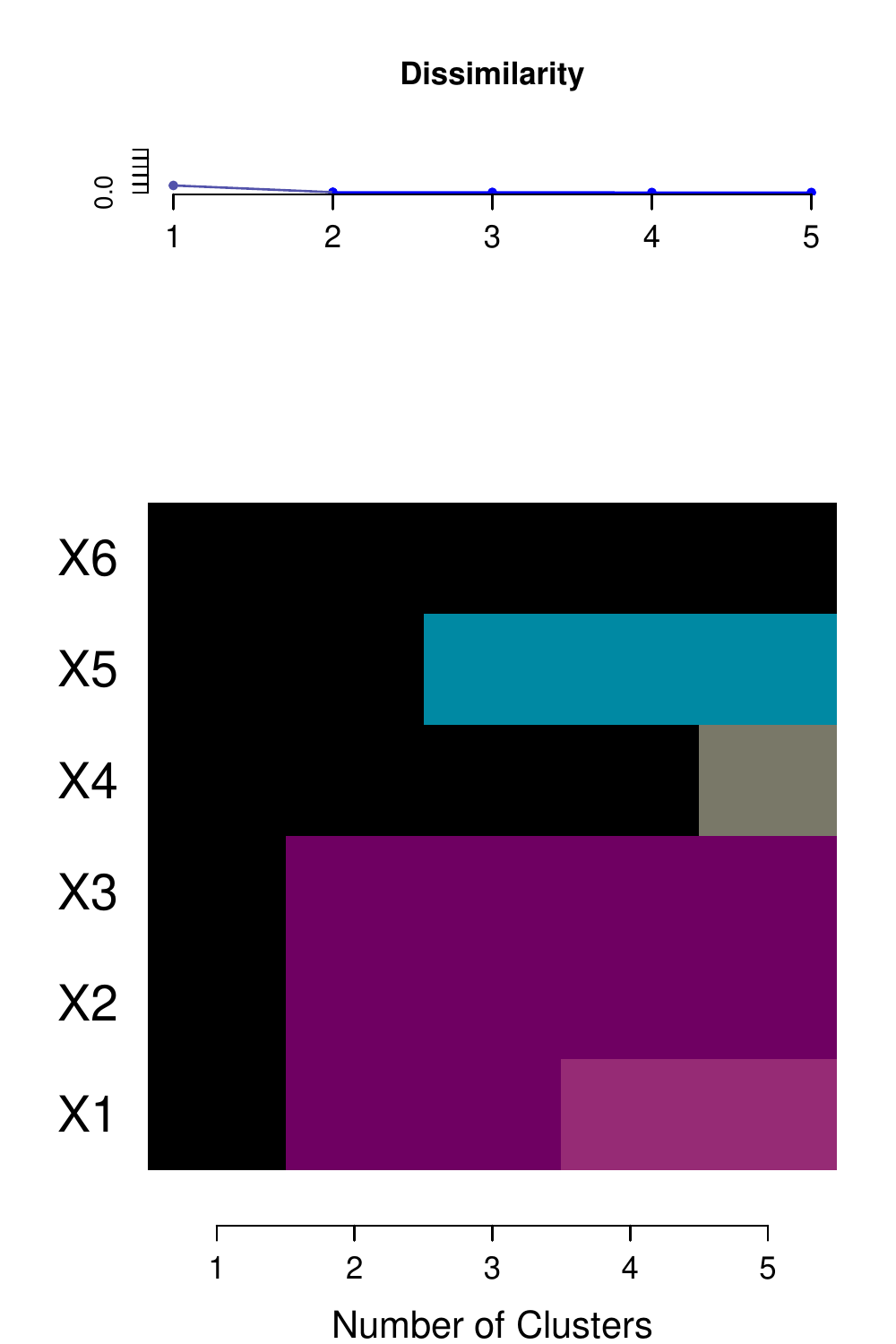}}
\caption{Clustering results for Experiment 1. Each figure represents the clustering 
results with different methods; the upper plot shows 
the minimum dissimilarity value and the lower plot represents which 
signals belong to the same cluster using colors.}\label{CR_E1}
\end{figure}
\textbf{Experiment 1}. Let $Z_1(t)$ and $Z_2(t)$ be two independent AR(2) latent processes with the same unimodal spectral density (peak at $2$ Hz). $\mathbf{X}(t)$ is a six-multivariate time series generated by  $X_i(t)=Z_1(t)+\varepsilon_i(t)$ 
for $i=1,2,3$, and $X_i(t)=Z_2(t)+\varepsilon_i(t)$ for $i=4,5,6$, where $\varepsilon_i(t)$ is white noise. 

We simulate time series of length $T=1000$, with a sampling frequency of $100$ Hz.
Figure \ref{CR_E1} shows the visualization plots for the clustering results. The HSM method identifies only 
one cluster that contains all six time series, while the HCC method divides the six time series into
 two clusters, $\{X_1,X_2,X_3\}$ and $\{X_4,X_5,X_6\}$.
All six time series in this case have the same spectral representation, which explains why the HSM method 
identifies only one cluster. 
However, there are two independent clusters which are only recovered by the HCC method.
The HSM method is not necessarily wrong; this method was proposed with different goals.

 \subsection{Comparison Between Coherence-Based Clustering Methods}
Now, we compare the HCC method with two other hierarchical clustering methods. 
The hierarchical methods receive as input the dissimilarity matrix $\mathbf{1}-\mathbf{C}(\omega)$, where $\mathbf{1}$ denotes the ones matrix.
The linkage functions considered are average (HAC) and complete (HMC). Notice that these linkage functions 
correspond to measure coherence between clusters as average and minimum coherence, 
respectively.

\noindent \textbf{Experiment 2}. Similar to Experiment 1, let $Z_1(t)$ and $Z_2(t)$ be two independent AR(2) latent processes with the same unimodal spectral density (peak at $2$ Hz) and let 
            $$A^T=
            \begin{pmatrix}
                 a_{1,1} & a_{2,1} &  a_{3,1} & a_{4,1} &  a_{5,1} & a_{6,1} \\
                a_{1,2} & a_{2,2} &  a_{3,2} & a_{4,2} &  a_{5,2} & a_{6,2} 
                \end{pmatrix} $$
be a coefficient matrix. Then, $\mathbf{X}(t)$ is a six-multivariate time series generated by  
$\mathbf{X}(t)= \mathbf{A}\mathbf{Z}(t)+\boldsymbol{\varepsilon}(t)$,
where $ \displaystyle \mathbf{Z}(t)=(Z_1(t), Z_2(t))^T$ and  
$\boldsymbol{\varepsilon}(t)$ is white noise. We simulate time 
series of length $T=1000$ with a sampling frequency of $100$ Hz 
and consider two cases for the coefficient matrix $A$, 
\begin{multicols}{2}
\small
\begin{itemize}
\item[Case 1:] $A^T=\begin{pmatrix}
                  1 & 1 & .2 & 0 & 0 & 0 \\
                  0 & 0 & 0 & 1 & 1 & .2 
                \end{pmatrix} $
\item[Case 2:] $A^T=\begin{pmatrix}
                 1 & .8 & .4 & 0 & .2 & .3 \\
                 0 & .2 & .1 & 1 & .8 & .2
                \end{pmatrix} .$

\end{itemize}
\end{multicols}
\normalsize

\begin{figure}
\centering
\subfigure[Case 1]{
\includegraphics[scale=.25]{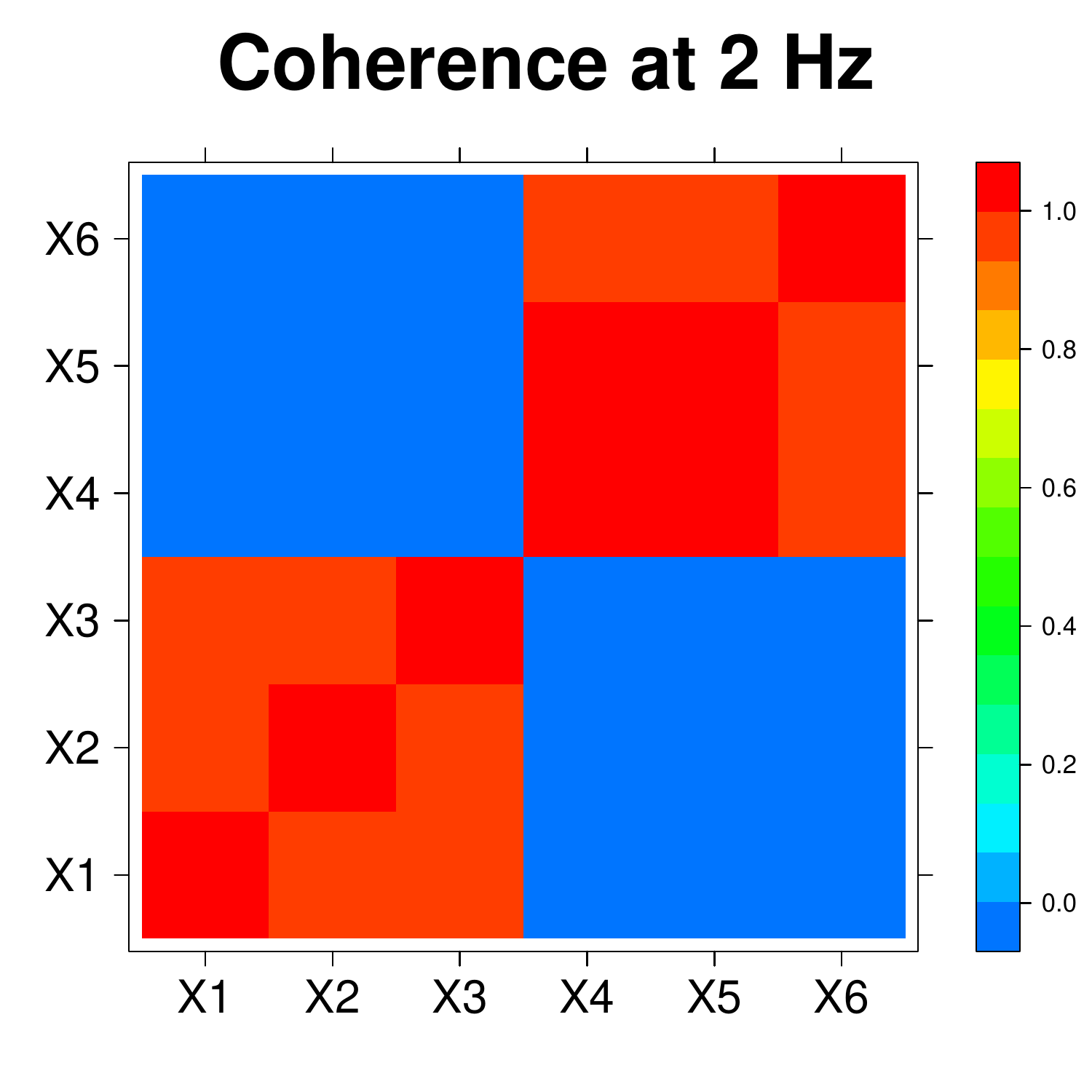}}
\subfigure[Case 2]{
\includegraphics[scale=.25]{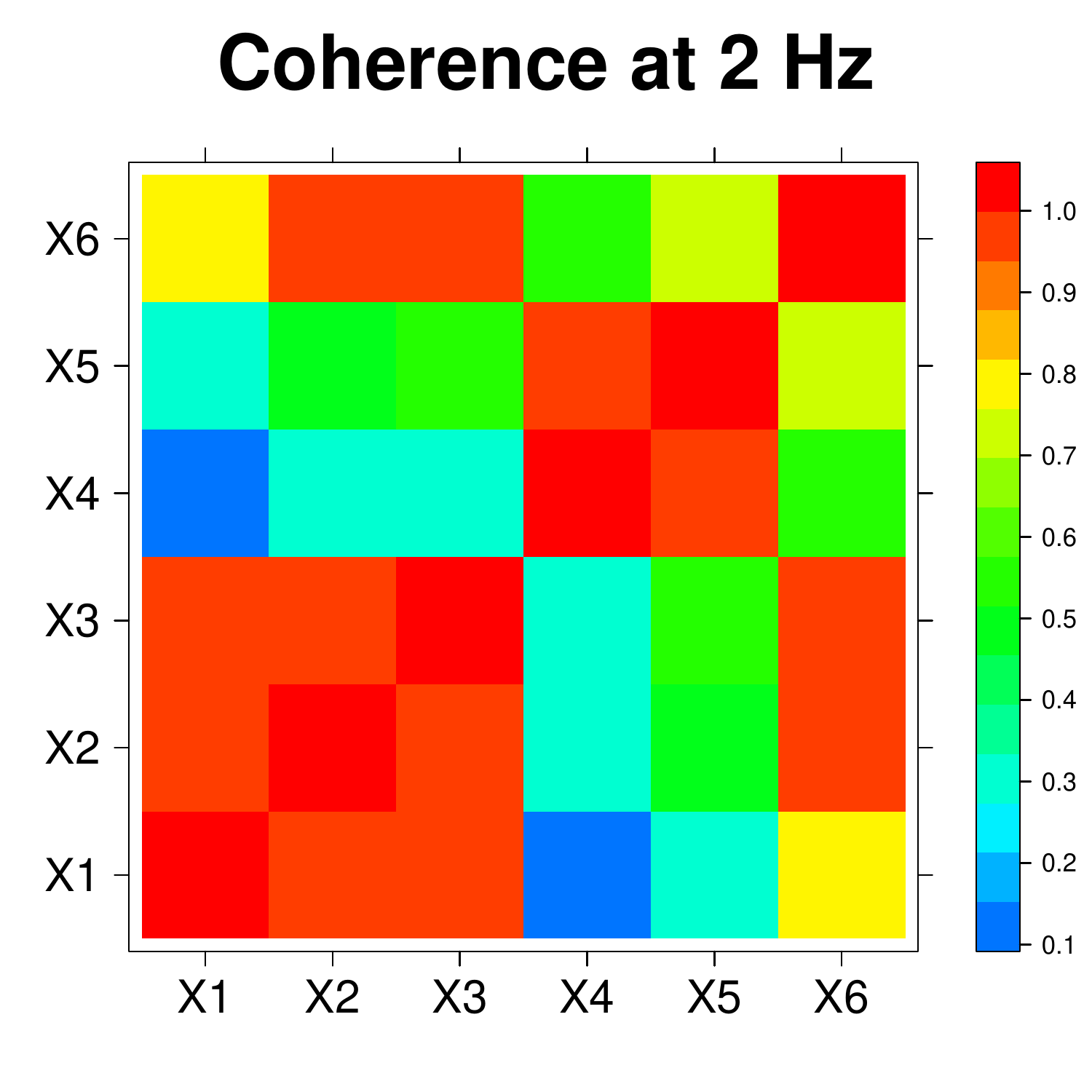}}
\caption{Estimated coherence of one simulation under Case 1 and Case 2 in Experiment 2.}\label{CohCases}
 \end{figure}

Figure \ref{CohCases} shows the estimated coherence matrix at $2$ Hz for one draw 
of each case, i.e., $\widehat{\mathbf{C}}(2)$.  Case 1 is an example of two well separated clusters, i.e.,
the coherence between any two members from different clusters is zero. In this case, 
$\{X_1(t),X_2(t)\}$ and $\{X_4(t),X_5(t)\}$ are highly correlated groups. 
$X_3(t)$ and $X_6(t)$ are less correlated with the members 
in these clusters, respectively. Case 2 is the most difficult case: $X_6(t)$ 
is only correlated with $X_5(t)$ but not with $X_4(t)$, and is perhaps 
more correlated with cluster $\{X_2(t),X_3(t)\}$. 
Compared to Case 1, Case 2 is more common in real applications.
 Our goal is to compare 
the clustering results using the different methods to show their advantages or disadvantages.
 
\begin{figure}
\centering
\subfigure[HCC] {\includegraphics[scale=.4]{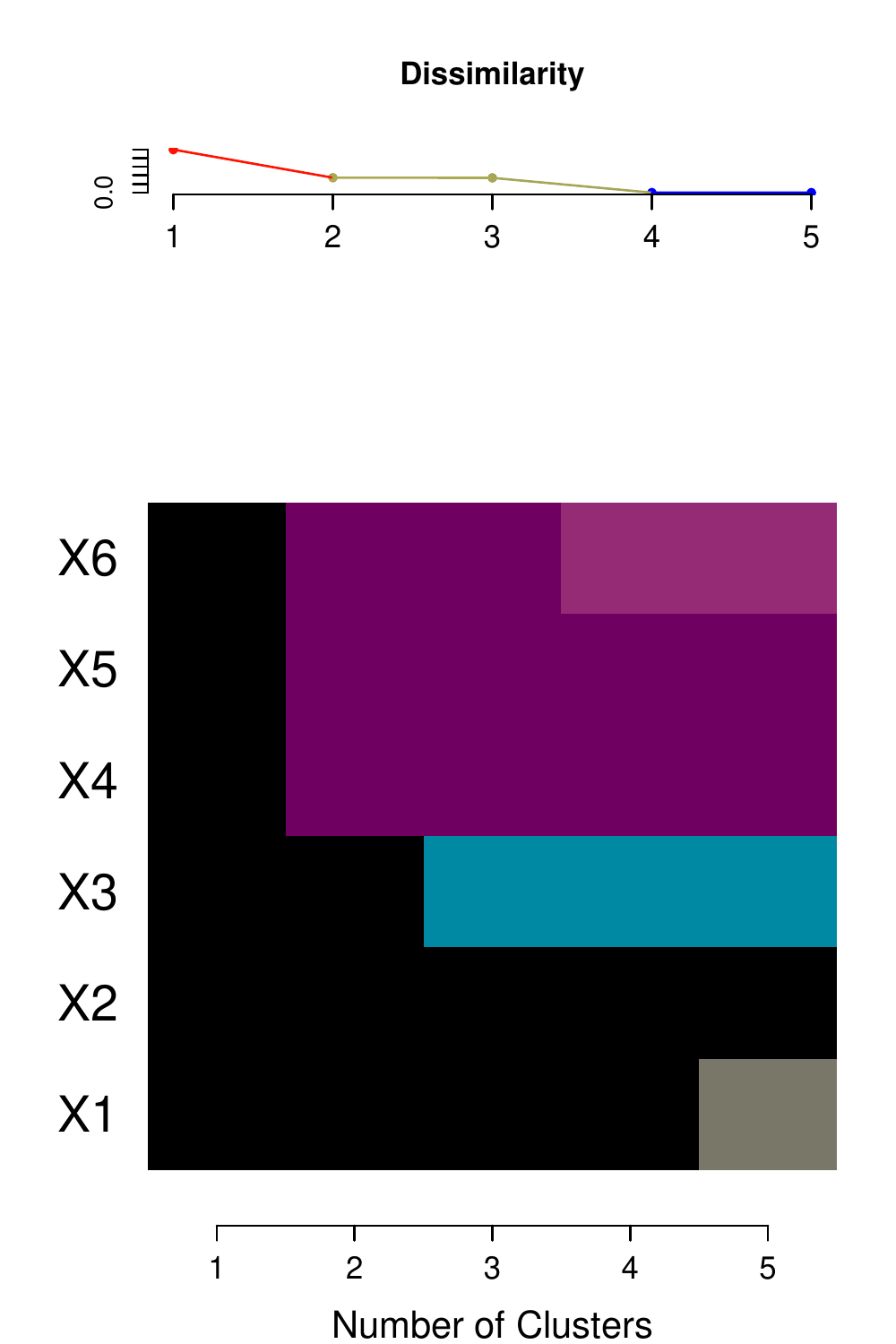}}
\subfigure[HAC] {\includegraphics[scale=.4]{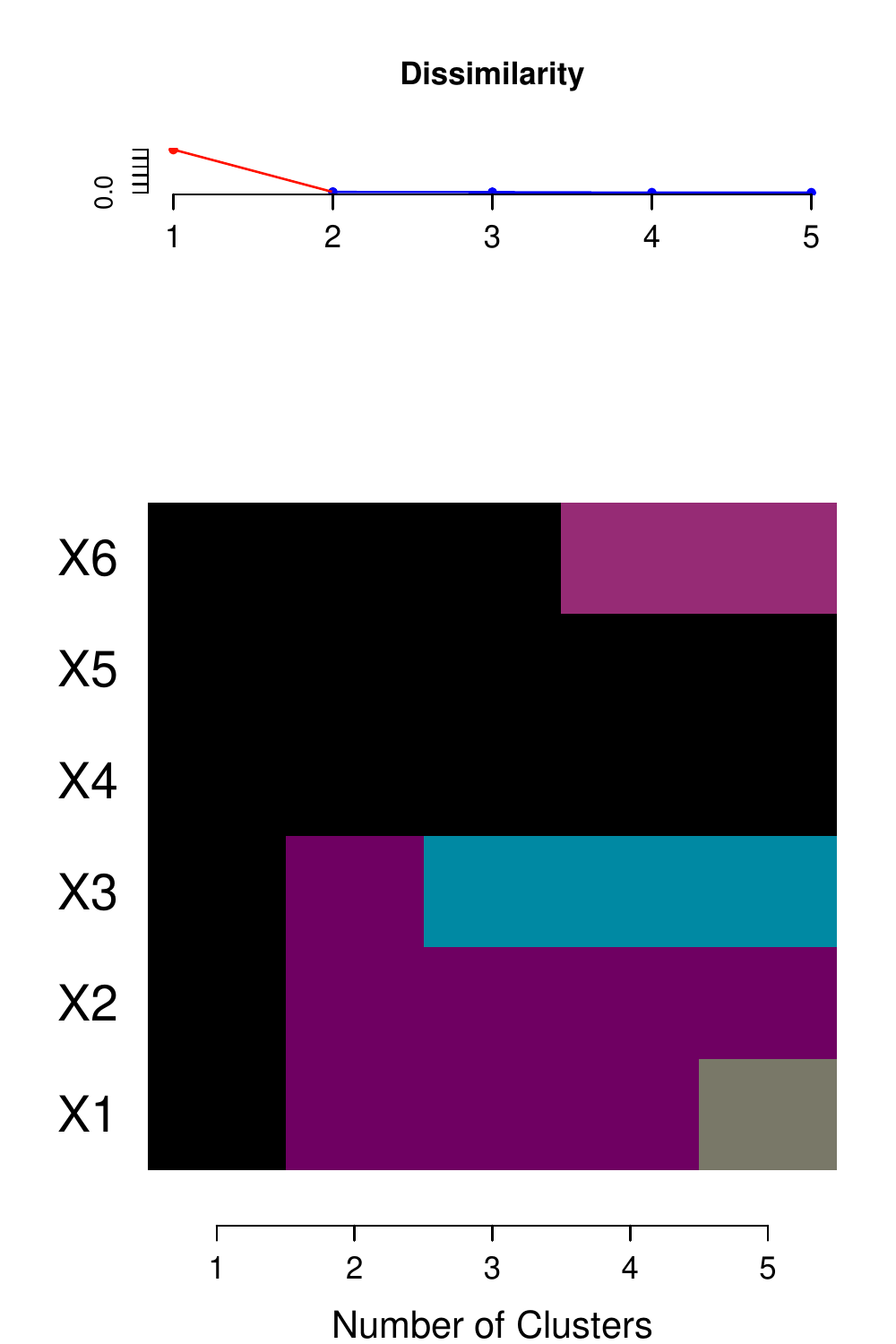}}
\subfigure[HMC] {\includegraphics[scale=.4]{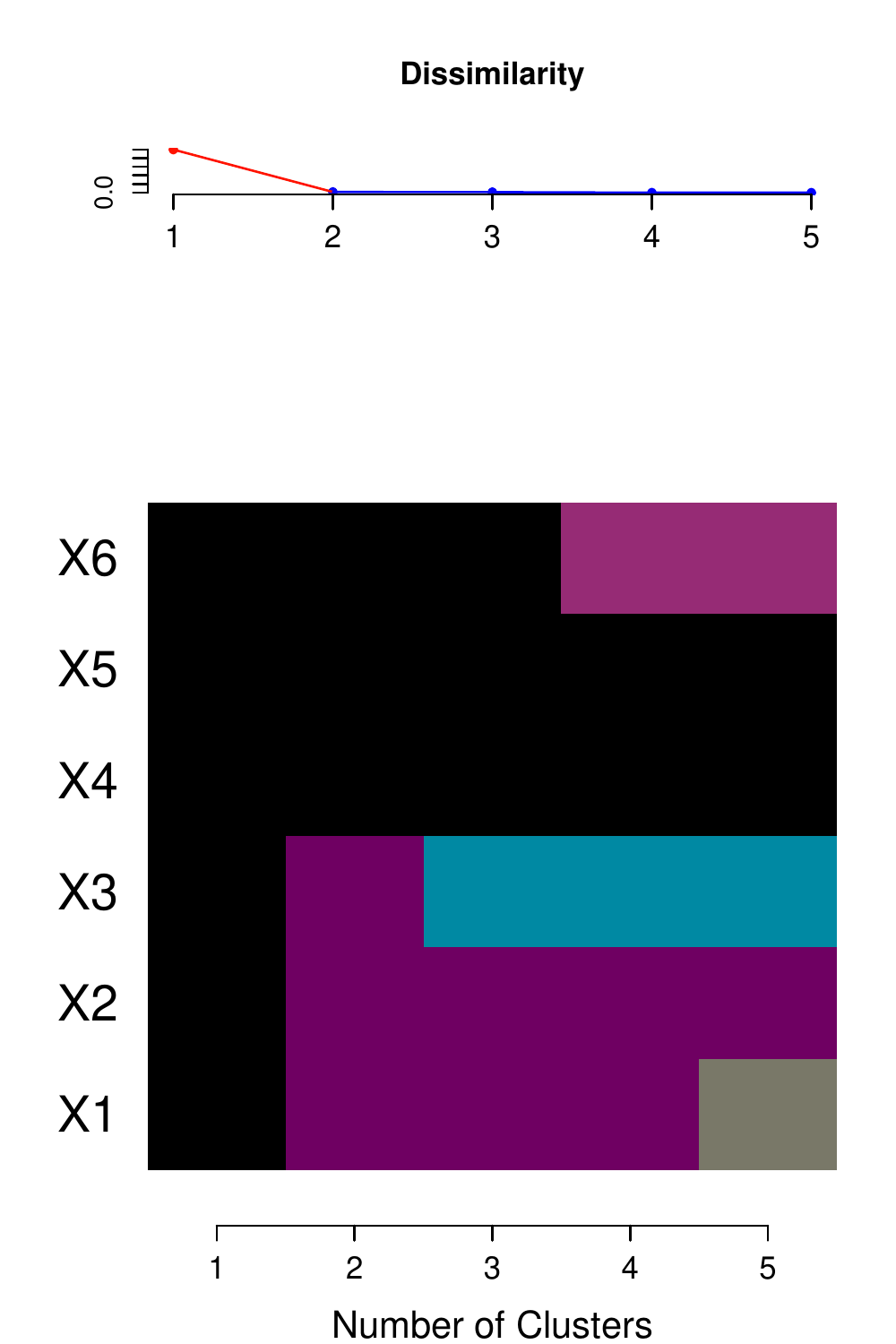}}
\caption{Clustering results for C1. Each figure represents the clustering results with different methods; the upper plot shows 
the minimum dissimilarity value and the lower plot represents which signals belong to same cluster using same colors. } \label{CR_C1}
\end{figure}

\begin{figure}
\centering
\subfigure[HCC] {\includegraphics[scale=.4]{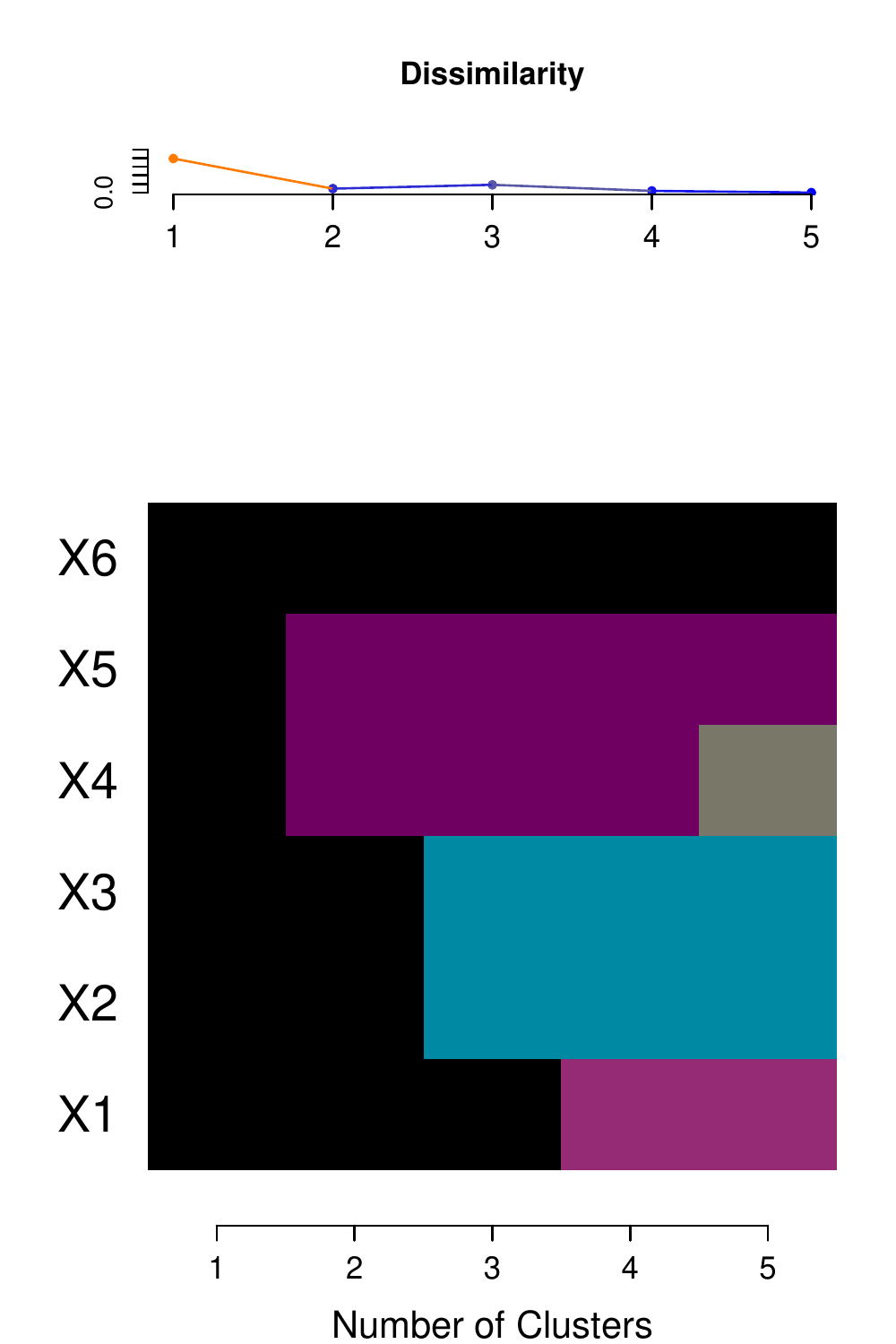}}
\subfigure[HAC] {\includegraphics[scale=.4]{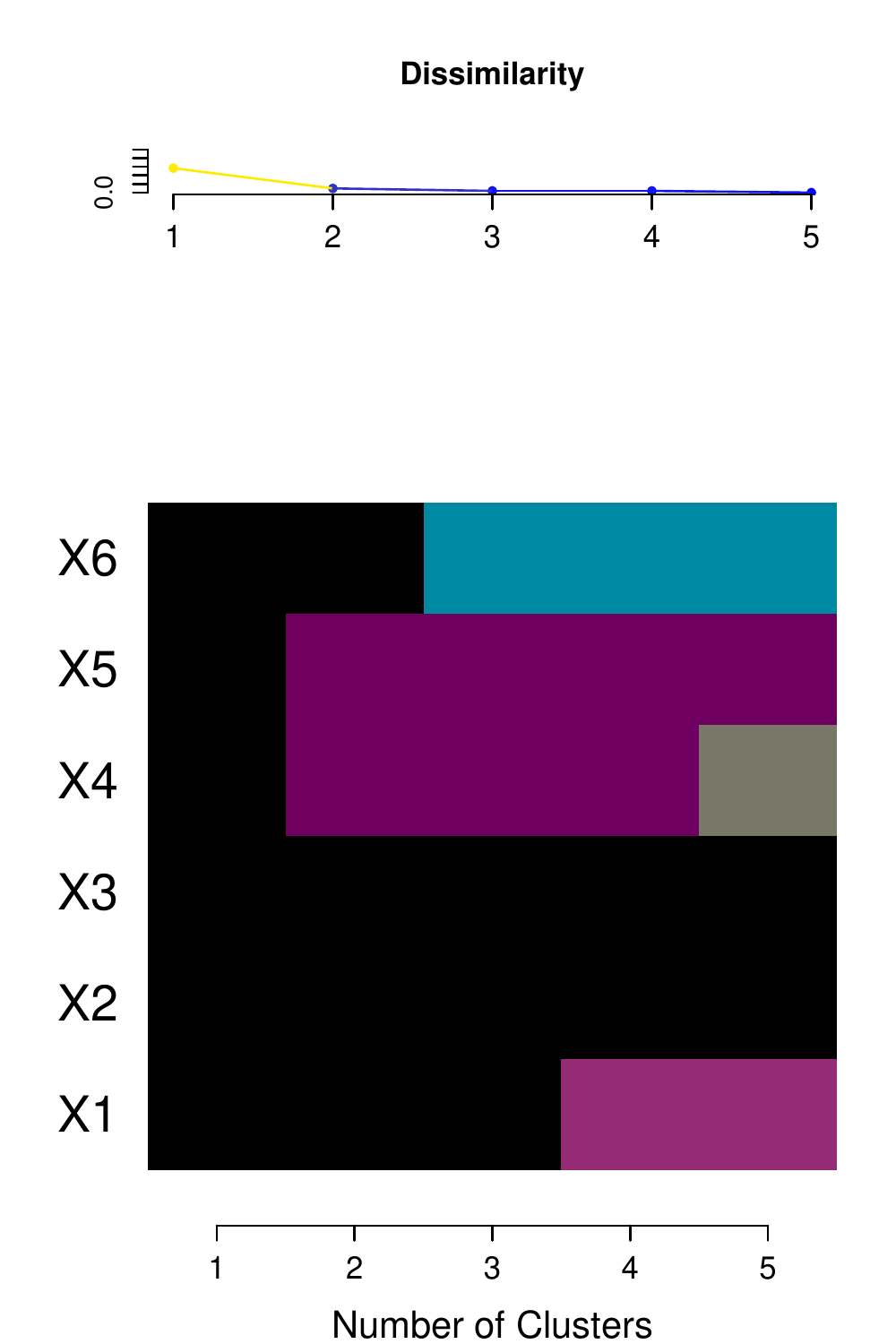}}
\subfigure[HMC] {\includegraphics[scale=.4]{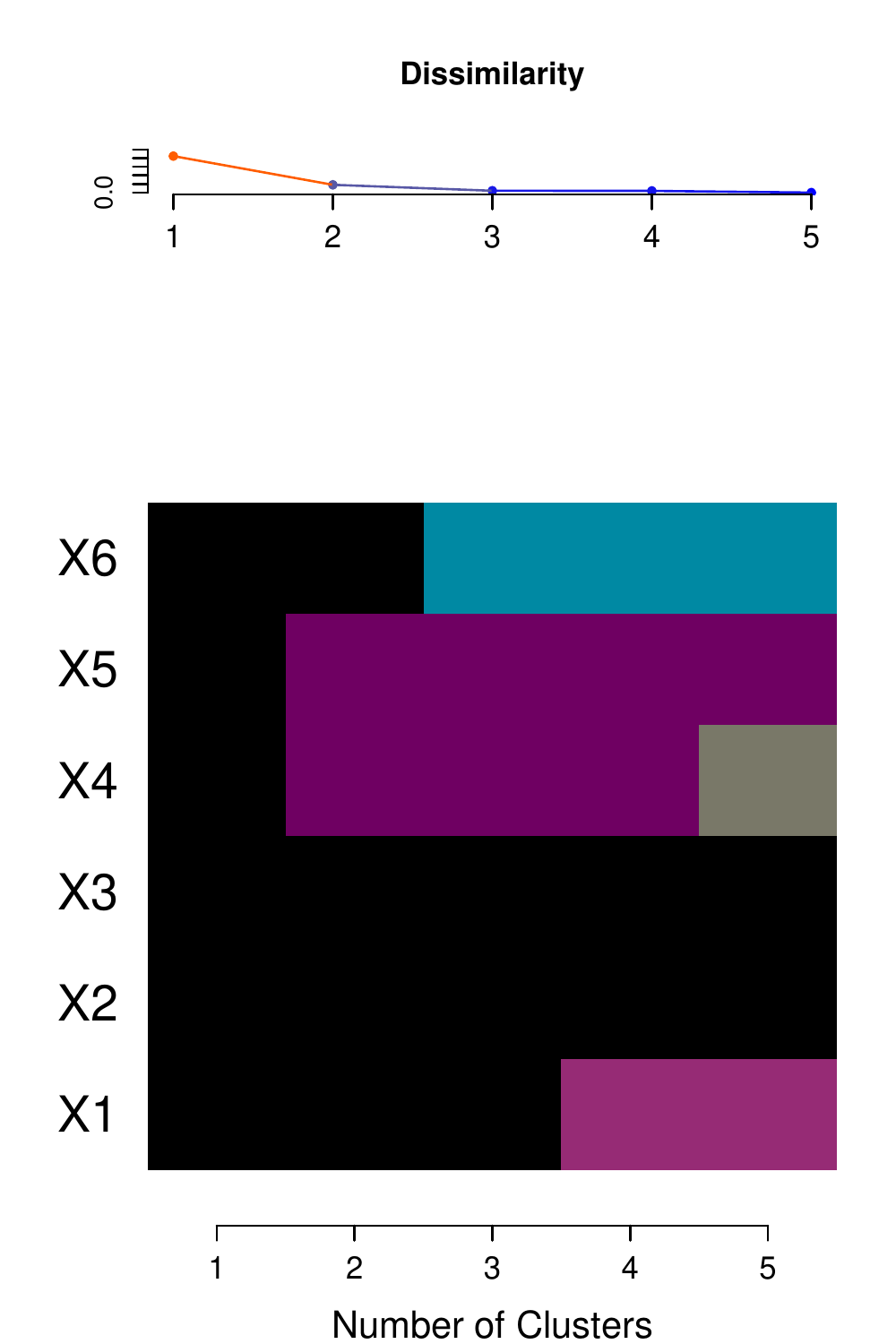}}
\caption{Clustering results for C2. Each figure represents the clustering results with different methods; the upper plot shows 
the minimum dissimilarity value and the lower plot represents which signals belong to same cluster using same colors.  } \label{CR_C2}
\end{figure}

Figures \ref{CR_C1} and \ref{CR_C2} show the clustering results for each case.  
Under these two cases, the three methods, HCC, HAC and HMC, perform equally well. All the three methods
detect the presence of two clusters. The resulting clusters 
correspond to the expected results. We perform this experiment multiple times, M=1000 
replicates, and the results are very consistent. In this sense, we did not find a substantial difference 
between the results from different methods under these simulation settings. However, the HCC method is preferable
over the others since it provides directly interpretable results.

Since we are motivated by the study of EEG signals, it is important to explore the \underline{scalability} of our method. 
In other words, which is the performance of the HCC method if we have a bigger number of time series or EEG channels. 
We consider the following example with 128 time series which is a usual size of a multichannel EEG.

\noindent \textbf{Experiment 3}. Let $Z_j(t)$ be an AR(2) latent process for $j=1,\dots,5$, with the unimodal spectral density concentrated around
$2,6,10,15$ and $40$ Hz, respectively. Each $Z_j(t)$ represents a latent process in the different frequency bands (delta, theta, alpha, beta, and gamma).
$\mathbf{X}(t)$ is a 128-multivariate time series generated by  
$$\mathbf{X}(t)= \mathbf{A}\mathbf{Z}(t)+\boldsymbol{\varepsilon}(t),$$
where $ \displaystyle \mathbf{Z}(t)=(Z_1(t), \ldots, Z_5(t))^T$ and 
$\boldsymbol{\varepsilon}(t)$ is white noise. The rows of the coefficient matrix $\mathbf{A}$ are built as follows:
$A_{l.}= c(1,.2,0,0,0)$ for $l=1,\ldots,25$, $A_{l.}= c(0,1,0,0,0)$ for $l=26,\ldots,50$, 
$A_{l.}= c(0,.2,1,0,0)$ for $l=51,\ldots,75$, $A_{l.}= c(0,0,0,1,0)$ for $l=76,\ldots,100$ and 
$A_{l.}= c(0,0,0,0,1)$ for $l=101,\ldots,128$.

\begin{figure}
\centering
\subfigure[Coherence matrix]{
\includegraphics[scale=.25]{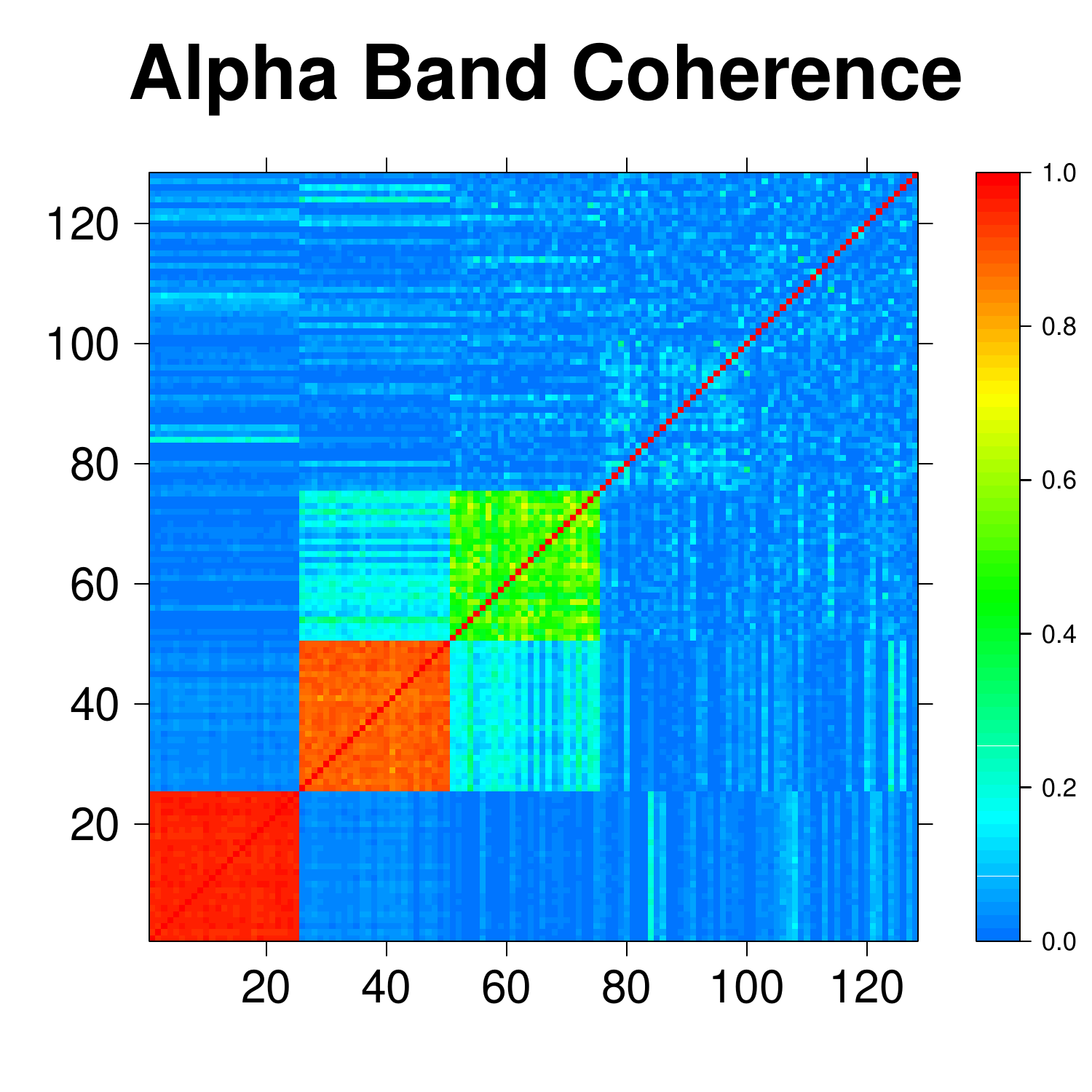}}
\subfigure[Coherence submatrix]{
\includegraphics[scale=.25]{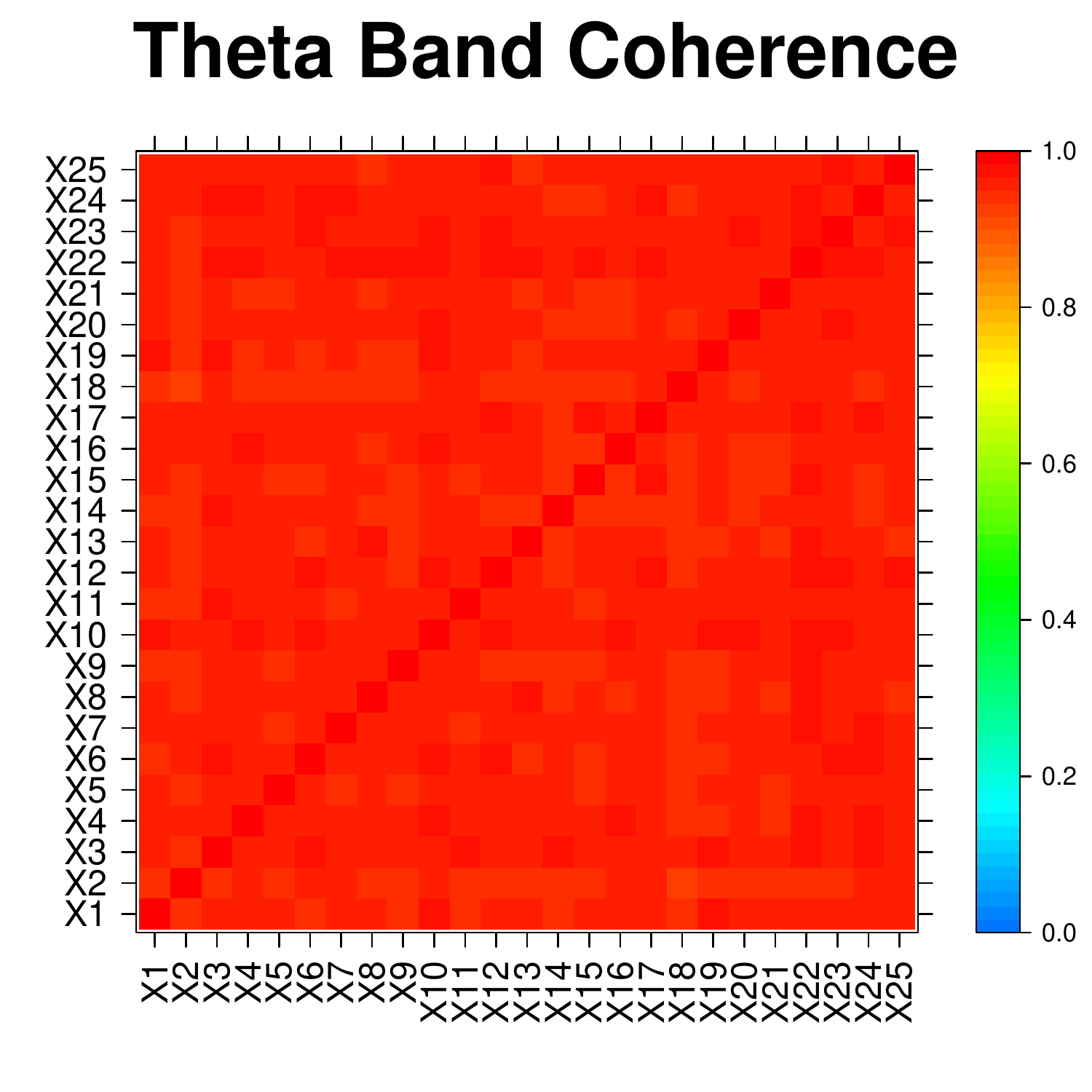}\includegraphics[scale=.25]{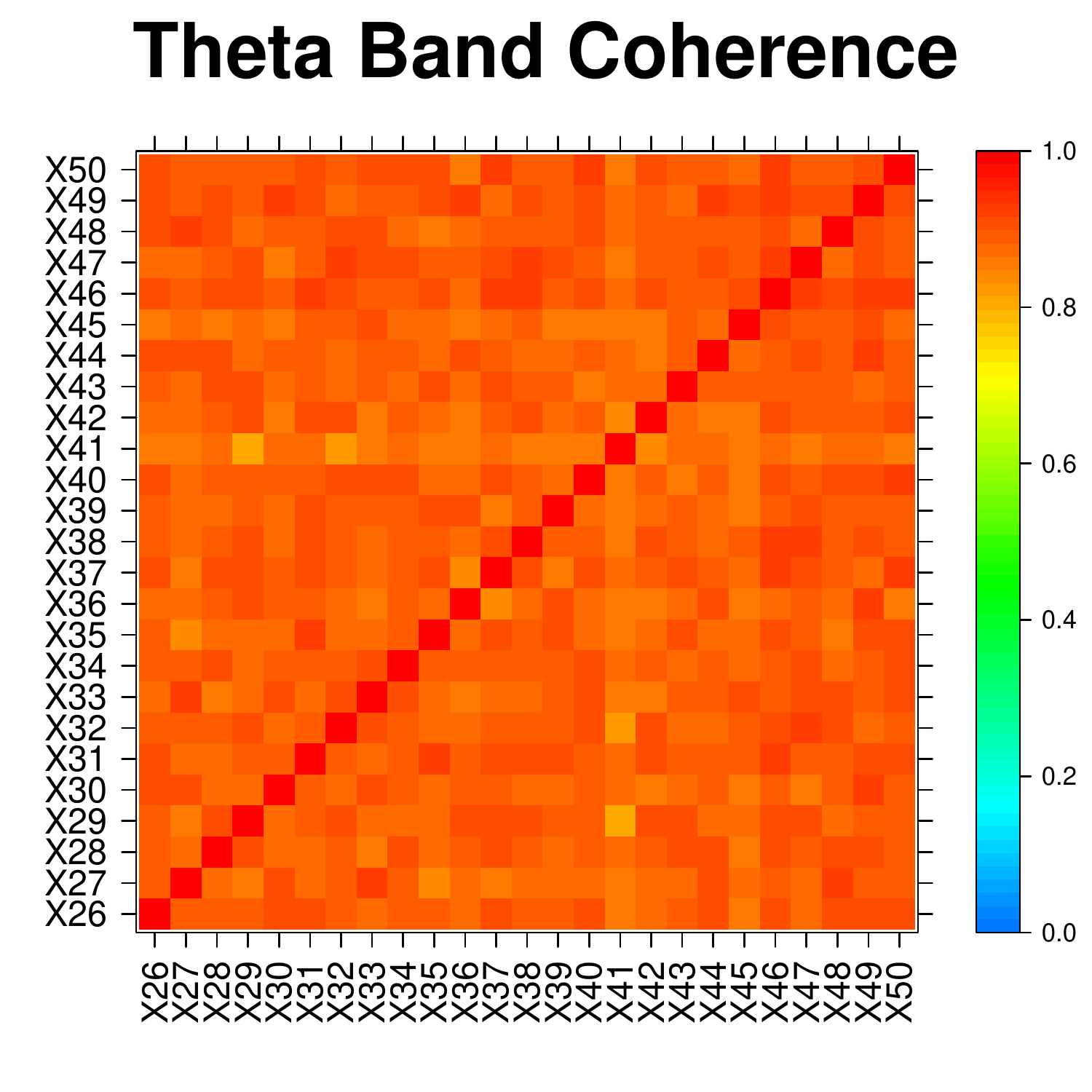}\includegraphics[scale=.25]{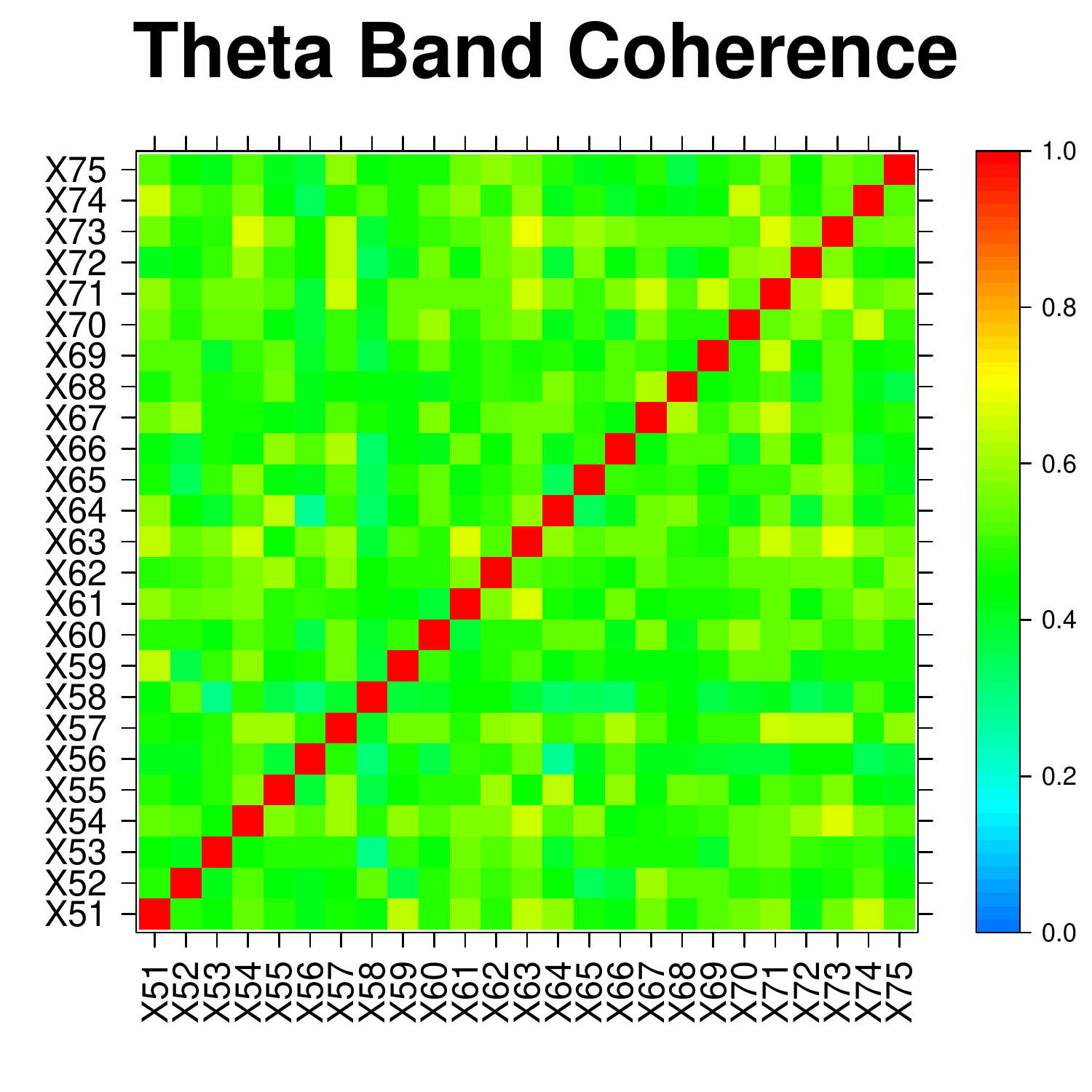}}
\caption{Estimated coherence of one simulation of Experiment 3.}\label{Coh128Series}
 \end{figure}

\begin{figure}
\centering
\subfigure[HCC] {\includegraphics[scale=.25]{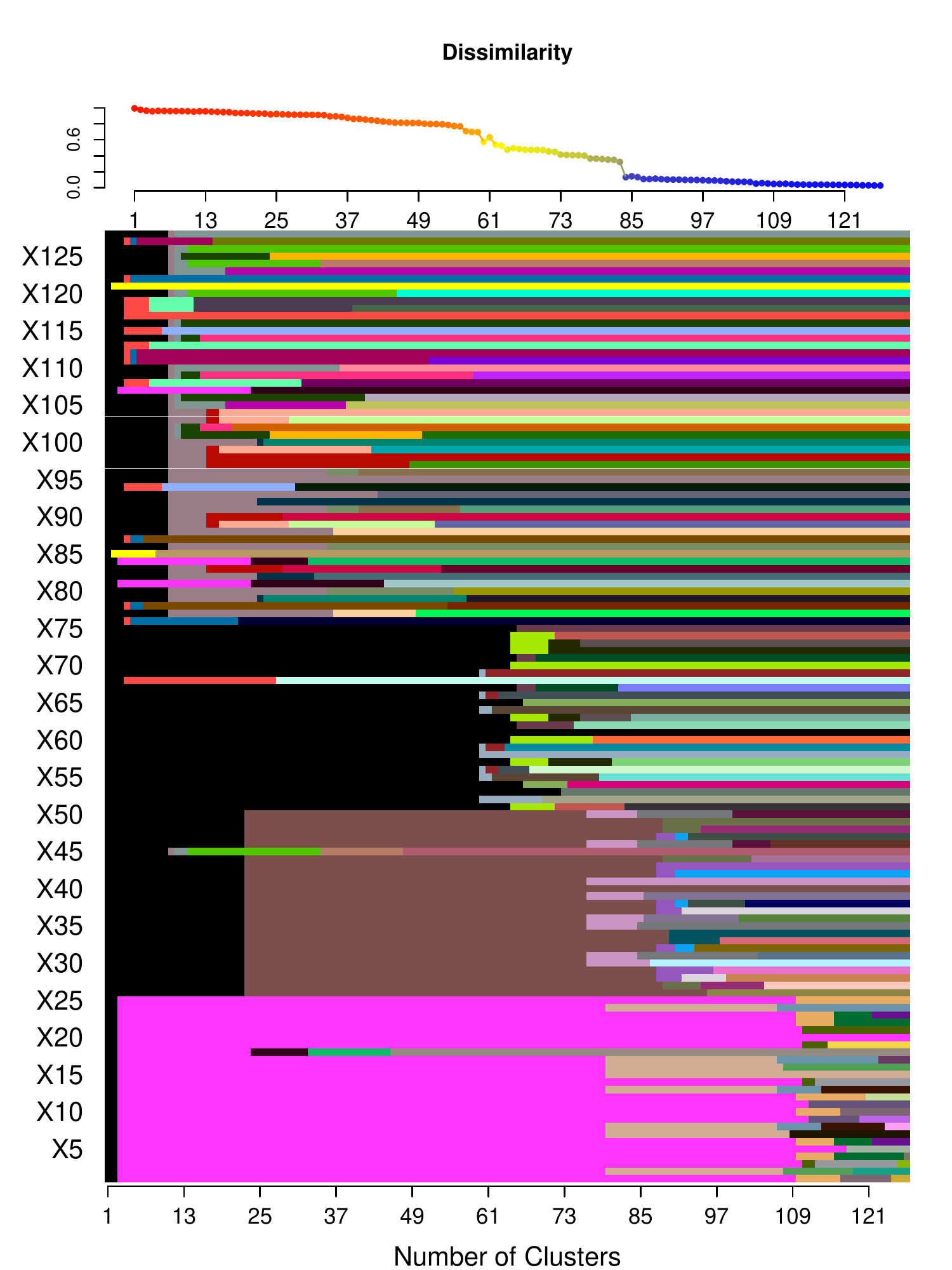}}
\subfigure[HAC] {\includegraphics[scale=.25]{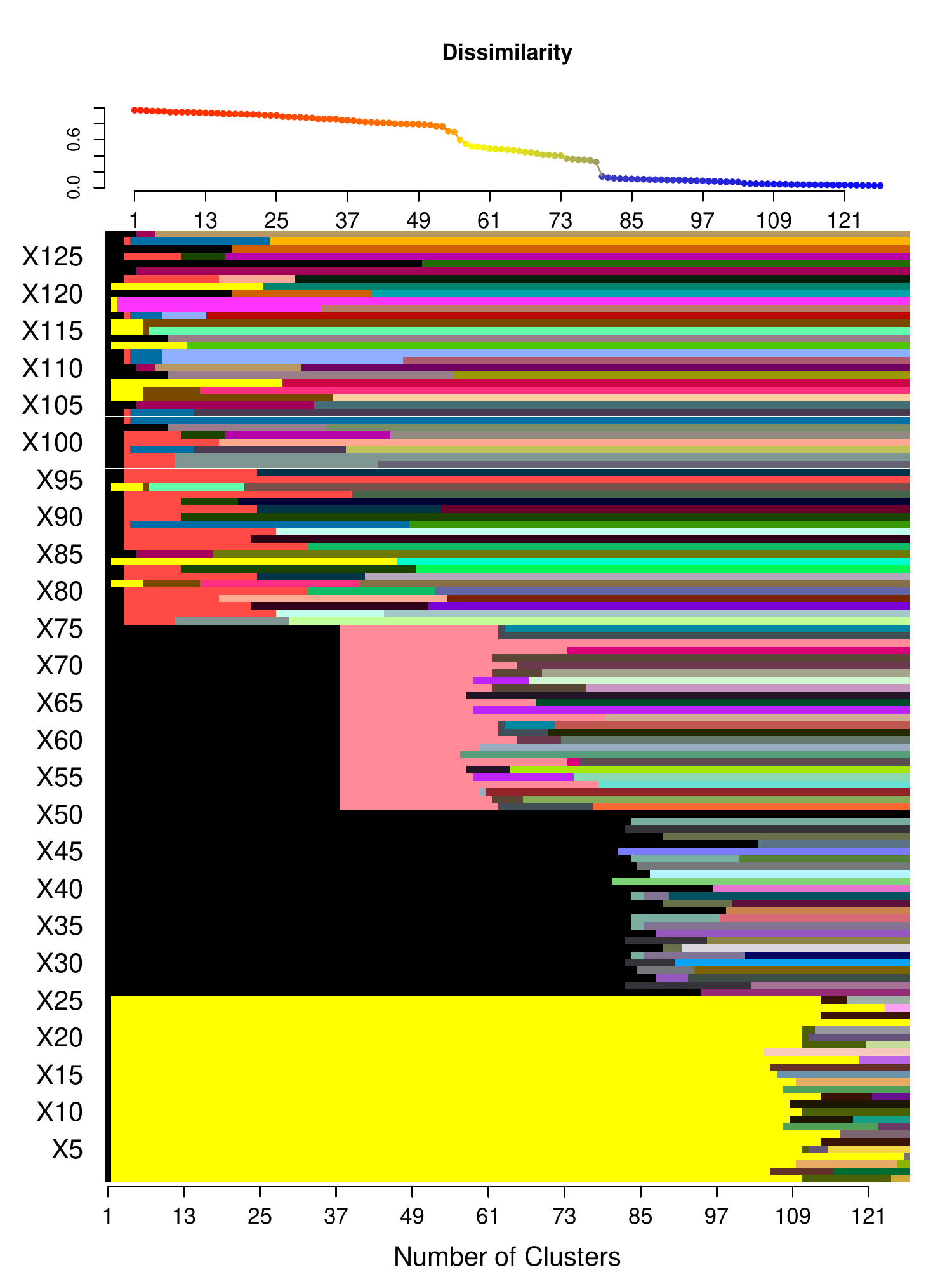}}
\subfigure[HMC] {\includegraphics[scale=.25]{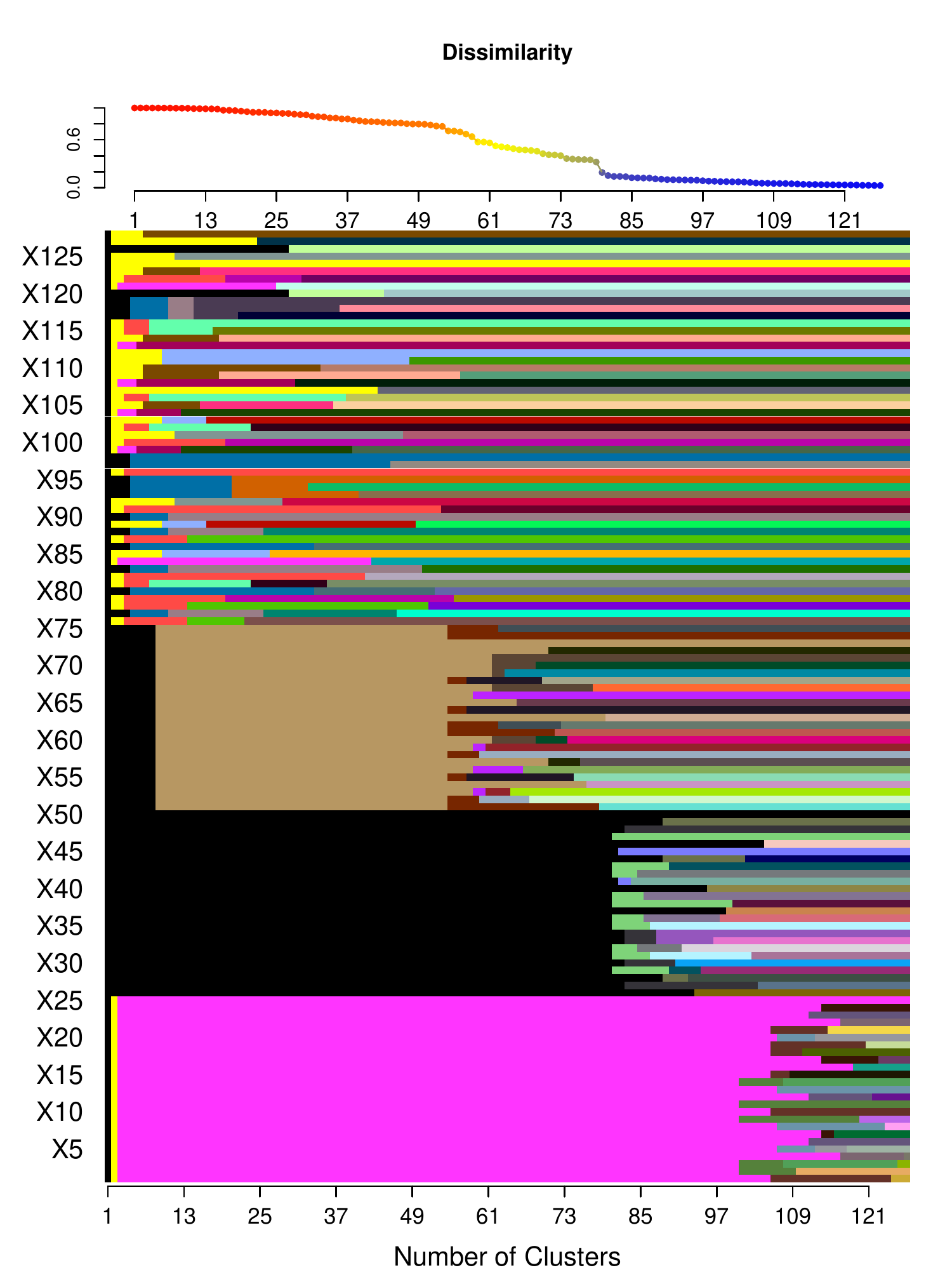}}
\caption{Clustering results for Experiment 3. Each figure represents the clustering results with different methods; the upper plot shows 
the minimum dissimilarity value and the lower plot represents which signals belong to same cluster using same colors.  } \label{CR_128Series}
\end{figure}

We simulate time series of length $T=1000$ with a sampling frequency of $100$ Hz.  Figure \ref{Coh128Series} shows the estimated 
integrated coherence on beta band. On beta band, there are two highly correlated clusters $C_1=\{X_1(t),\ldots ,X_{25}(t)\}$ 
and $C_2=\{X_{26}(t),\ldots ,X_{50}(t)\}$. There is a third cluster $C_3=\{X_{51}(t),\ldots ,X_{75}(t)\}$ which has lower within correlation.
Figure  \ref{CR_128Series} shows the clustering result for experiment 3. Overall, the three methods show similar results. These three main clusters 
were recovered by the HCC, HAC and HMC methods. One difference is that if we underestimate the number of clusters, 
the HAC method will merge (when $k=37$) $C_3$ and $C_2$ before the HCC method (when $k=23$). On the other hand, the HMC method 
will merge (when $k=8$) $C_3$ and $C_2$ after the HCC method. This reflects that average 
coherence could overestimate the within cluster correlation. In contrast, cluster coherence measures more reasonable the within cluster 
dependency without being too restrictive as HMC. Here, we show one simulation of this experiments but the clustering results were very consistent through 
different replicates, similar to Experiment 2. 

We consider a model of spatially correlated time series. 
In \textbf{Experiment 2} and \textbf{Experiment 3}, the coefficient matrix values were fixed values.
In \textbf{Experiment 4}, the coefficients vary depending on the spatial location
of the signal, $X_s(t)$, with respect to the spatial location of the latent process, $Z_{s_i}(t)$.
This case is a realistic model for EEG data.

\begin{figure}
\centering
\subfigure[EEG channels \label{SimEEG1}]{\includegraphics[scale=.32]{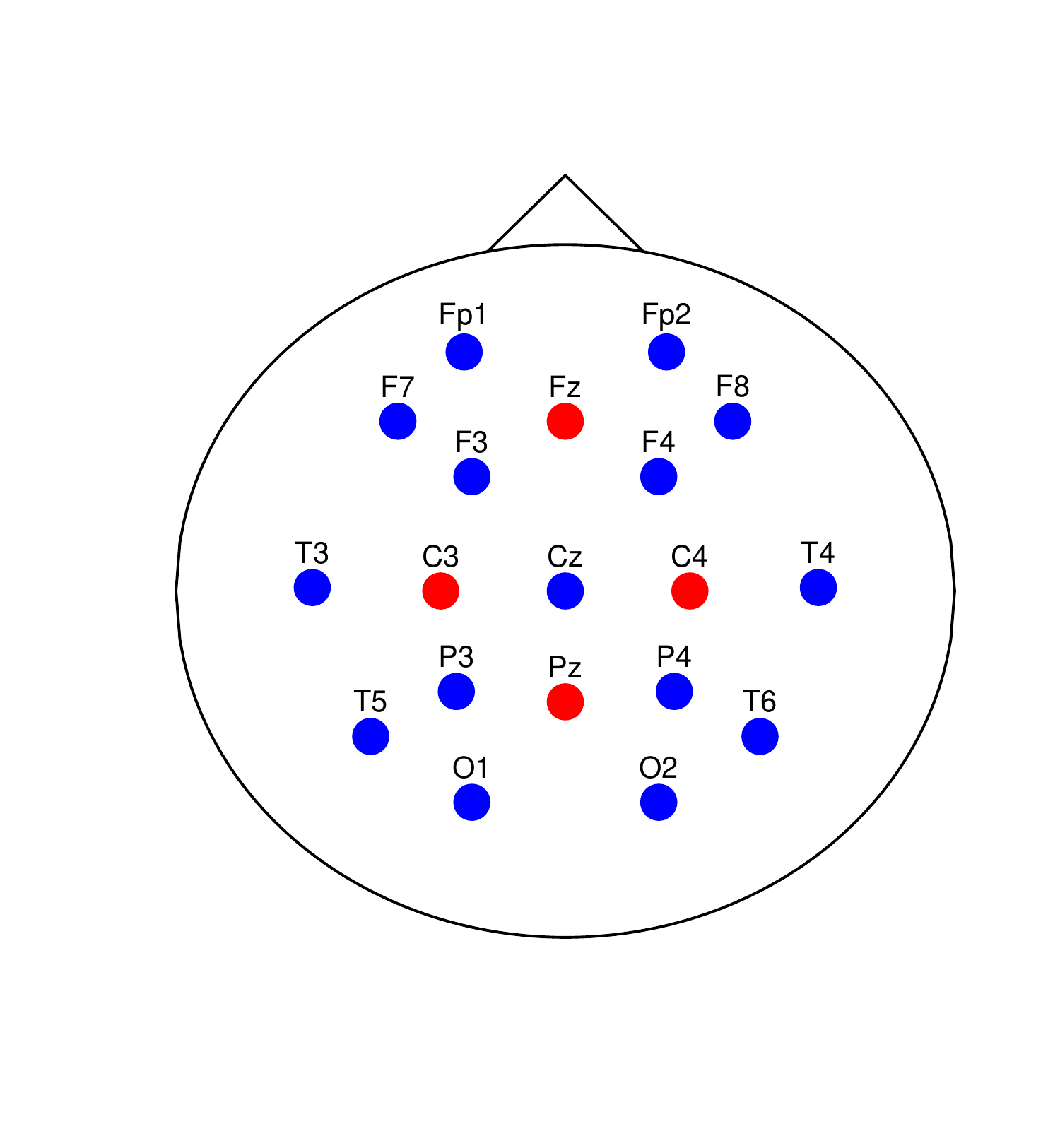}}
\subfigure[Integrated Coherence\label{SimEEG3}]{\includegraphics[scale=.32]{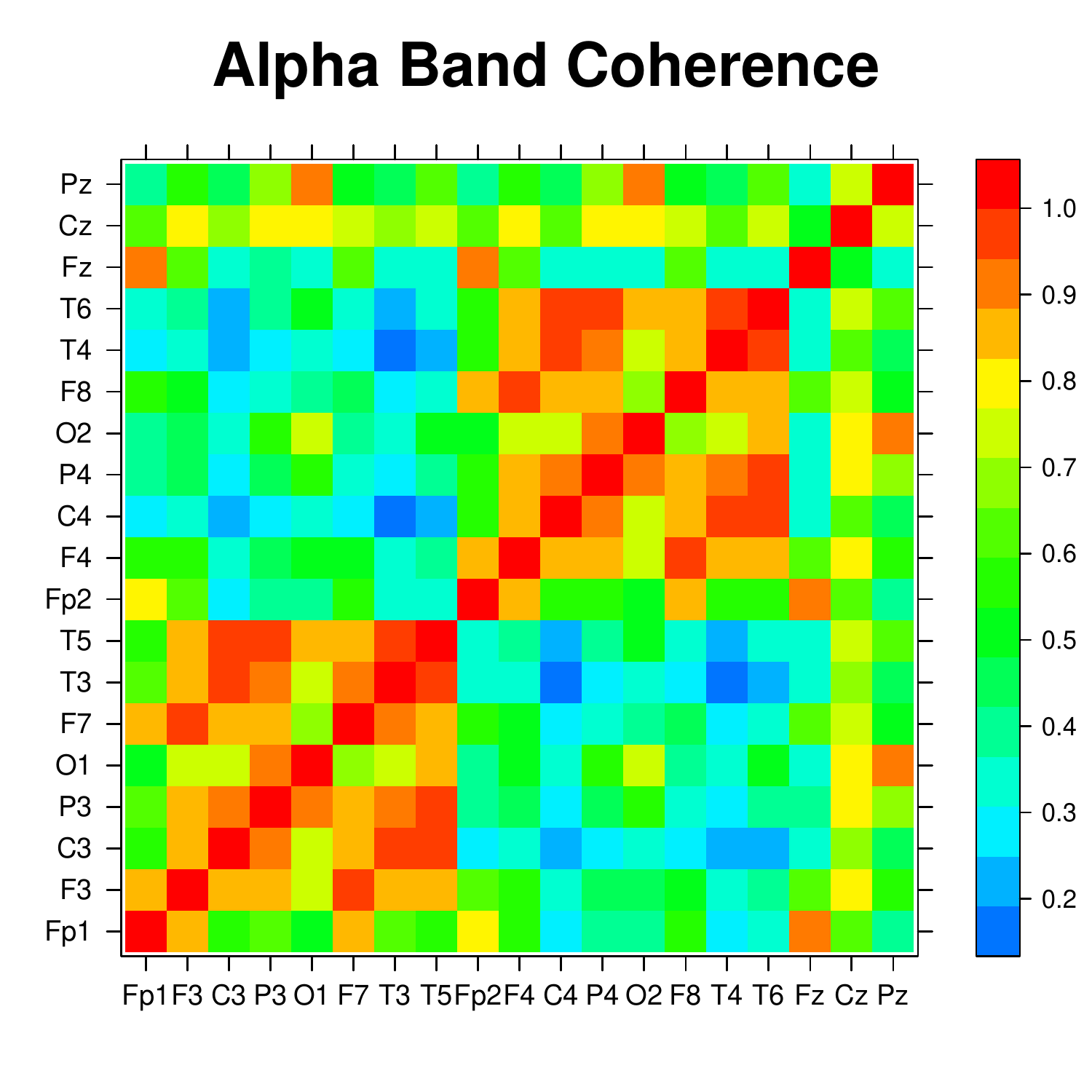}}
\subfigure[Scree plot \label{MVE3}] {\includegraphics[scale=.3]{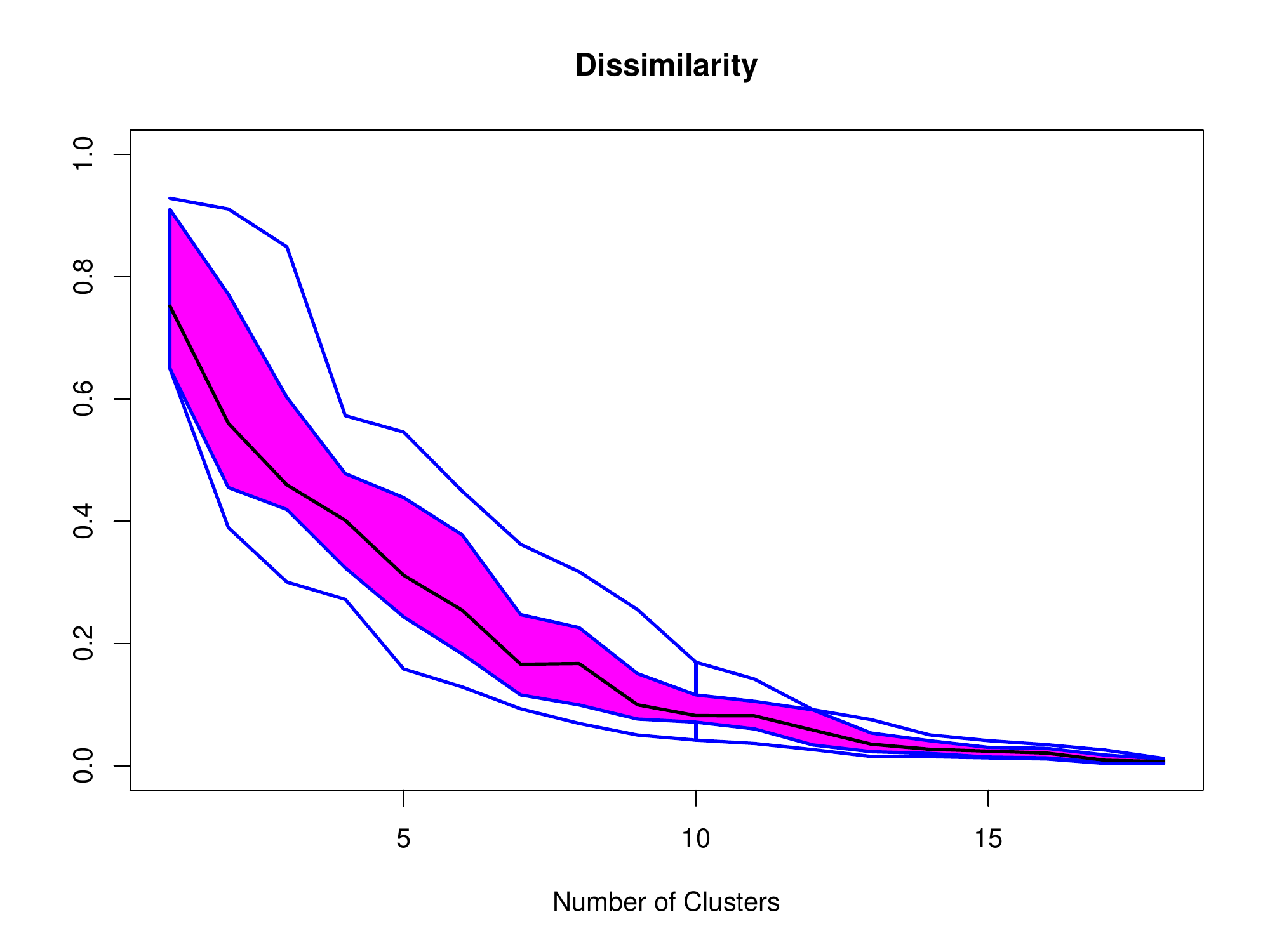}}
\caption{(a) Location of the EEG channels on the scalp (2D projection). Location of latent signals are highlight in red. (b) Integrated coherence on the alpha band (average over the 100 replicates). (c) Functional box plot of the scree plot obtained from 100 replicates.}
\end{figure}

\noindent \textbf{Experiment 4}. Consider a multichannel EEG data with $19$ channels. 
Let $X_s(t)$ be the EEG signal at channel $s$. Figure \ref{SimEEG1} shows the 
spatial location of the channels in a $2D$ projection. 
We assume that $\{X_s(t)\}$ is generated by a mixture of independent time series $Z_{s_i}(t)$, $i=1,\ldots,4$, 
located at $C3$, $C4$, $Pz$ and $Fz$ channels, respectively. 

To simulate $X_s(t)$: First, we simulate four independent AR(2).  
$Z_{s_1}(t)$ and $Z_{s_2}(t)$ are located at $C3$ and $C4$ channels 
and have a spectral density with power concentrated at 9 Hz. 
 $Z_{s_3}(t)$ and $Z_{s_4}(t)$ are located at $Pz$ and $Fz$ 
 channels and have a spectral density with power concentrated 
 at 10 Hz. Then, we compute the mixture as follows,
$$X_s(t)=\sum_{i=1}^{4} a(s,s_i) Z_{s_i}(t),$$ 
with $a(s,s_i)=\exp\left(-\frac{|| s-s_i ||}{\kappa}\right)$, where $\kappa = 1/3$. We consider  $T=1000$ time
points and $M=100$ replicates of this 
experiment.

We applied the three clustering methods to the simulated EEG signals for the alpha band, 8-12 Hz. 
Figure \ref{SimEEG3} shows the integrated coherence on the alpha band,
averaged over the 100 replicates. There are two highly correlated clusters located on 
different sides of the brain, the left central with the left temporal channels, and the right central with the right temporal
channels. We expect that channels located near to 
same source will be clustered together. 
Figure \ref{MVE3} presents the functional boxplot of the scree plot obtained by the HCC method. This plot suggests the 
presence of five or six highly correlated clusters. 
\begin{figure}
\centering
\subfigure[HCC] {\includegraphics[scale=.35]{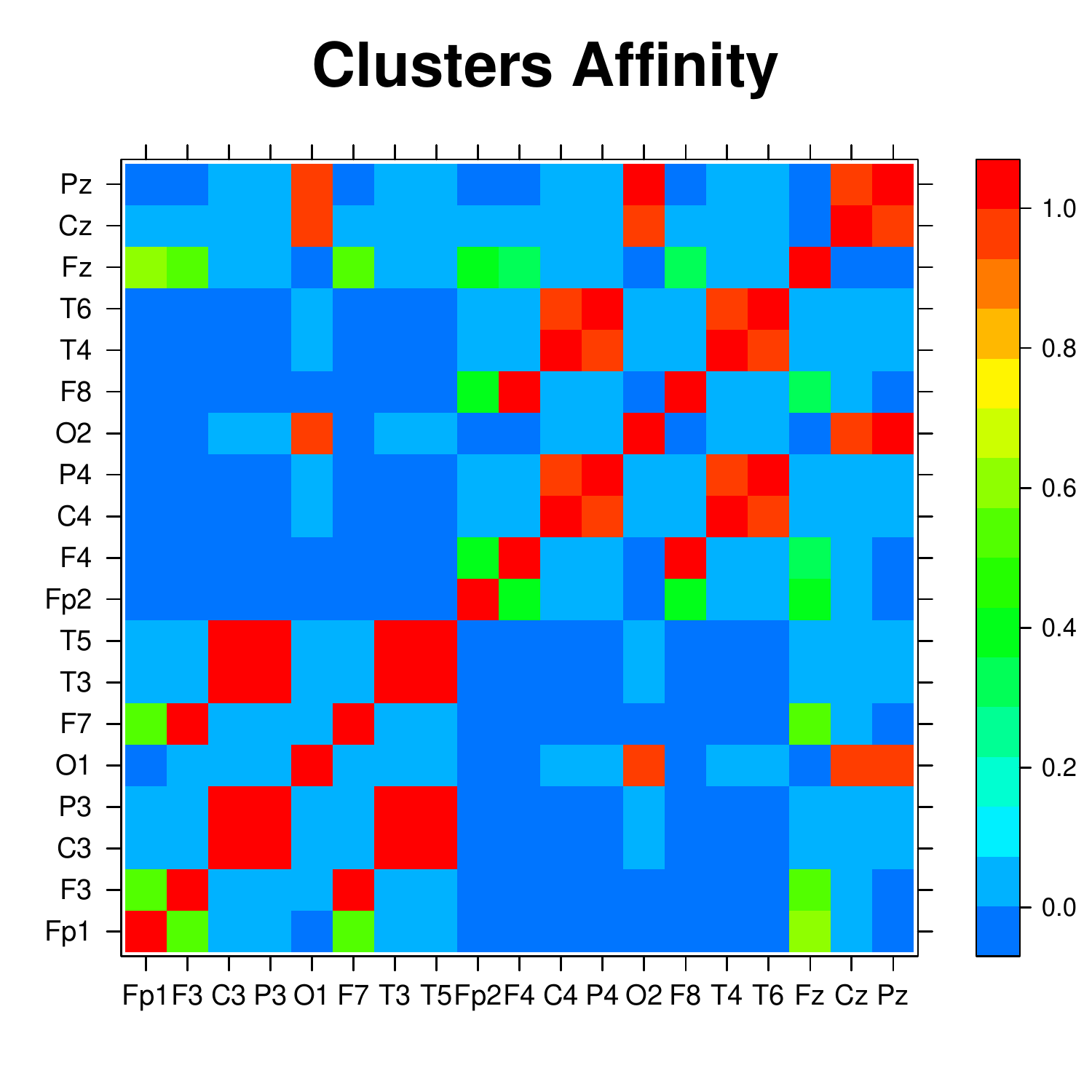}}
\subfigure[HAC] {\includegraphics[scale=.35]{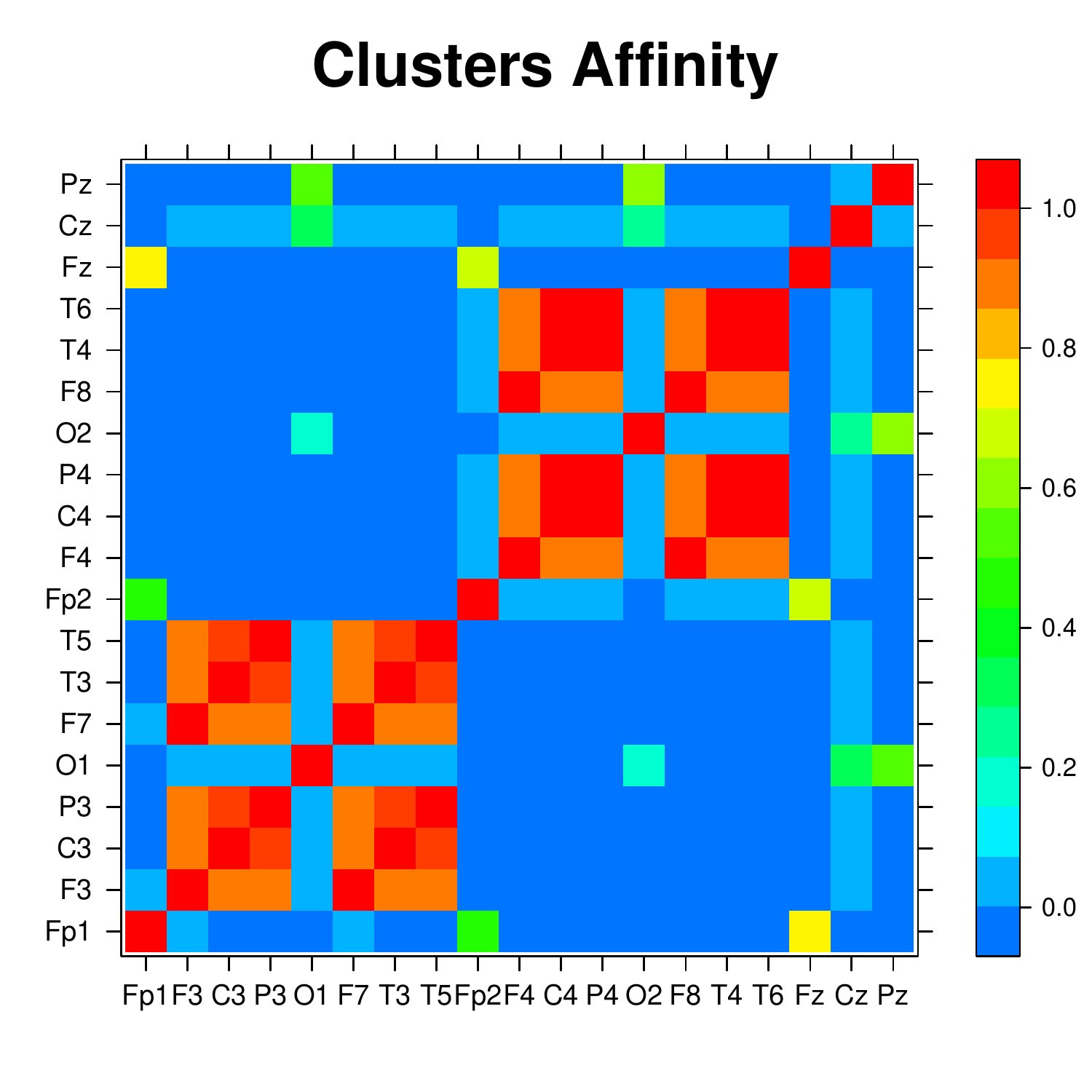}}
\subfigure[HMC] {\includegraphics[scale=.35]{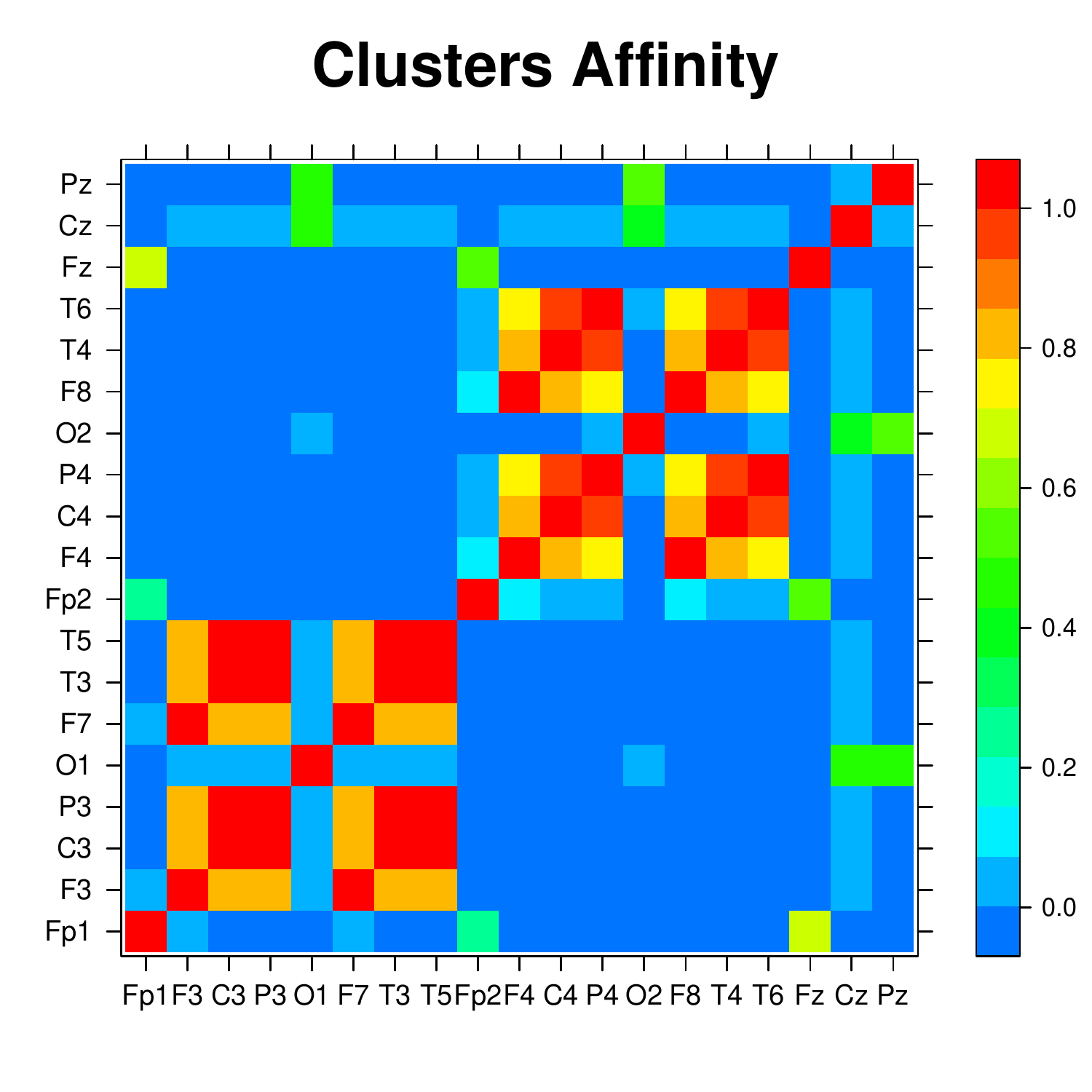}}
\caption{Affinity matrix over 100 replicates with 6 clusters.} \label{CR_100_EEG}
\end{figure}

\begin{figure}
\centering
\subfigure[HCC\label{CR_EEG4}] {\includegraphics[scale=.32]{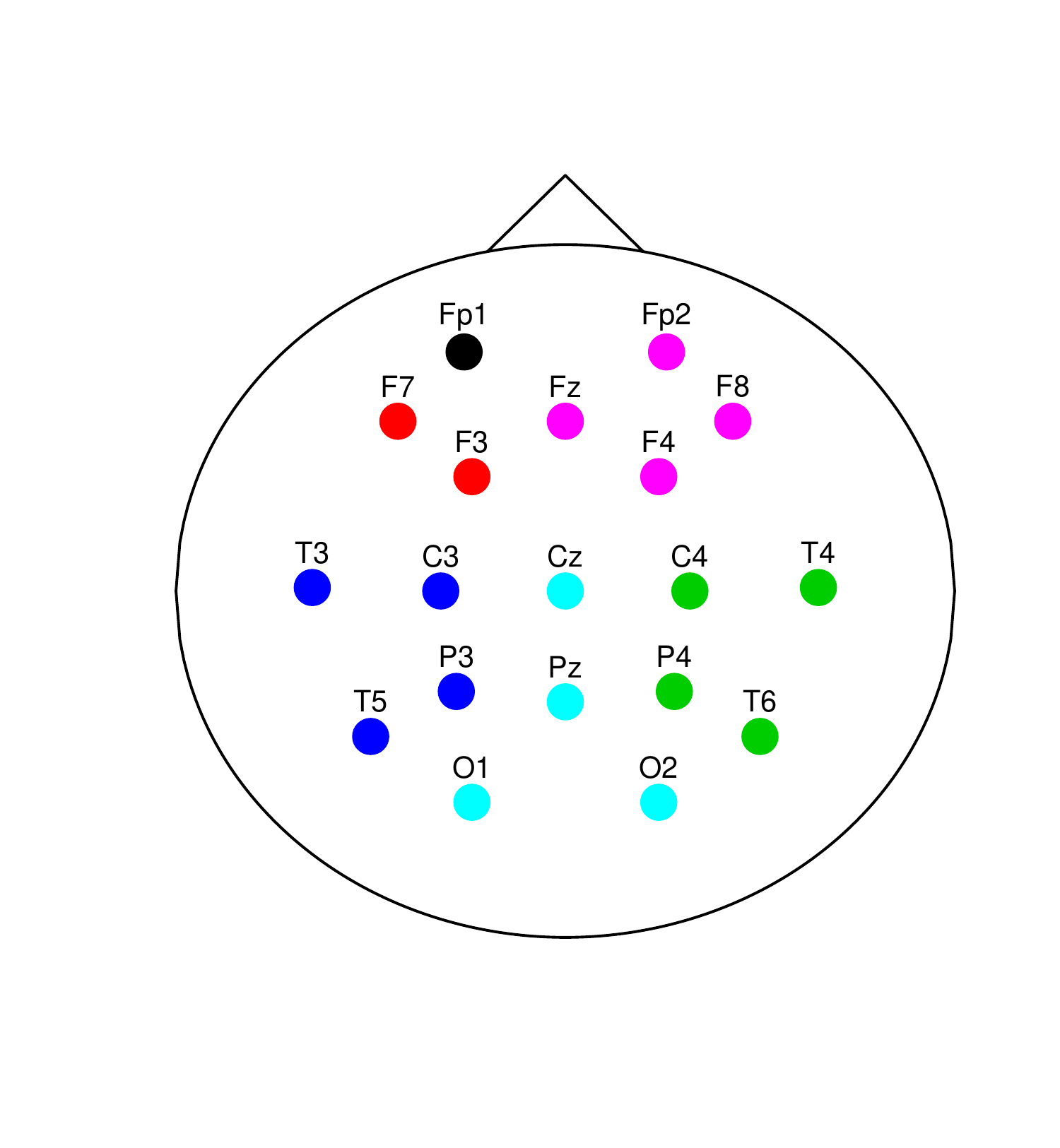}}
\subfigure[HAC\label{CR_EEG5}] {\includegraphics[scale=.32]{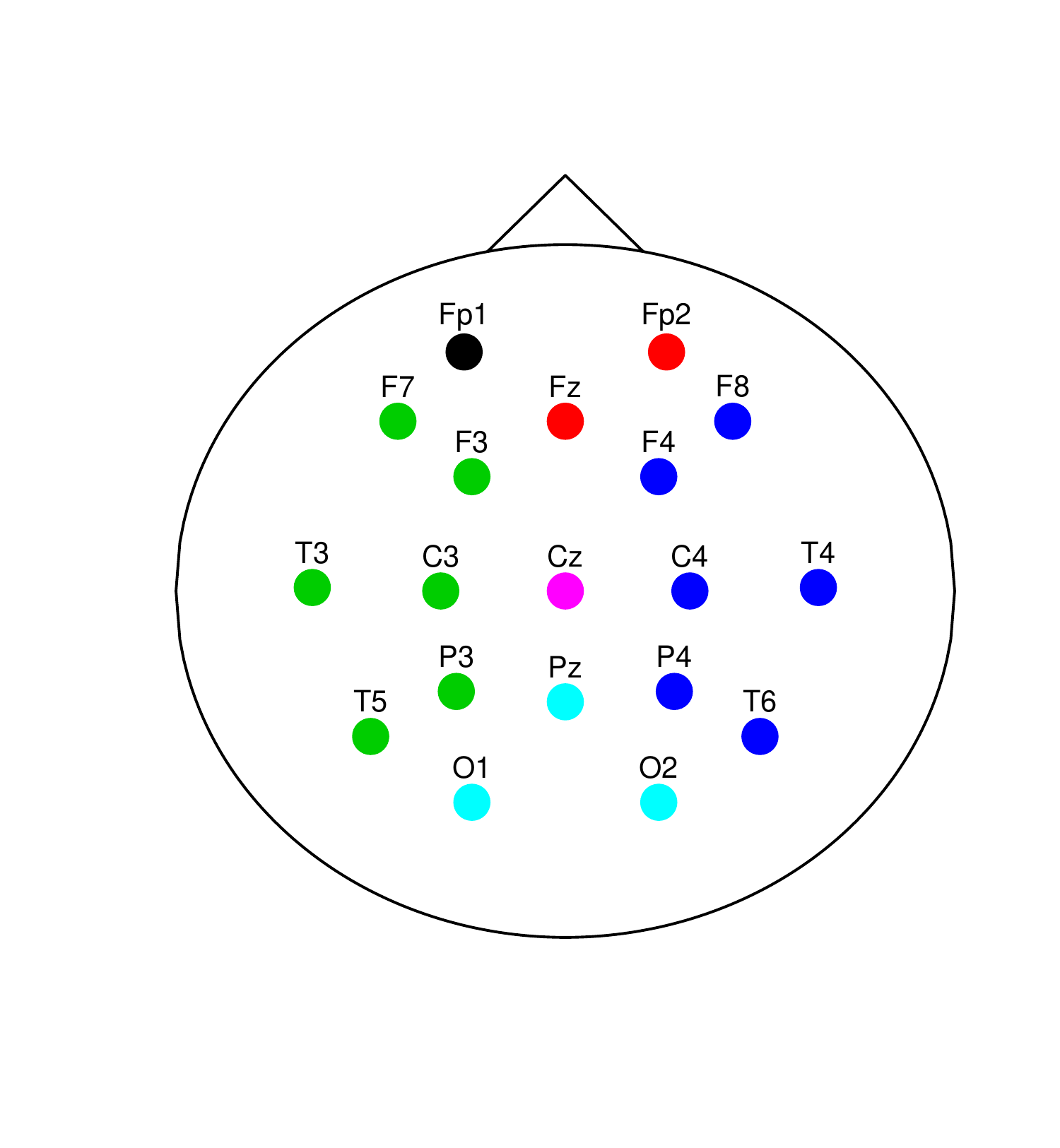}}
\subfigure[HMC\label{CR_EEG6}] {\includegraphics[scale=.32]{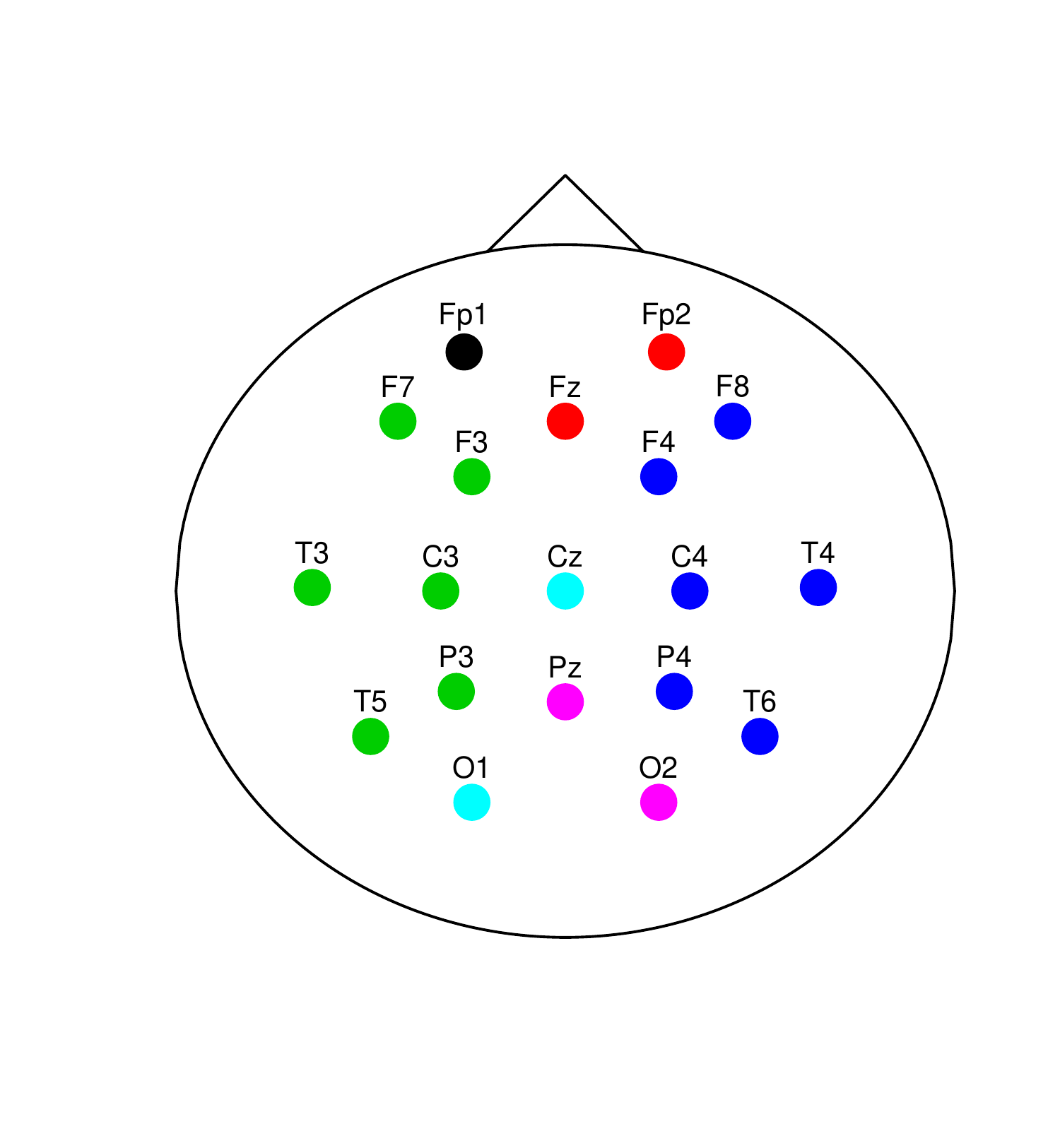}}
\caption{Cluster locations on the scalp by choosing 6 clusters. Different clusters are represented by colors.} \label{CR_EEG}
\end{figure}

To compare the clustering results, we fix the number of clusters to six for all of the methods and
plot the affinity matrices. The affinity matrix $A$ is defined as follows, let
$A_{ij}^m=1$ if channels $i$ and $j$ are in the same cluster at replicate $m$, then $A_{ij}=\sum_{m=1}^M A_{ij}^m / 100 $. 
Figure \ref{CR_100_EEG} presents the affinity matrix over 100 replicates for the three methods.
In general, the clustering results are very consistent; the HMC method has the most variability.
The HAC and HMC methods tend to have two large clusters and isolate the channels $O1$, $O2$ and $Cz$. 
In contrast, the HCC method tends to have six clusters with similar size.

The clustering results between replicates differ little; we consider replicate 85 as a representative example.
This replicate corresponds to the median curve in the functional boxplot (see Figure \ref{MVE3}). 
In this case, HAC method assigns $Cz$ to one single cluster and the HMC assigns $\{Cz,O1\}$ to one cluster. 
In contrast, the HCC method assigns $\{Cz,Pz,O1,O2\}$ together in one cluster, which seems more reasonable. 
Another difference is the assignment of $\{F4,F8\}$. The HAC and HMC methods assigns these channels to the same cluster as 
$\{C4,T4,P4,T6\}$, while the HCC asignes them to the same cluster as $\{Fz,Fp2\}$. These two different clustering results 
are an example of the advantage of the HCC method, since $\{F4,F8\}$ are strongly correlated with $\{C4,T4\}$ but weakly correlated with $\{P4,T6\}$. 
Therefore, the resulting clusters from the HCC method are stronger correlated than those from the other methods.

\subsection{Effect of the Physiological artifacts in Coherence-Based Clustering}

Finally, we explore the possible effect of physiological artifacts in the HCC and HAC clustering methods. 
The presence of physiological artifacts on EEG data is very common. In fact, raw EEG data usually is very noisy and some preprocessing of the EEG data is needed. 
Physiological artifacts on EEG data can be produced by eye movement, muscular movement, technical problems, etc. 
Among  these causes the ocular movement or eye blinking is the most present. According to \citet{Viqueira13}, 
the electrical activity related to eye blinks is registered in the frontal region of the brain. 
\begin{figure}
\centering
\subfigure[Physiological artifact \label{Cont1}] {\includegraphics[scale=.25]{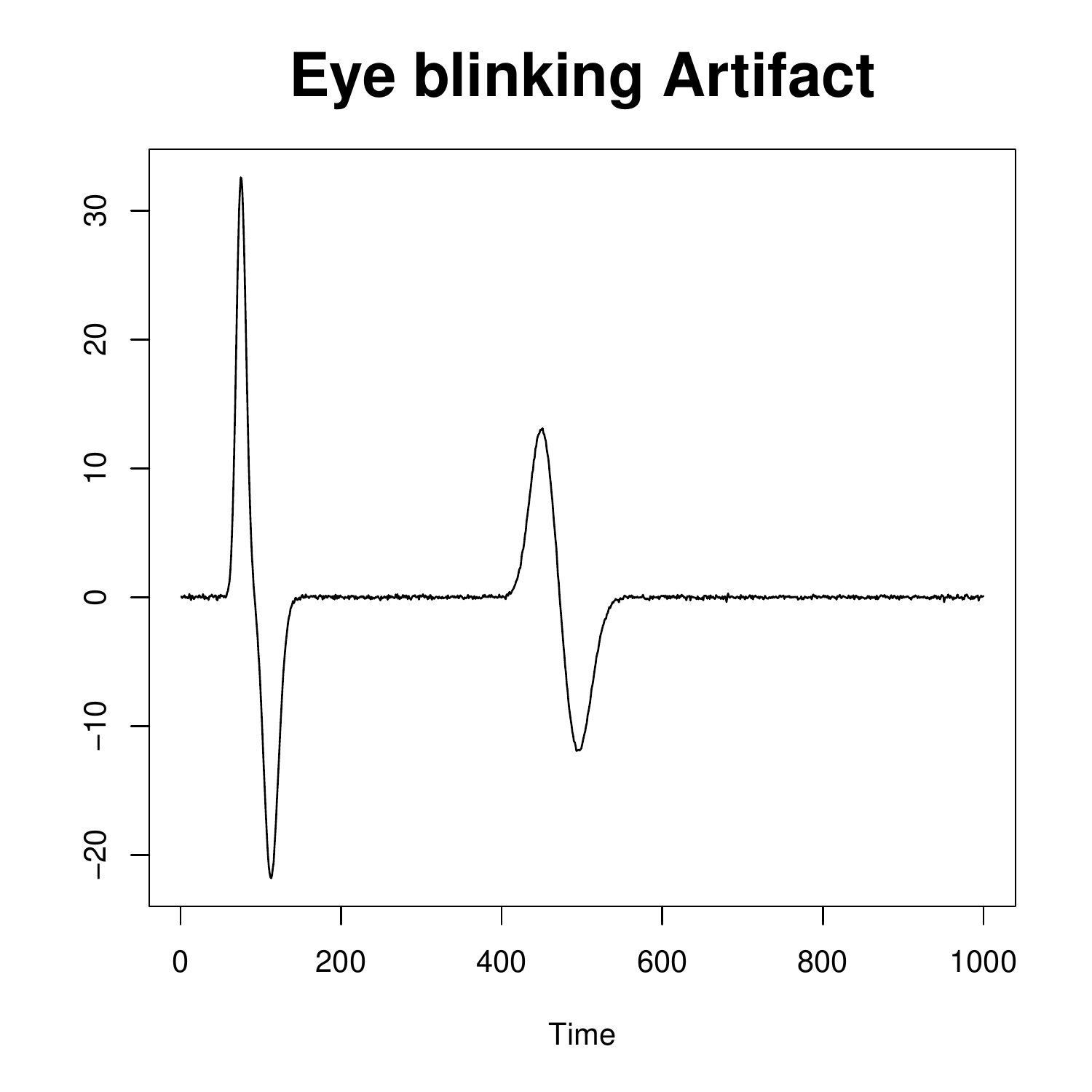}}
\subfigure[Original EEG signals  \label{Cont2}] {\includegraphics[scale=.27]{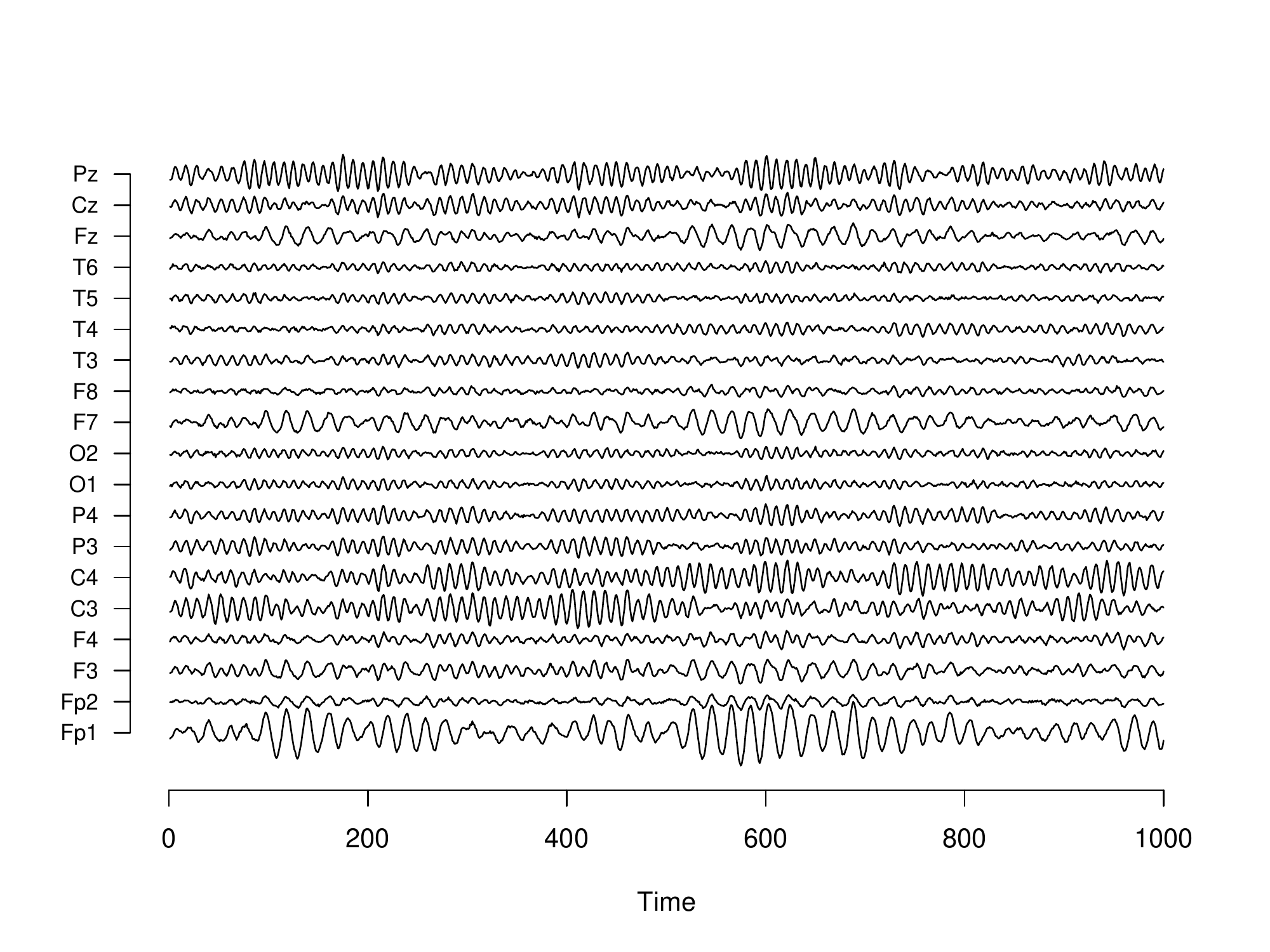}}
\subfigure[Contaminated EEG signals  \label{Cont3}] {\includegraphics[scale=.27]{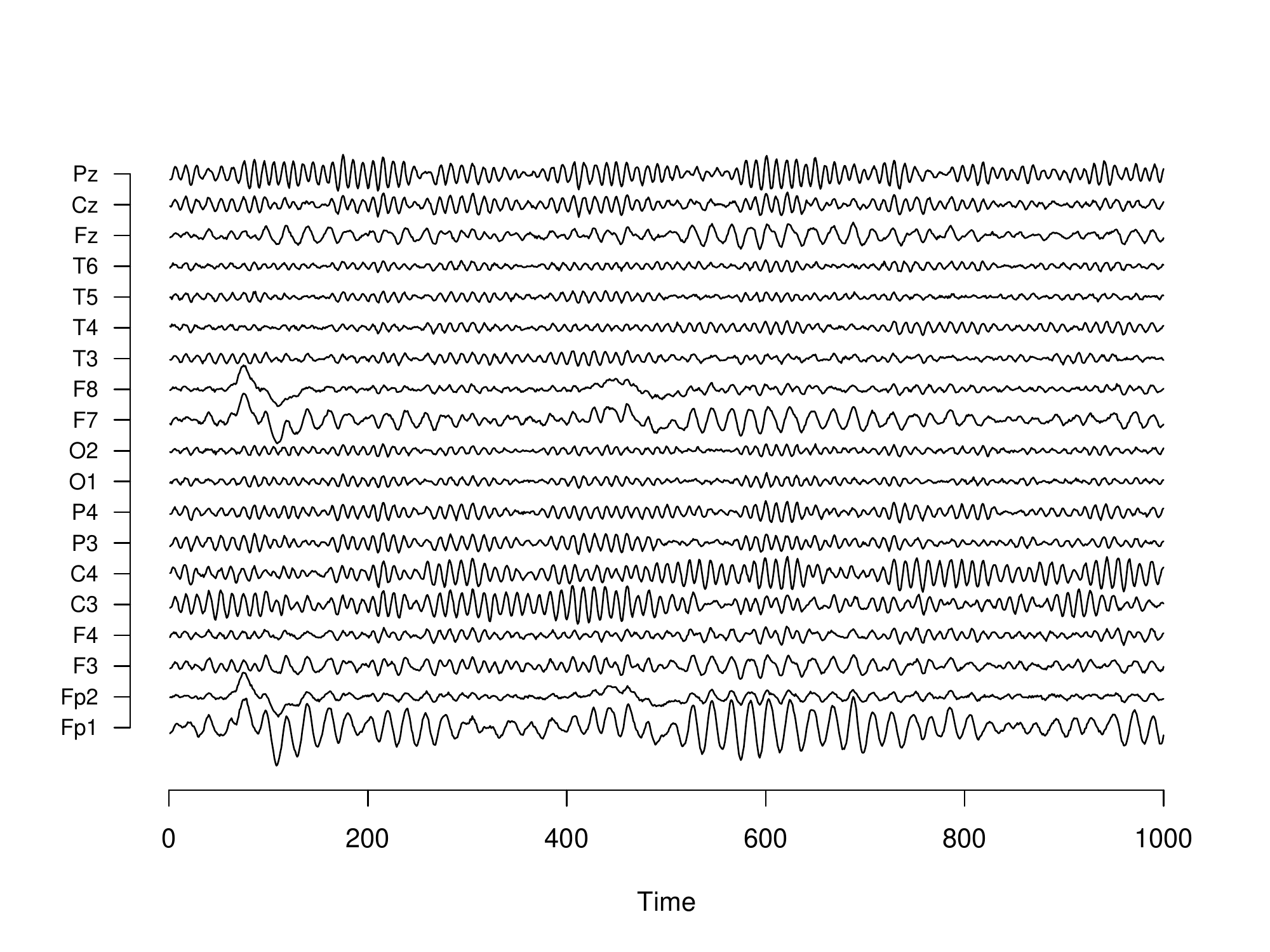}}
\caption{Contamination of EEG signals by the eye blinking artifact.} 
\end{figure}

\begin{figure}
\centering
\subfigure[Original EEG signal]{\includegraphics[scale=.35]{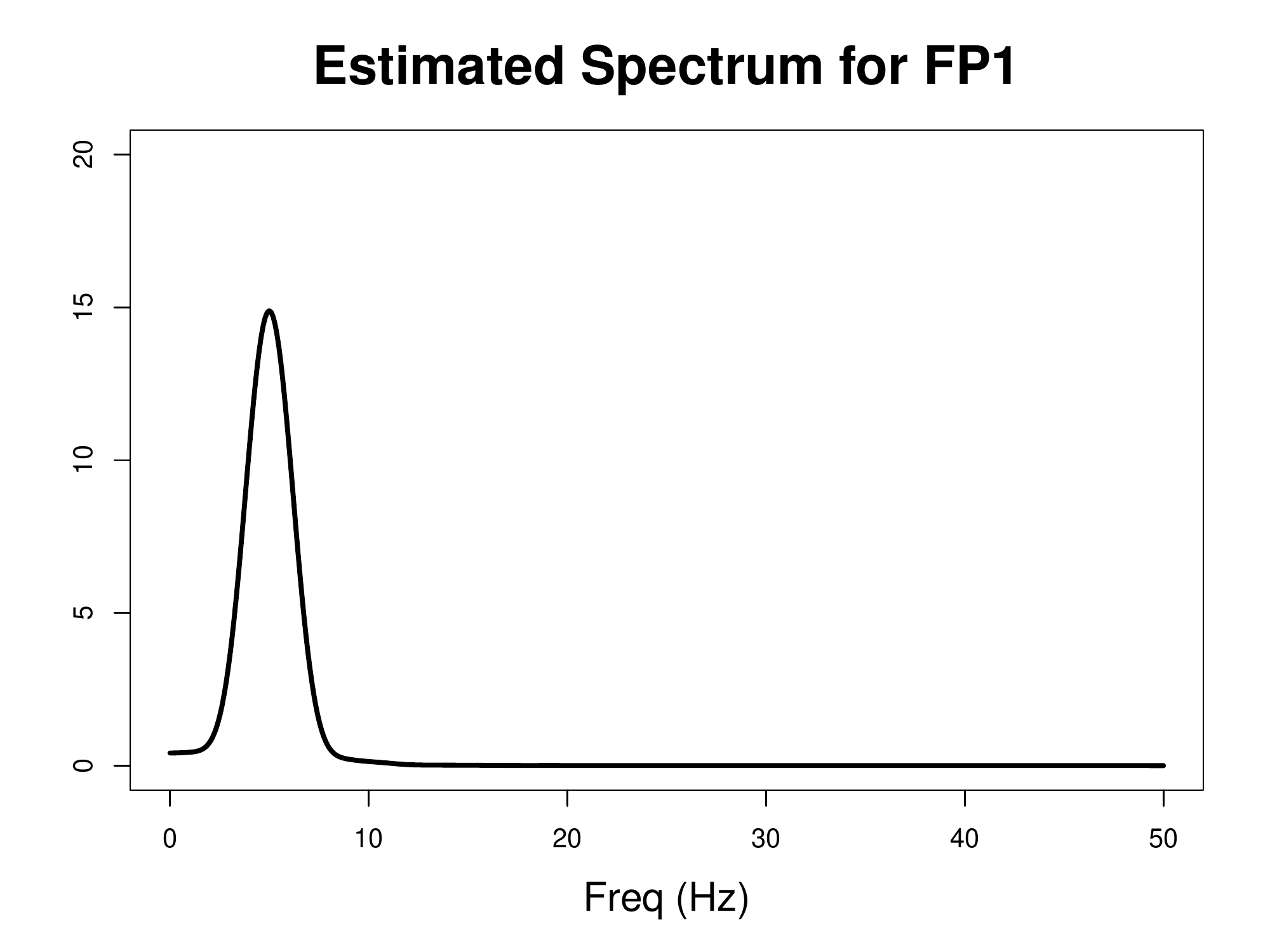}}
\subfigure[Contaminated EEG signal]{\includegraphics[scale=.35]{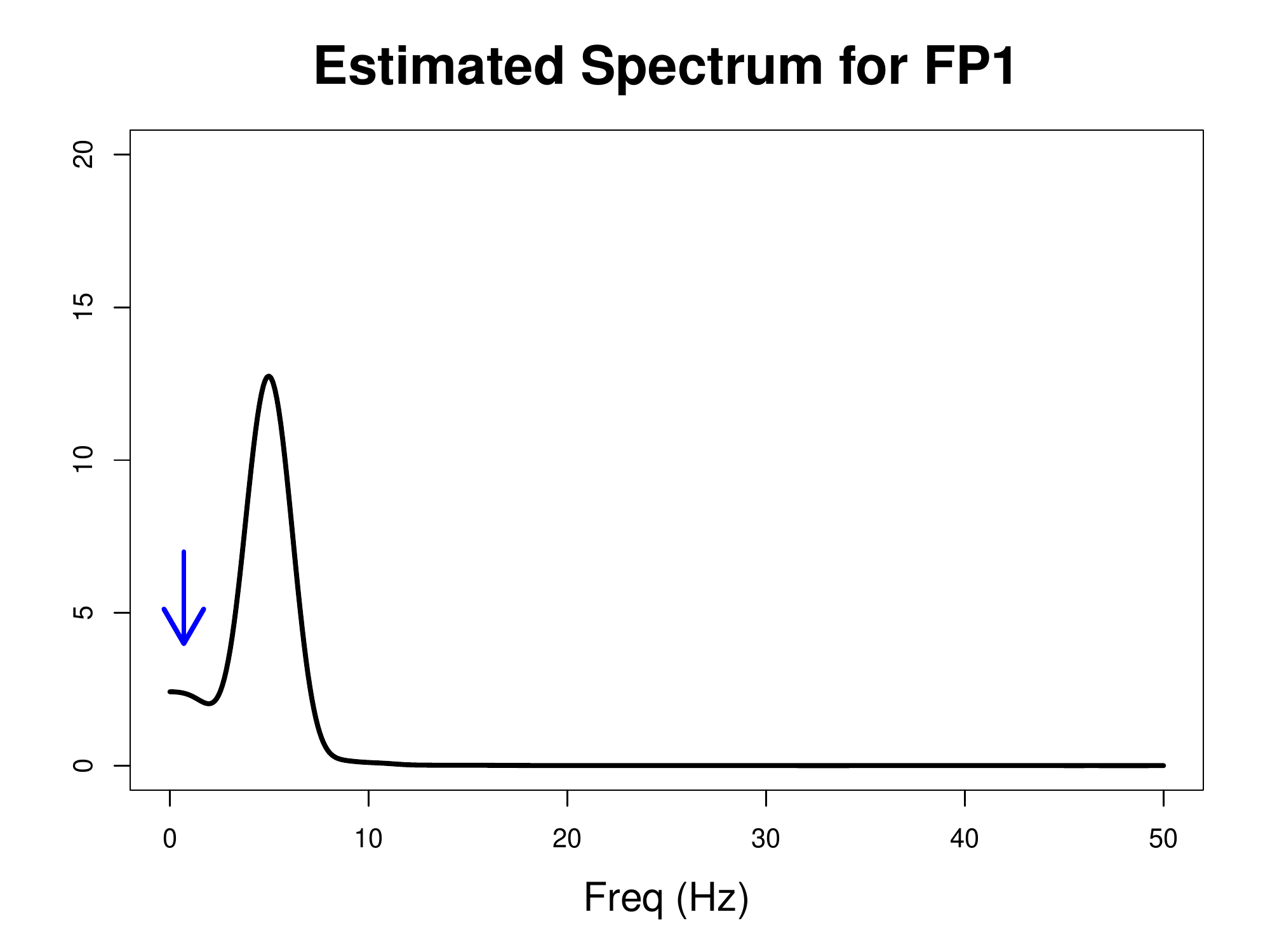}}
\caption{Estimated spectra of $FP1$ channel without artifact and with artifact.} \label{Cont4}
\end{figure}

We consider an illustration of this situation with simulation setting similar to \textbf{Experiment 3}. We keep $Z_{s_1}(t)$, $Z_{s_2}(t)$ and $Z_{s_3}(t)$ at the corresponding channels $C3$, $C4$ and $Pz$. We add 
$Z_{s_4}(t)$, $Z_{s_5}(t)$ located at $FP1$ and $FP2$ channels 
with spectral density with power concentrated at 5 Hz.  
The eye blinking artifact shape was generated as the difference between two gamma functions. Figure \ref{Cont1} shows the eye blinking signal, this shape emulates the physiological shape with additive noise \citep[see ][]{Viqueira13}. We simulate one EEG signal with 19 channels as in \textbf{Experiment 3} and then we contaminate channels $FP1$, $FP2$, $F7$ and $F8$. Figures \ref{Cont2} and \subref{Cont3} show the EEG simulated signals without and with the contamination, respectively. The eye blinking artifact has an effect on lower frequencies due to the high amplitude in less than a second, see Figure \ref{Cont4}.

\begin{figure}
\centering
\subfigure[HCC - Original EEG signals]{\includegraphics[scale=.29]{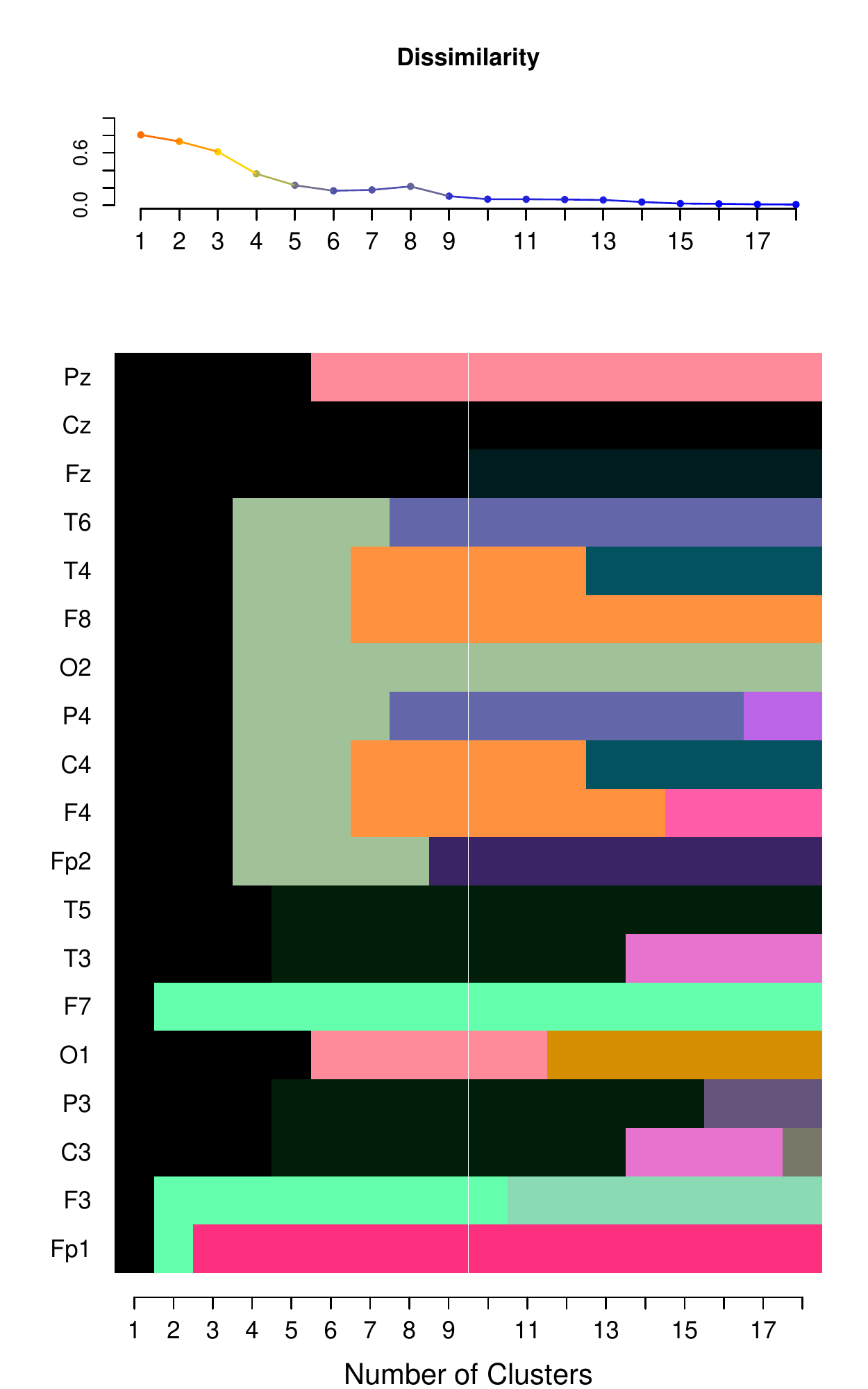}}
\subfigure[HCC - Contaminated EEG signals]{\includegraphics[scale=.29]{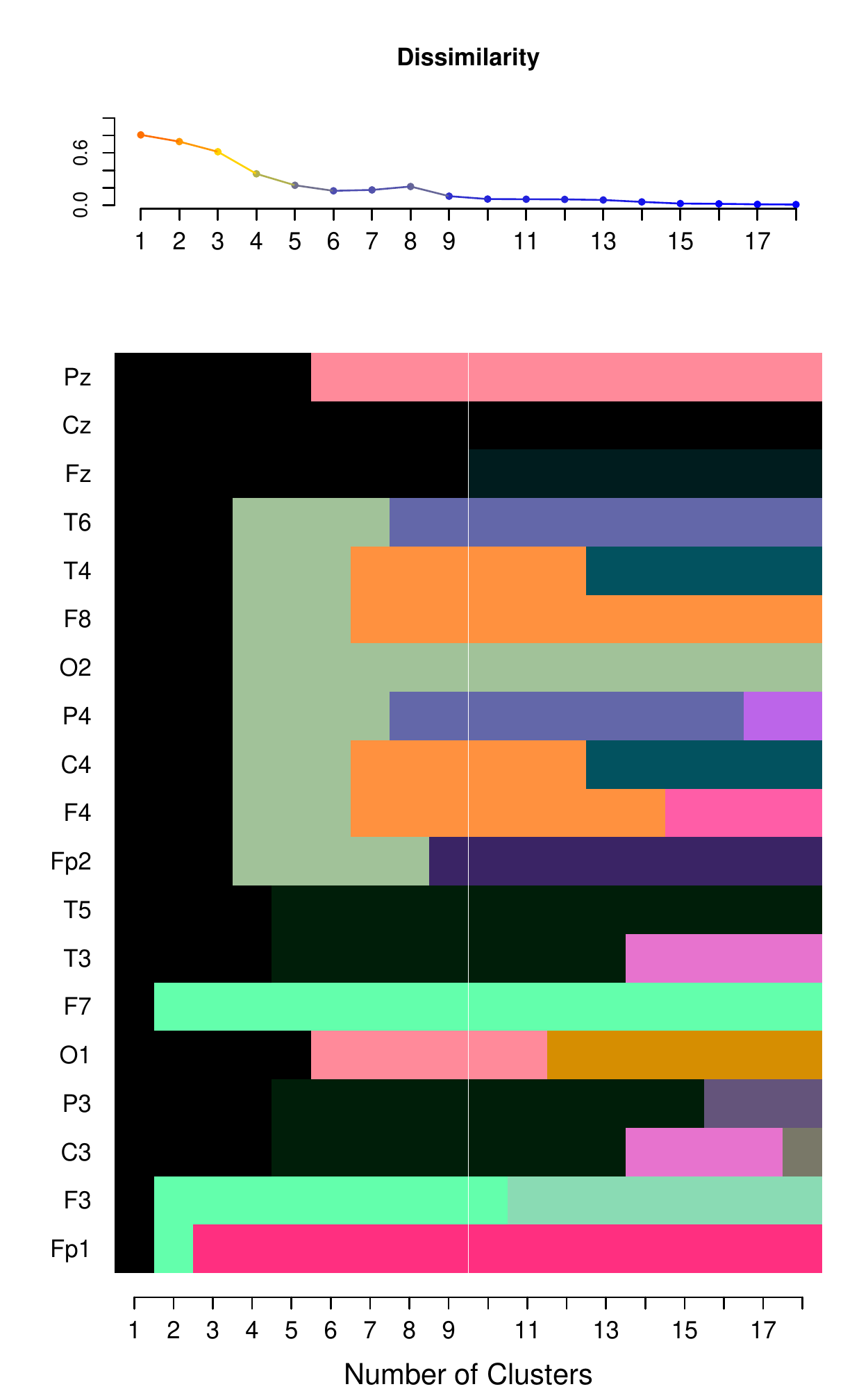}}
\subfigure[HAC - Original EEG signals]{\includegraphics[scale=.29]{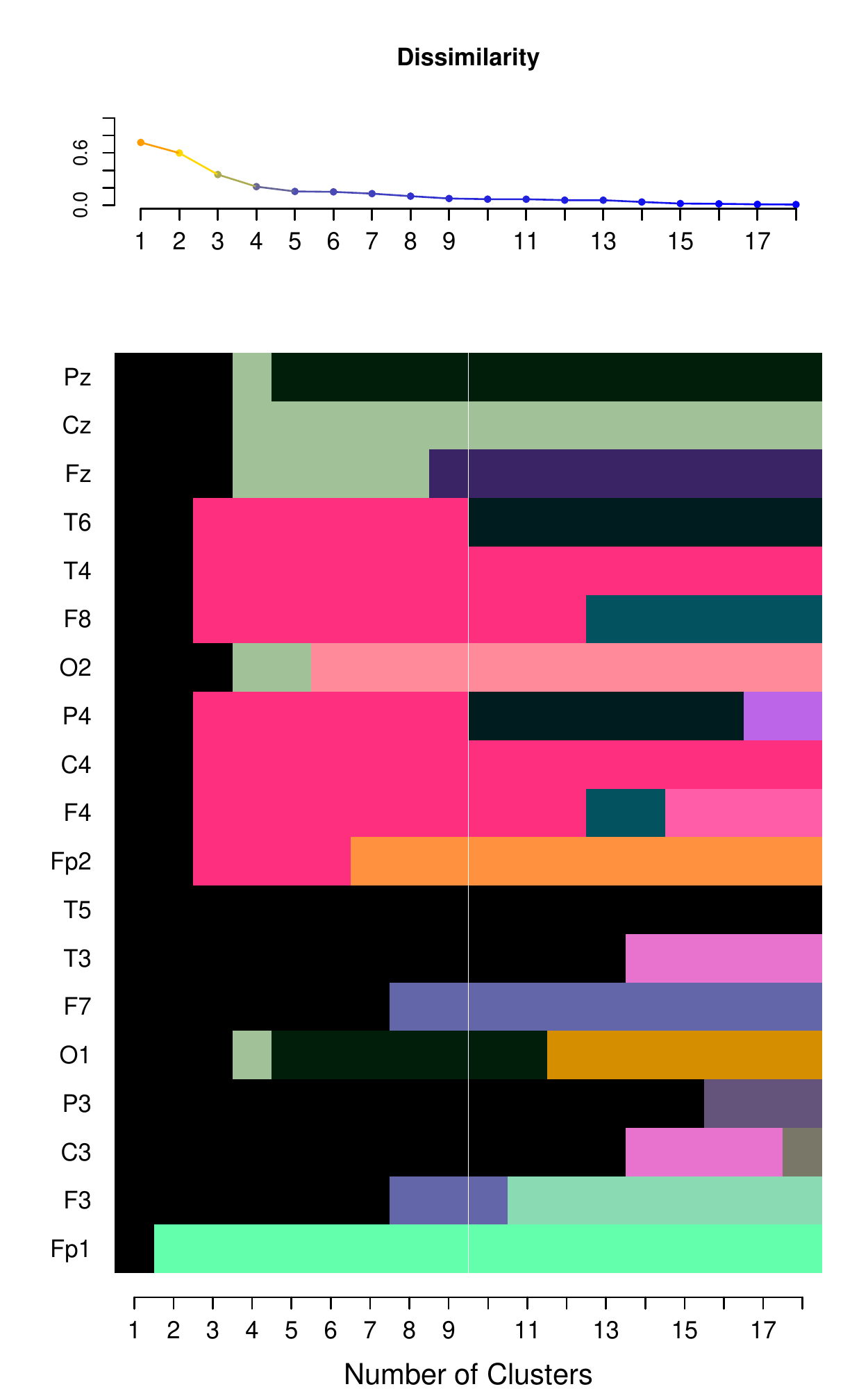}}
\subfigure[HAC - Contaminated EEG signals]{\includegraphics[scale=.29]{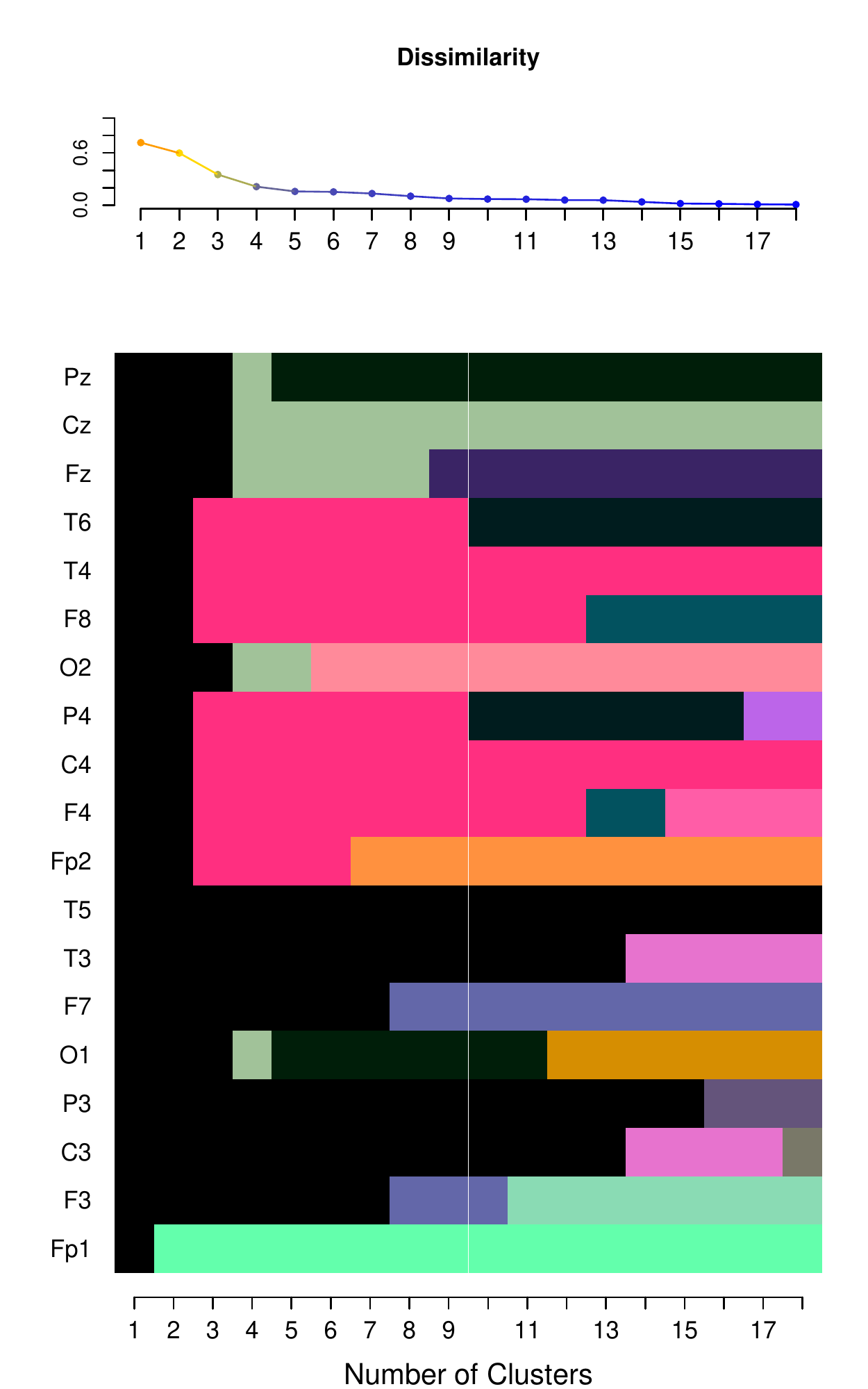}}
\caption{Visualization tool of clustering results on alpha bandwith original EEG signals and contaminated signals.} \label{Cont5}
\end{figure}

\begin{figure}
\centering
\subfigure[HCC - Original EEG signals]{\includegraphics[scale=.29]{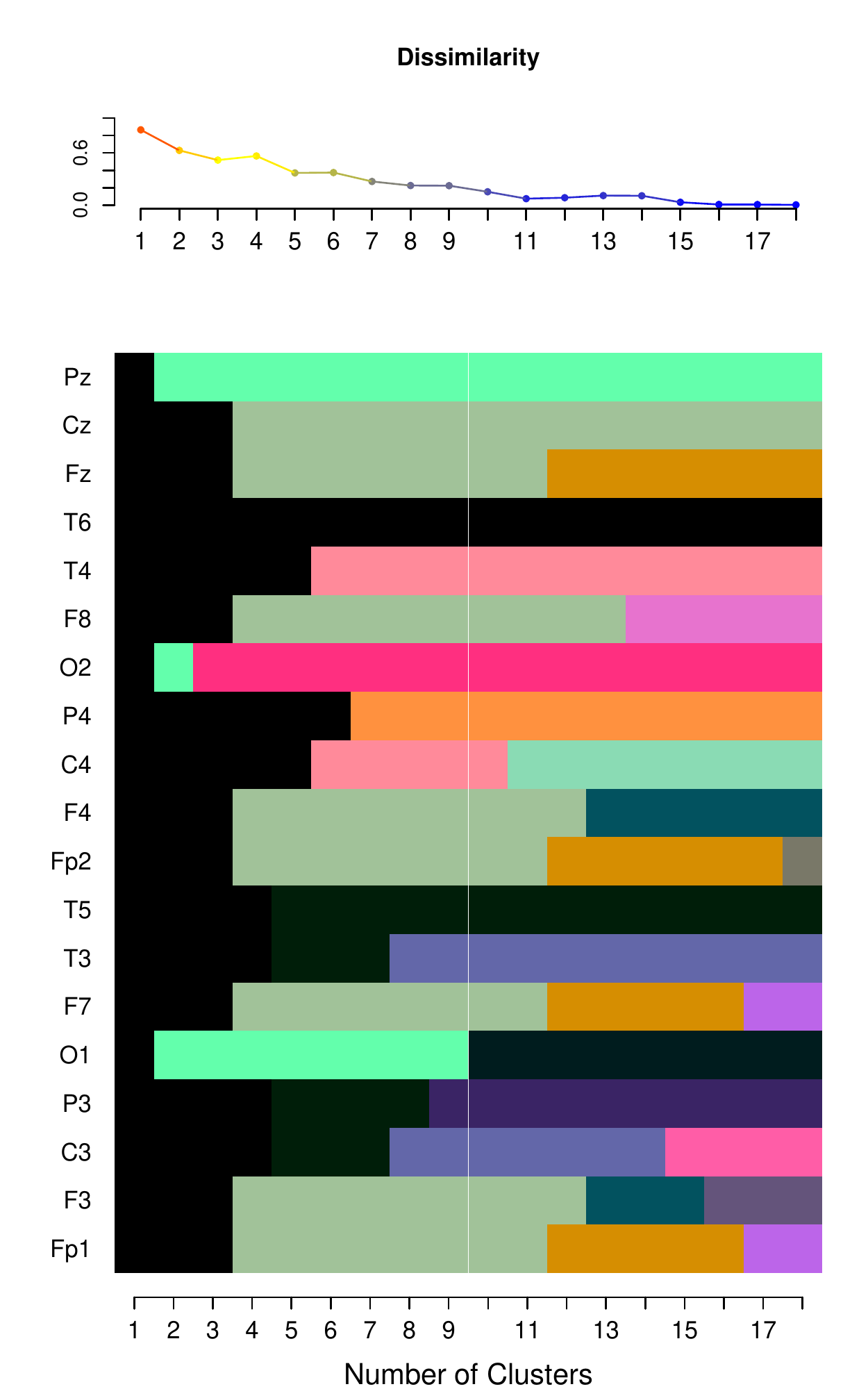}}
\subfigure[HCC - Contaminated EEG signals]{\includegraphics[scale=.29]{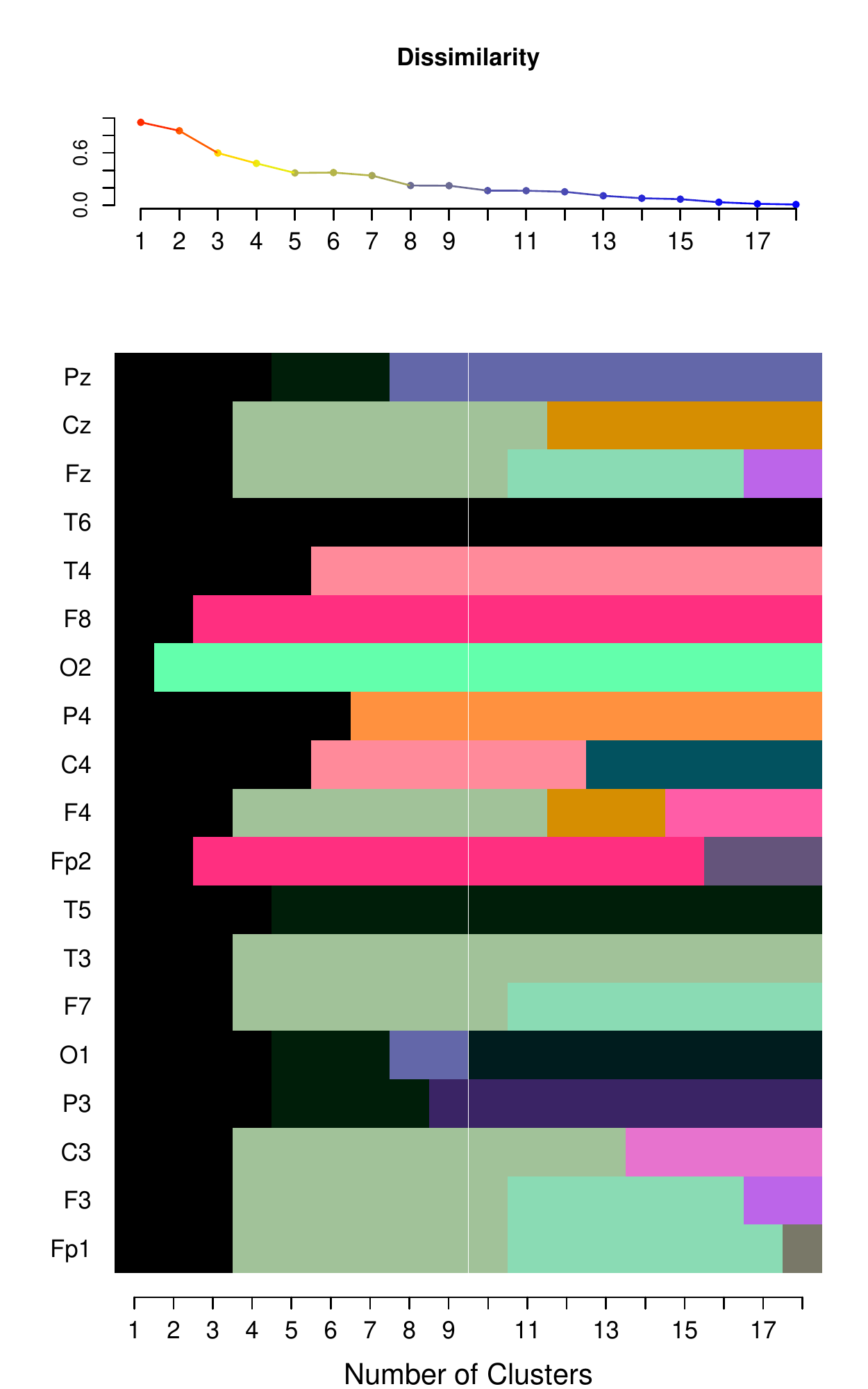}}
\subfigure[HAC - Original EEG signals]{\includegraphics[scale=.29]{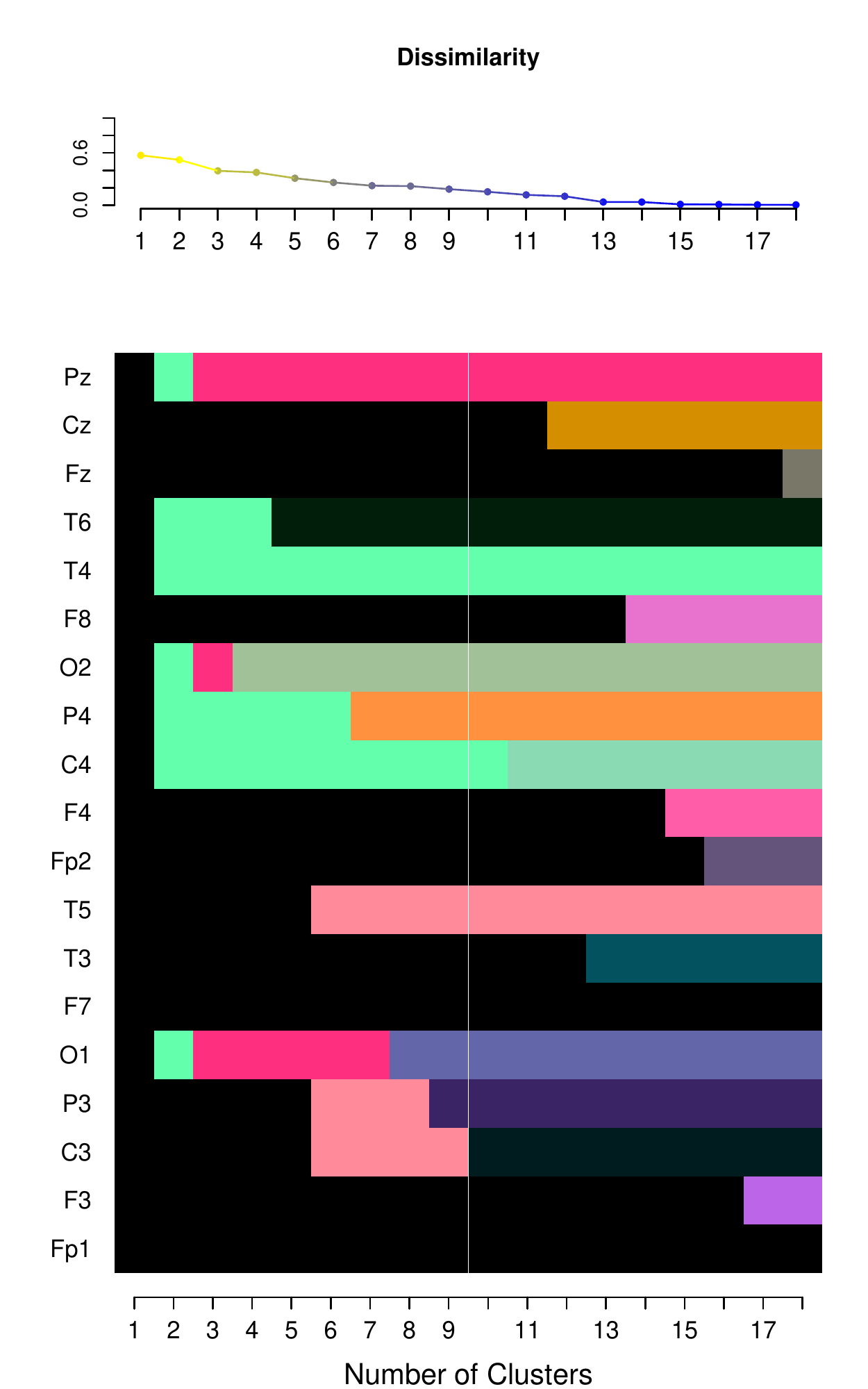}}
\subfigure[HAC - Contaminated EEG signals]{\includegraphics[scale=.29]{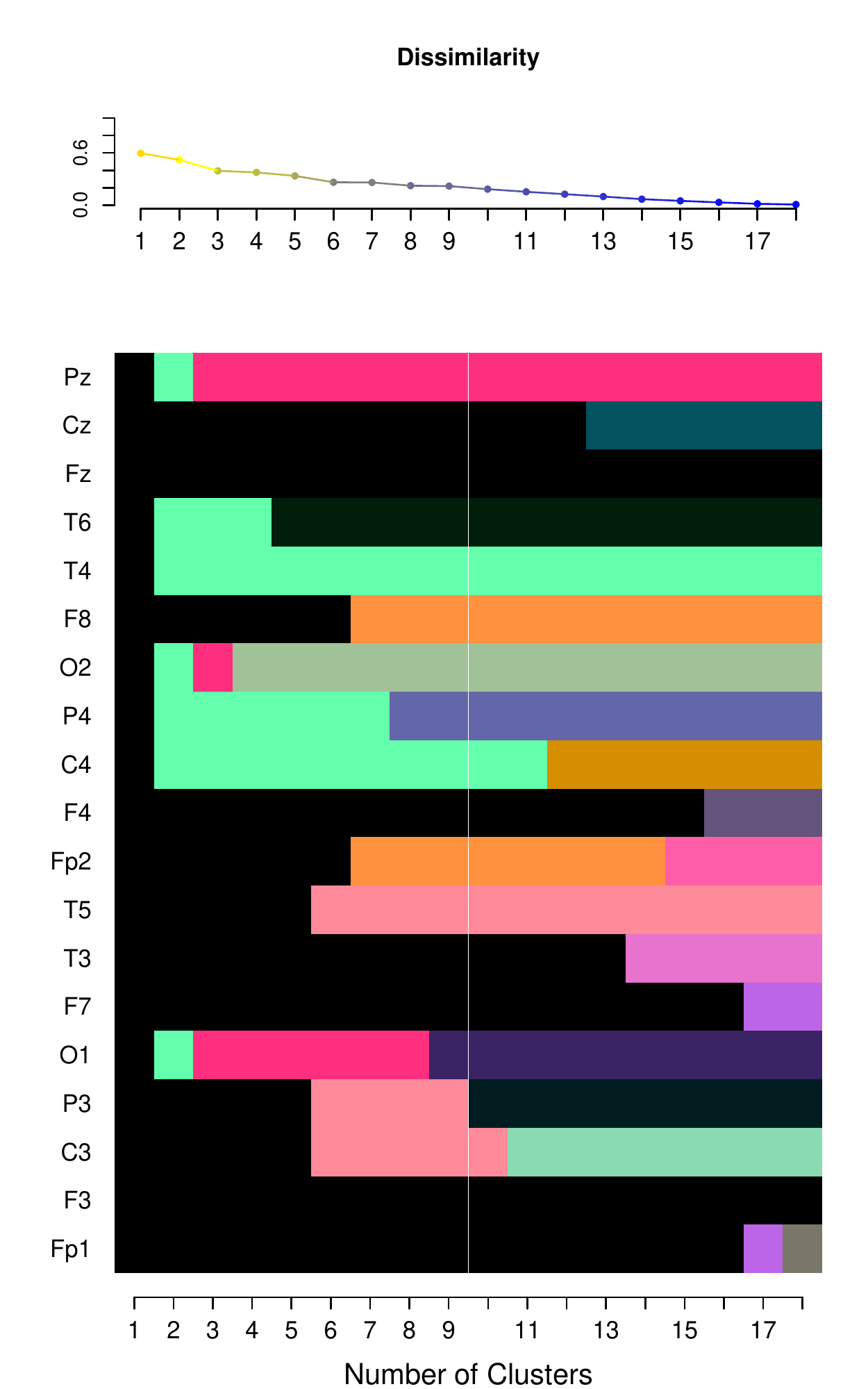}}
\caption{Visualization tool of clustering results on theta band with original EEG signals and contaminated signals.} \label{Cont6}
\end{figure}

We applied HCC and HAC clustering methods (HMC was omitted since results are similar to HAC) in the alpha and theta bands. Figure \ref{Cont5} shows the clustering results with the original (simulated) EEG signals and the contaminated with eye blinking artifact on alpha band. We observed that there is no significant effect of the eye blinking artifact in the clustering results on this frequency bands. 
However, the theta band clustering is different. Figure \ref{Cont6} shows the clustering results on theta band. If we consider 
five clusters, for both methods, the eye blinking artifact will separate $FP2$ and $F8$ from the rest of the frontal channels. It seems that the presence of the eye blinking artifact could produce its own cluster in low frequencies in this case. A more exhaustive simulation study needs to be performed to have a more general conclusion. But, 
we need to know that artifacts could affect the obtained clusters when using coherence based clustering methods.

\section{Data Analysis Using the HCC-Vis toolbox}
We developed the HCC method to cluster signals from many channels in a brain network.
Then, the identified clusters represent connected brain regions. 
We studied an EEG recorded from a patient of Dr. Malow 
(neurologist formerly at the University of Michigan) during an epileptic seizure.
The main interest is to identified connectivity on high frequency bands. 
Therefore, we present the results obtained by the HCC method on the alpha and beta bands.
To visualize the clustering results we developed
the HCC-Vis, a Shiny app (RStudio) \url{https://carolinaeuan.shinyapps.io/hcc-vis/}.

\subsection{Analysis of Epileptic EEG seizure}\label{Seizure1}
\begin{figure}
\centering
\includegraphics[width=8cm, height=5cm]{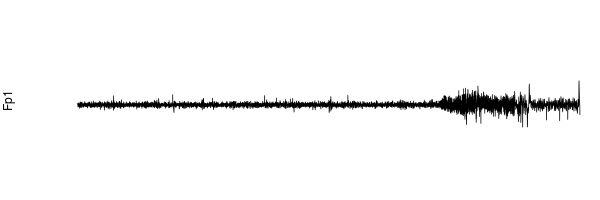}
\includegraphics[width=8cm, height=5cm]{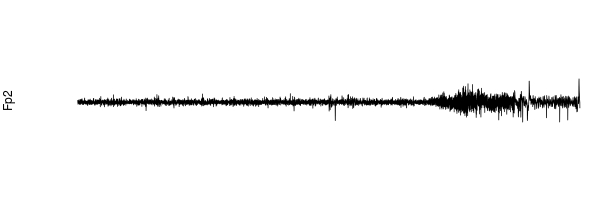}
\vspace{-1.5cm}\\
\includegraphics[width=8cm, height=5cm]{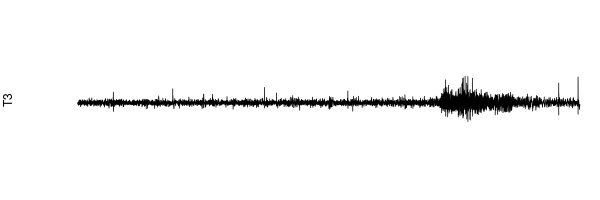}
\includegraphics[width=8cm, height=5cm]{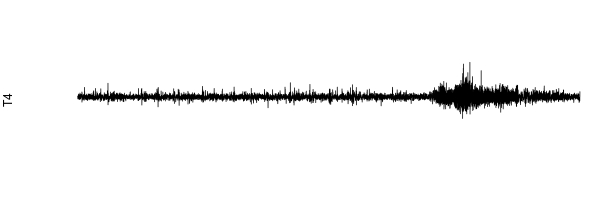}
\caption{EEG traces from channels Fp1 and Fp2 located in the frontal region (left and right) 
and T3 and T4 located in the temporal region (left and right).}\label{EEGtraces}
\end{figure}

Dataset corresponds to EEG data recorded from a female patient during spontaneous epileptic 
seizure. The recording lasted for 500 seconds and was digitized at 100 Hz. 
The data array has 21 channels with 19 bipolar scalp electrodes placed according 
to the 10-20 system and two sphenoidal electrodes placed intracranially at the base 
of the temporal lobe. Figure \ref{EEGtraces} shows four of the 21 EEG signals. The seizure onset was 
recorded around 340 seconds. The seizure begin 
in the left temporal region of the brain, where channel $T3$ was located. 
Therefore, our analyses used $T3$ as the main focal point (main node). We address the following questions: 
(1) Which channels are connected to $T3$ during different seizure phases, such as before and after the seizure? 
(2) Does the connectivity structure between channels differ between the alpha activity and the beta activity? 
(3) Does seizure modify brain connectivity? 

To answer these questions, we divide the EEG recording into disjointed 
10-second segments, and applied the HCC method to each segment.  
Additionally, we show the results from the HSM method to 
complement the interpretation of the HCC results, and from the HAC method 
to compare with the HCC the results. Here, we show the results for time segment 5, 
before seizure (40 to 50 seconds); time segment 35, early seizure (340 to 350 seconds);
and time segment 38, middle seizure (370 to 380 seconds). 

\begin{figure}
\centering
\subfigure[Before \hspace{2cm} - \hspace{2cm} Early \hspace{2cm} - \hspace{2cm} Middle Seizure]{\includegraphics[scale=.25]{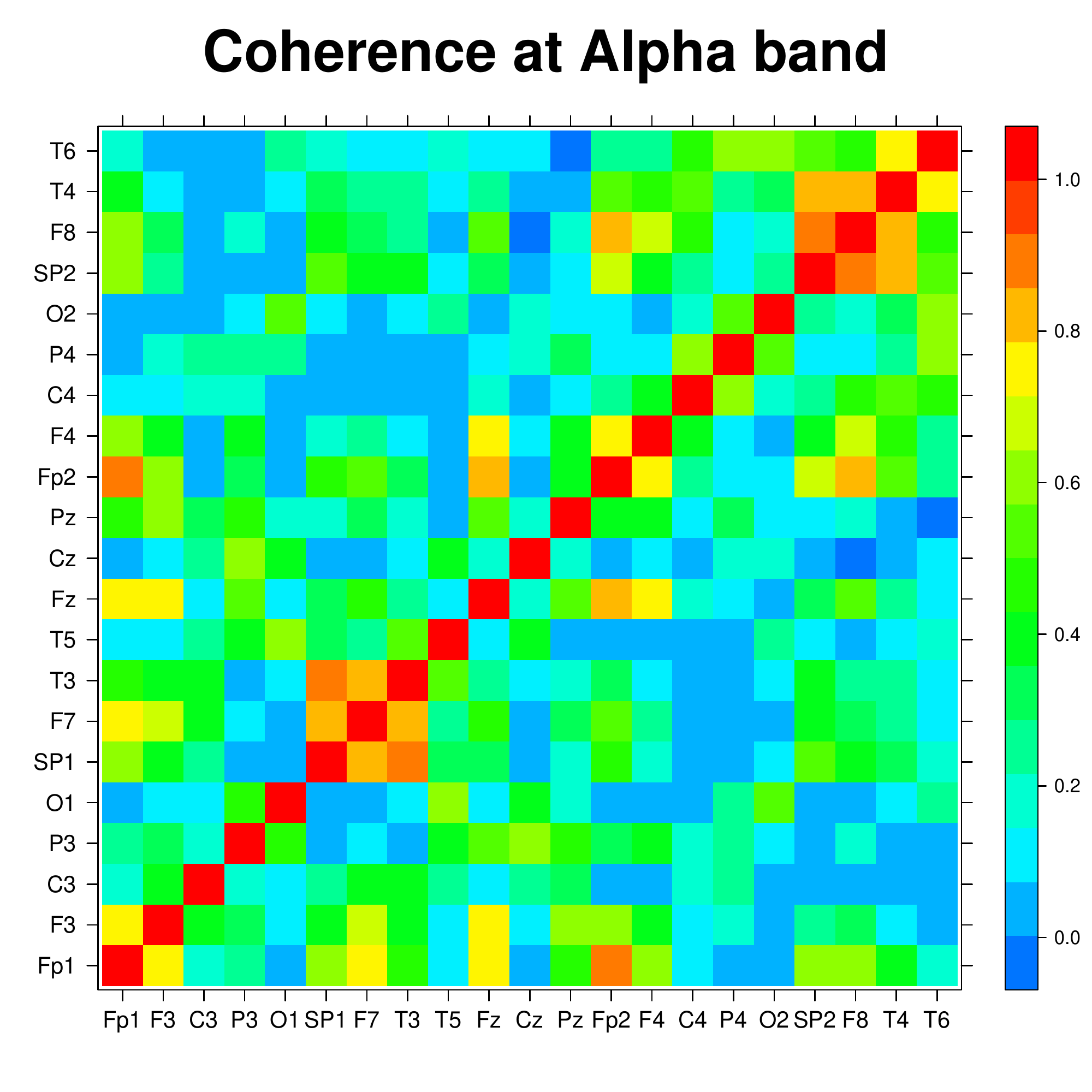}\includegraphics[scale=.25]{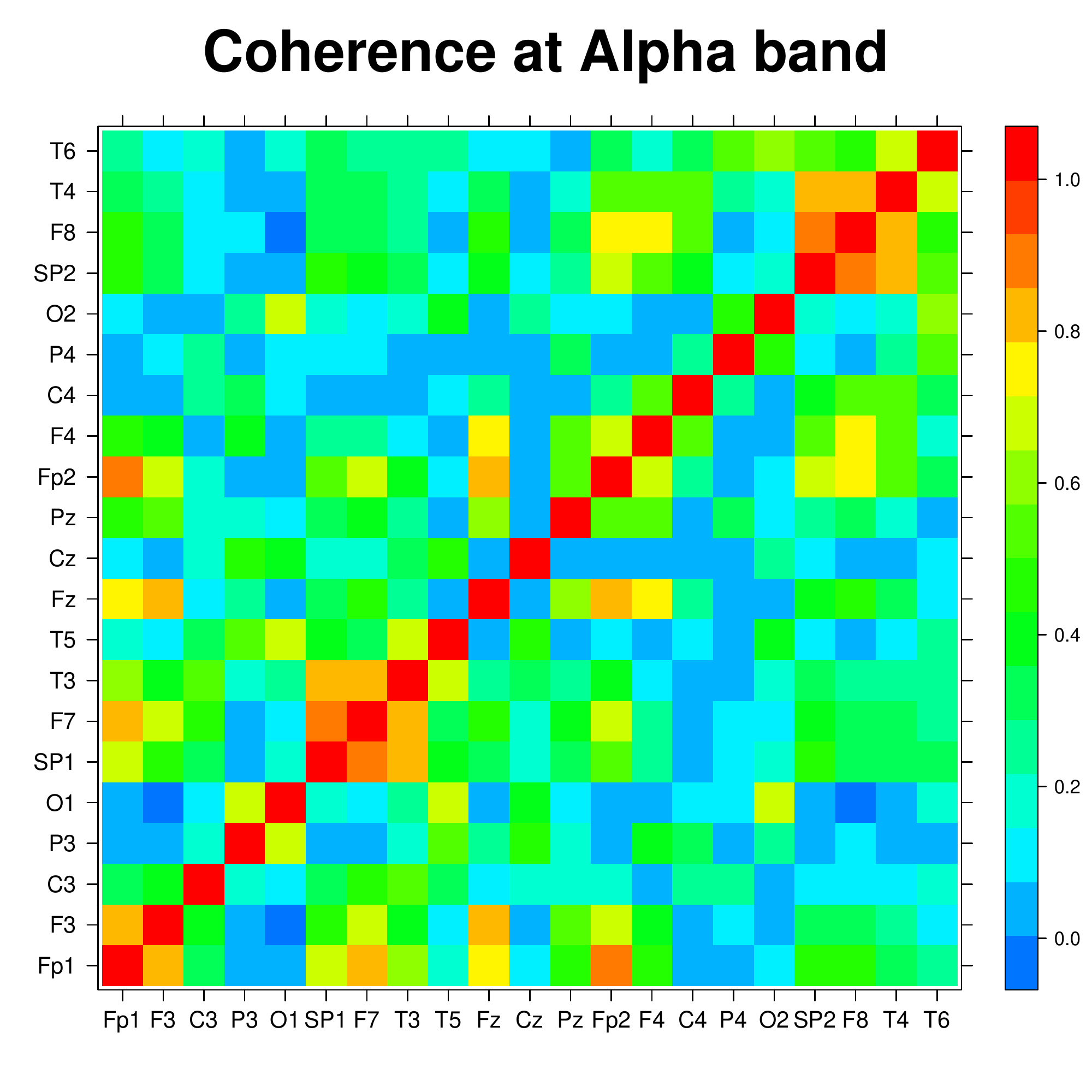}
\includegraphics[scale=.25]{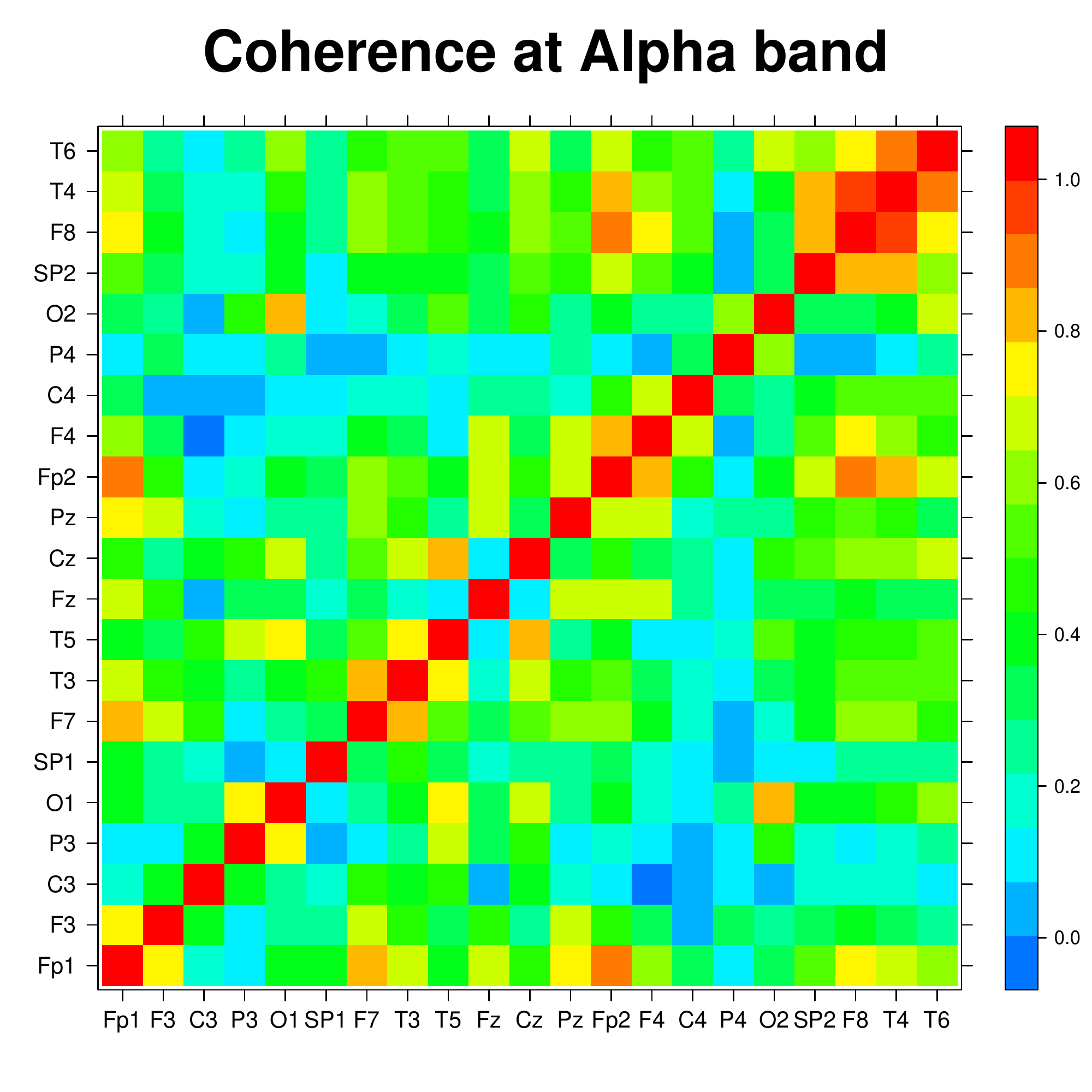}}\\
\subfigure[Before \hspace{2cm} - \hspace{2cm} Early \hspace{2cm} -  \hspace{2cm} Middle Seizure]{\includegraphics[scale=.25]{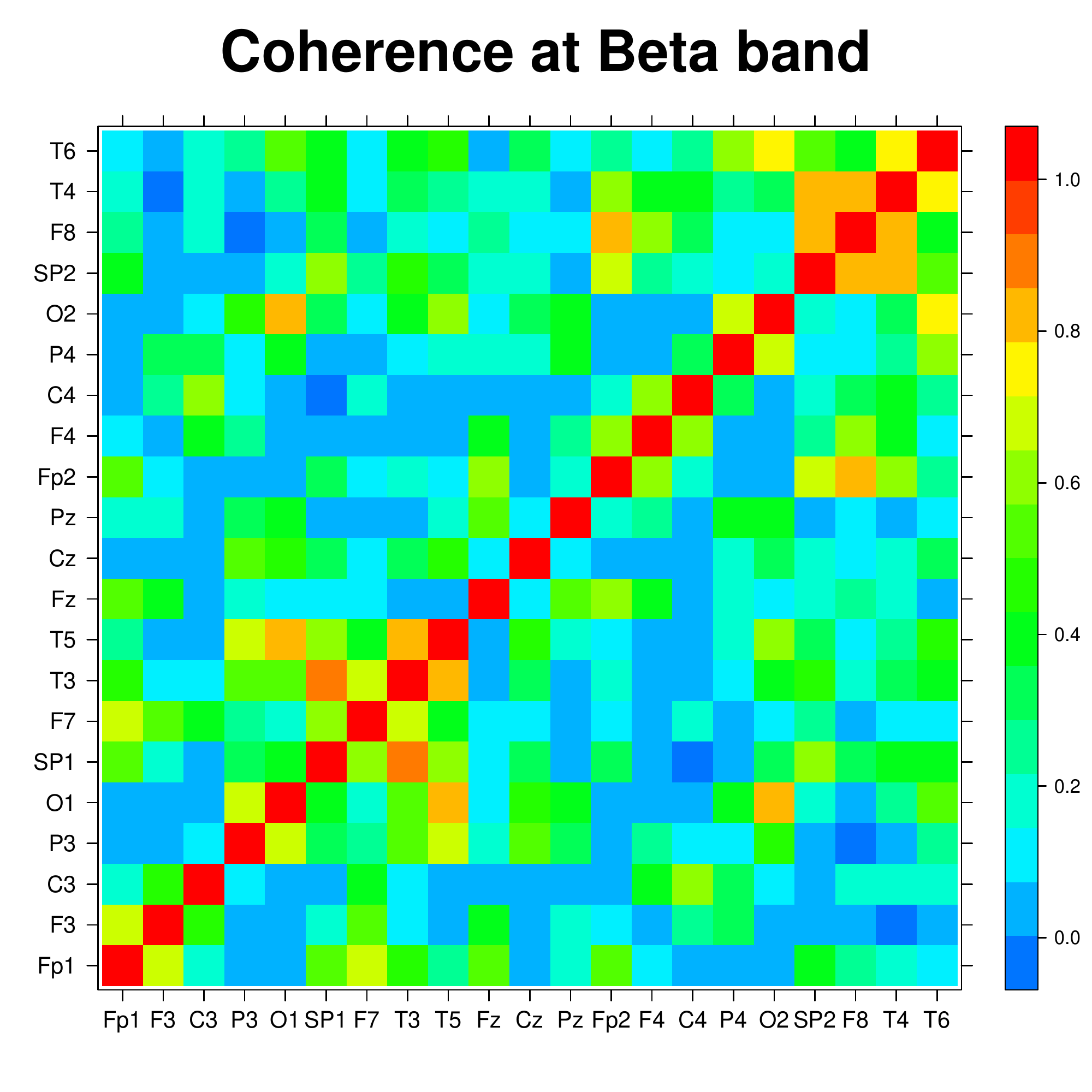}
\includegraphics[scale=.25]{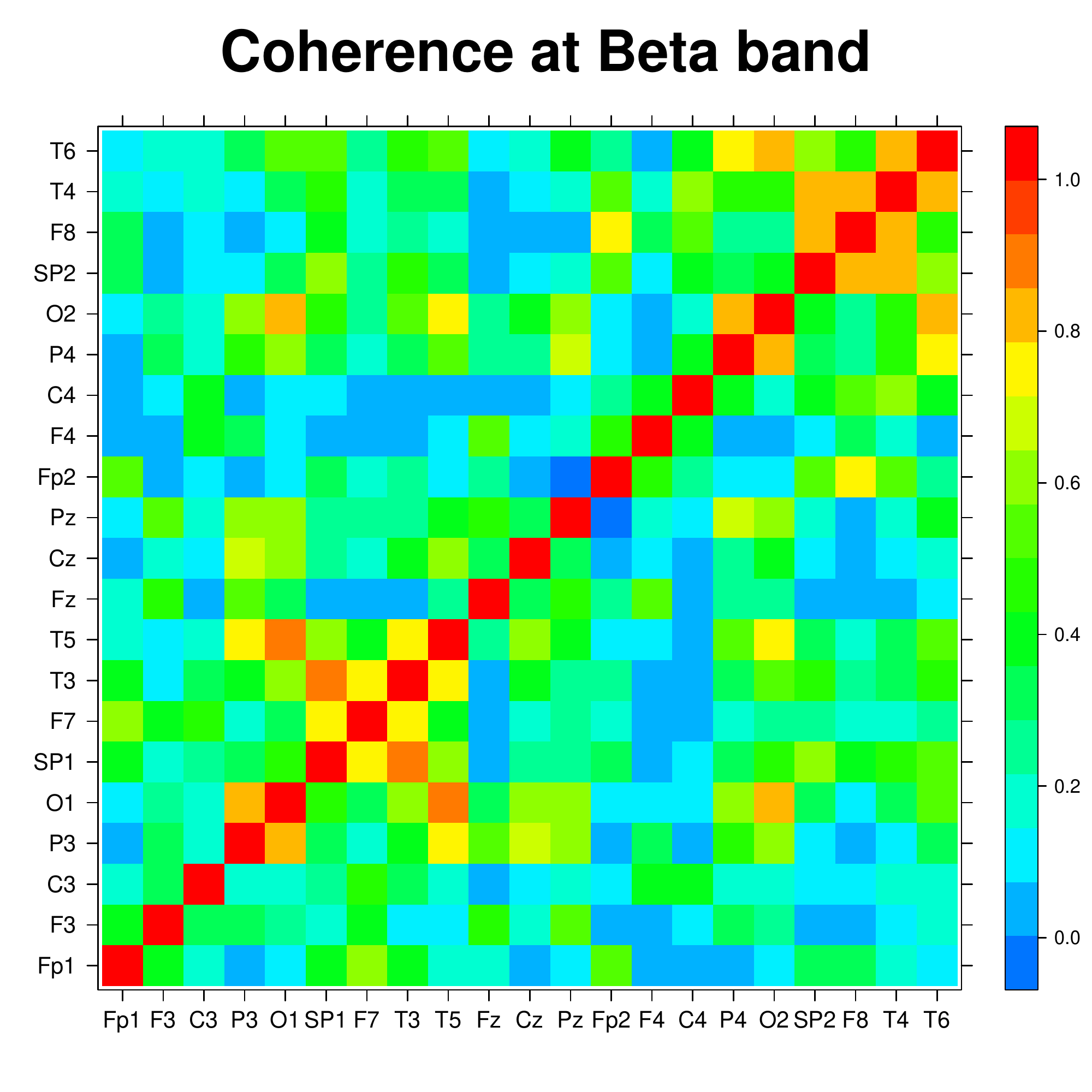}\includegraphics[scale=.25]{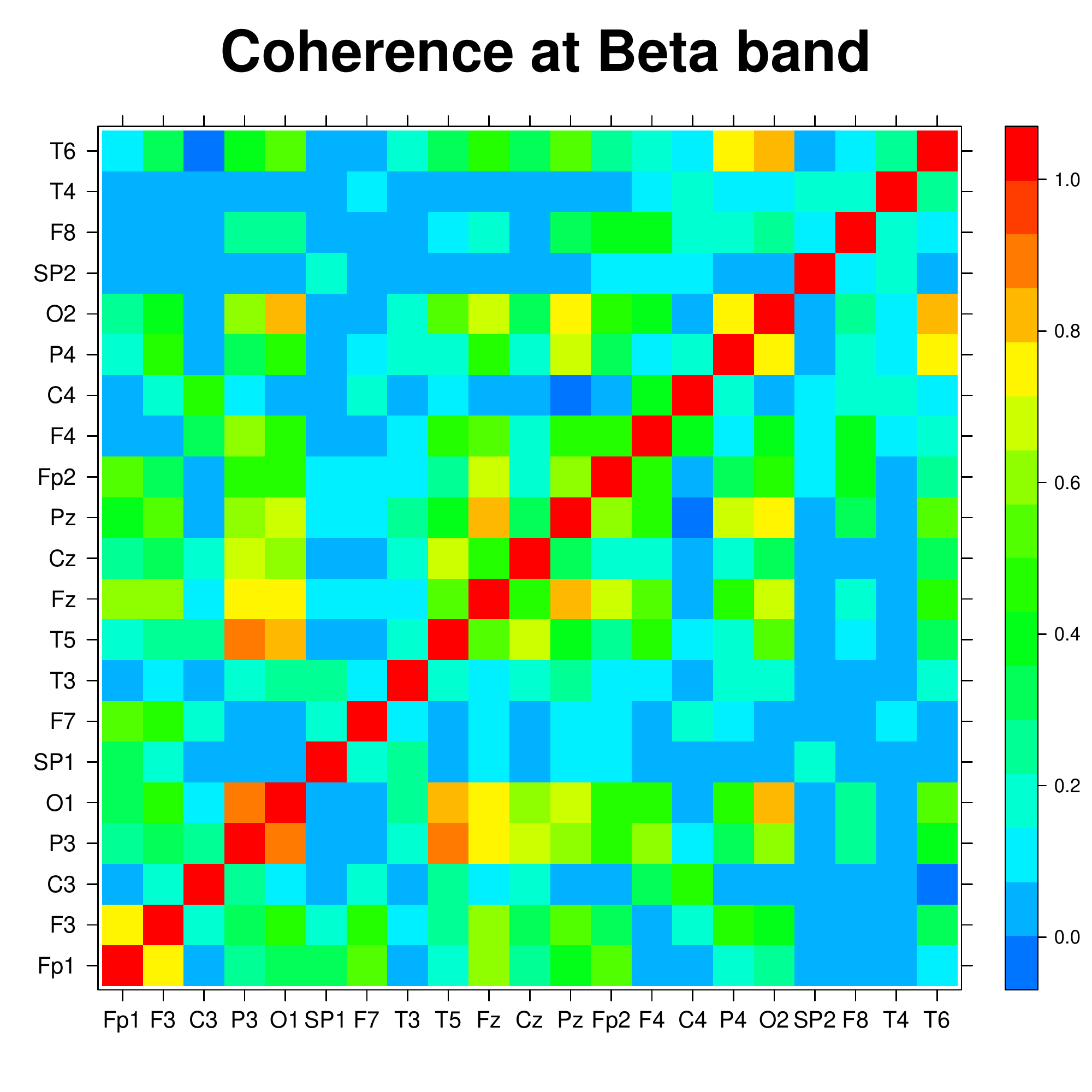}}
\caption{Integrated coherence between EEG channels for different states, before seizure, early seizure and middle seizure on  (a) alpha band and (b) beta band. }\label{SeizureCoh}
\end{figure}

Figure \ref{SeizureCoh} shows the integrated coherence on alpha and beta bands for the three scenarios, 
before, early and middle seizure. The coherence (or integrated coherence) differs between frequency 
bands and scenarios. Some channels, e.g., as $F7$ and $SP1$, are highly connected with $T3$ in each scenario 
but on different scales. We transformed these matrices into $1-integrated~ coherence$ to initialize the dissimilar 
matrix for HCC and HAC clustering methods. 

\begin{figure}
\centering
\subfigure[\label{ShiSei1}]{ \includegraphics[scale=.3]{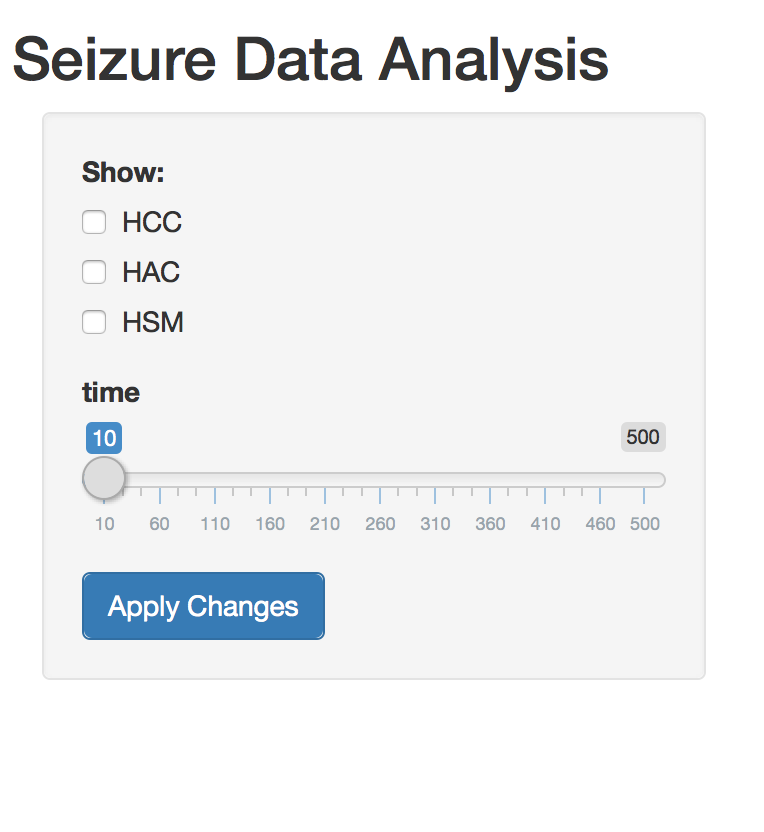}}
\subfigure[\label{ShiSei2}]{ \includegraphics[scale=.25]{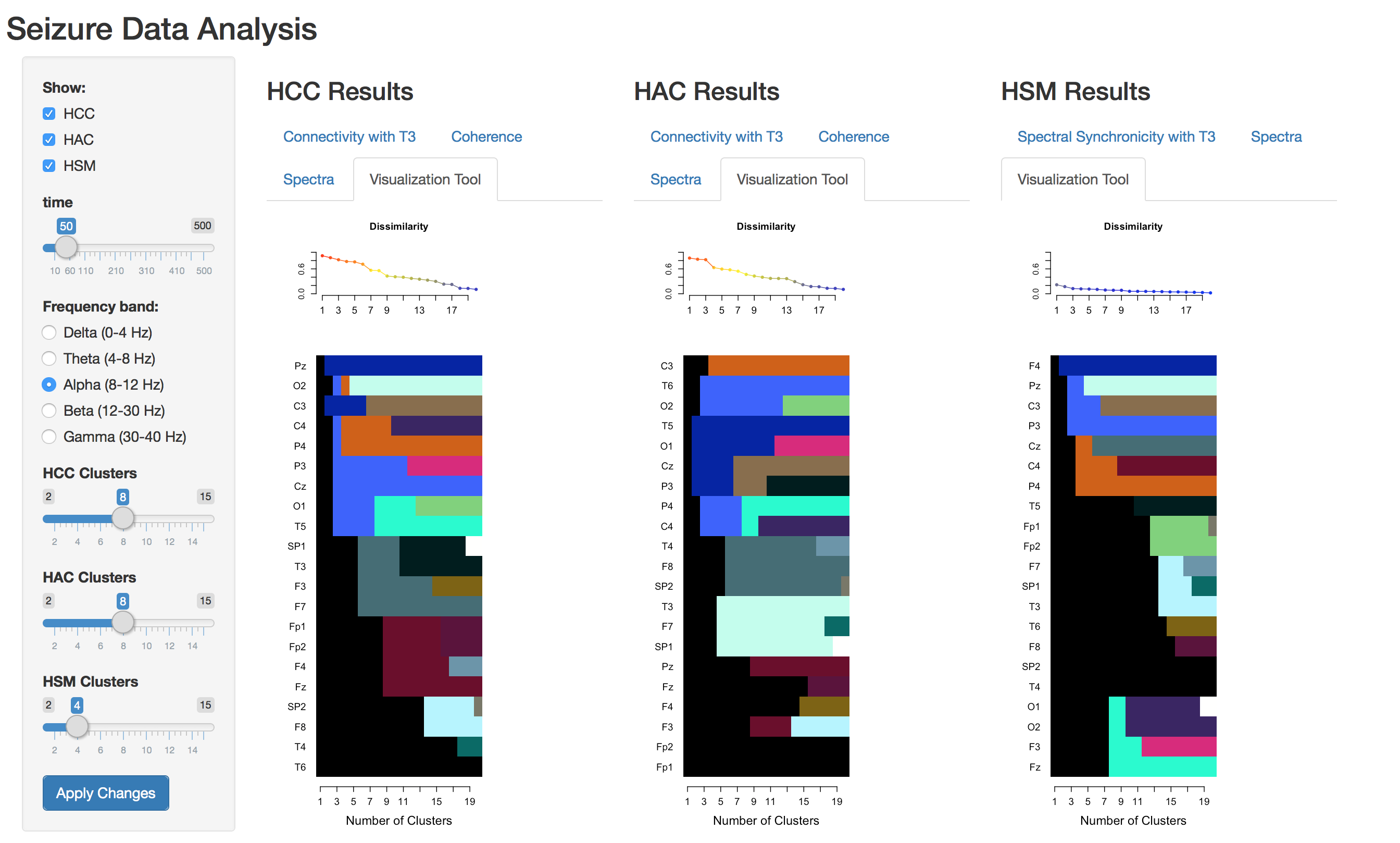}}
\caption{HCC-Vis. (a) Initial screen. (b) Visualization Tools for clustering on alpha band.}
\end{figure}

Figure \ref{ShiSei1} shows the initial screen of the HCC-Vis. We select the three clustering methods to be displayed and time 50 seconds (before seizure) that correspond to before seizure escenario. 
Then, we select the alpha band to execute the clustering methods. Figure \ref{ShiSei2} shows the visualization tools to decide the number of clusters and to observe the clustering dynamics between different clustering methods. We repeat this for early seizure (340 to 350 seconds)
and middle seizure (370 to 380 seconds) in alpha and beta bands.
In general, the clustering results from different methods, scenarios, and frequency bands 
differ in some cases but are similar in others. The clustering dynamics 
between the HCC method and the HAC method are different 
for all of the channels. HSM method identifies few clusters; this means that many channels 
capture activity in similar frequency bands and they are spectrally synchronized. Our goal is 
to identify connectivity based on coherence, so that we will focus on the HCC and HAC clustering results.     

Now, we show the clustering results from $T3$ in the following sense. 
T3 is represented by the yellow dot in each figure.  
If a channel belongs to the same cluster with $T3$, it is represented 
in red, otherwise it is in blue. Note that this does 
not mean that we just have two clusters. In fact, the number of clusters  
for HCC and HAC is fixed to $8$ on the alpha band and beta band for all scenarios based on the scree plot
criterion. 
The HSM method is computed over the whole frequency range $0-50$ Hz and 
$3$ clusters were selected in this case for all scenarios. 
The assigned number of clusters were based on the visualization tool; 
we also consider the same number of clusters for HCC and HAC to 
make comparisons easier. We can also explore other options in the 
HCC-Vis to produce more robust conclusions.
\begin{figure}
\centering
\subfigure[Before Seizure - Alpha band]{\includegraphics[scale=.2]{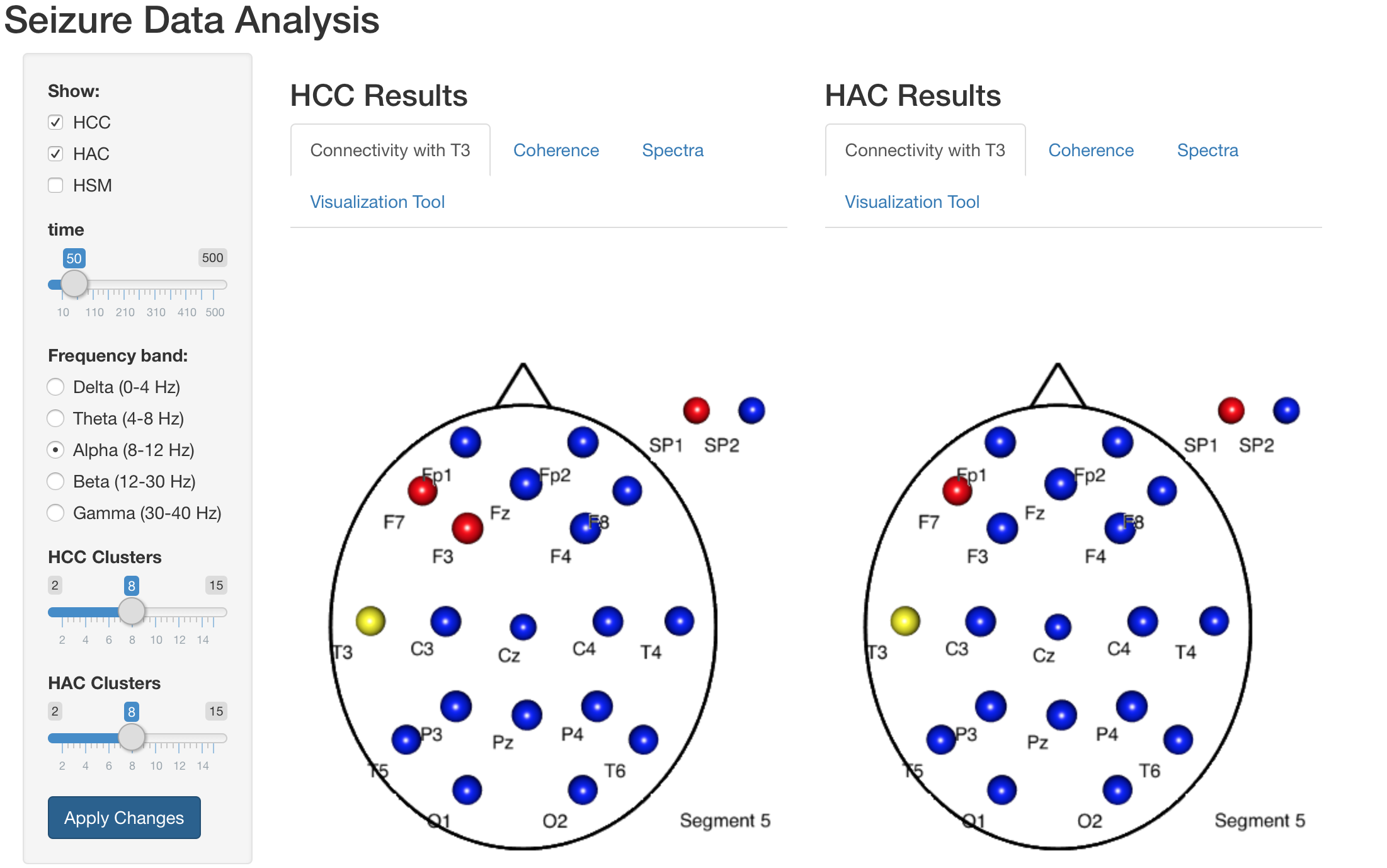}}
\subfigure[Early Seizure - Alpha band]{\includegraphics[scale=.2]{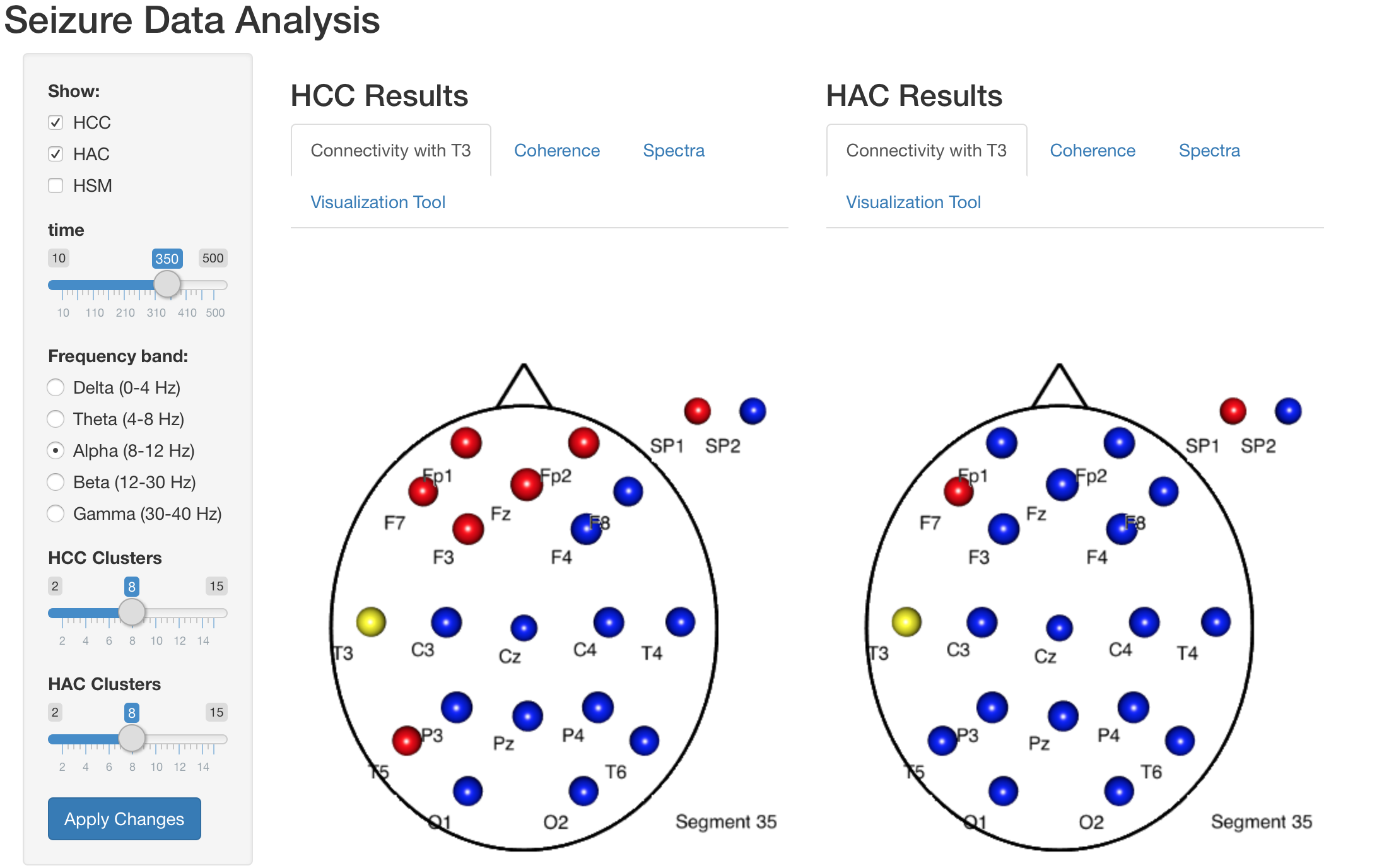}}
\subfigure[Middle Seizure - Alpha band]{\includegraphics[scale=.2]{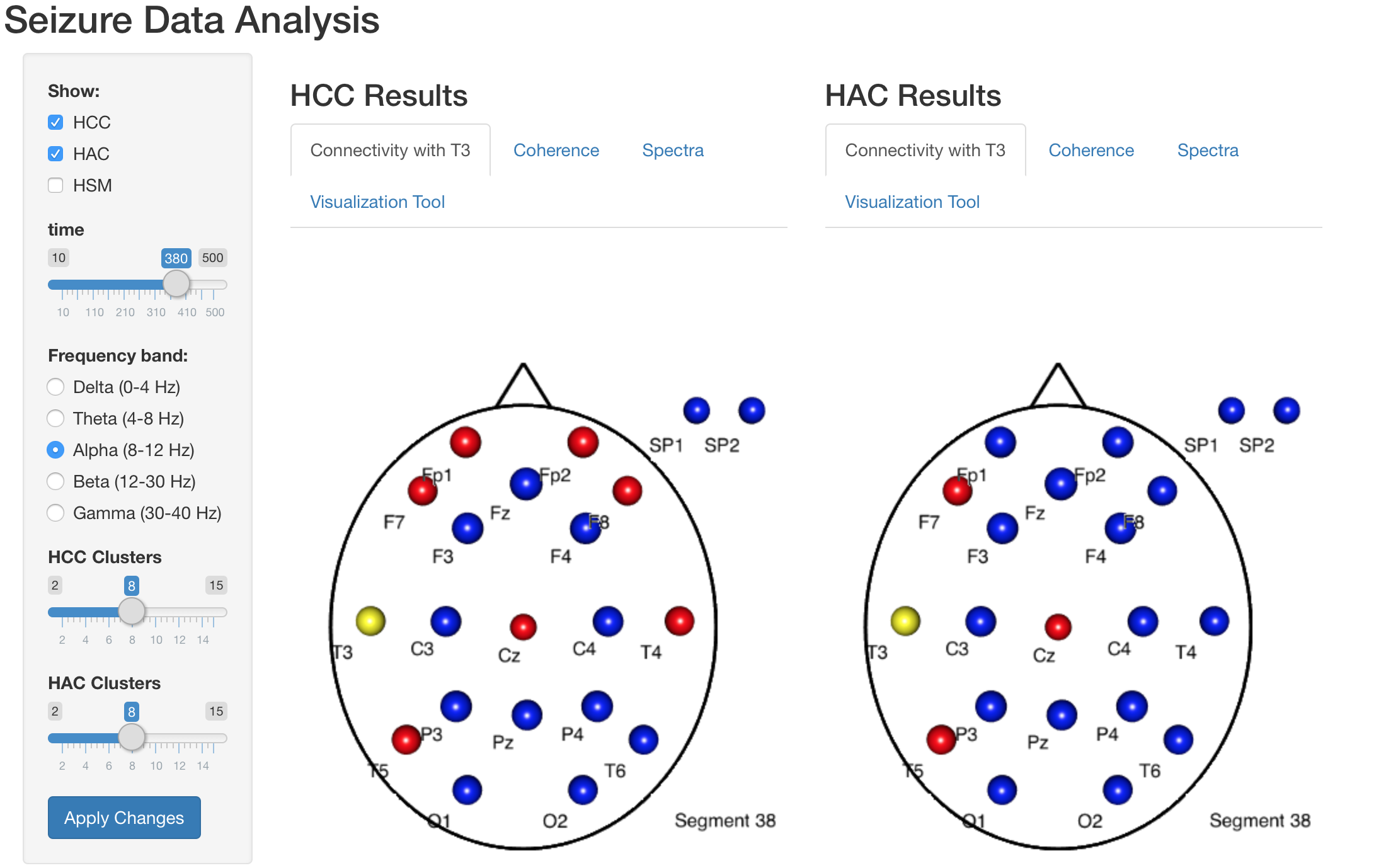}}
\caption{Connectivity with $T3$ based on coherence in the alpha band. In each case, 
the HCC results are on the left side and the HAC results are on the right side. 
This analysis was produced using the HCC-Vis (\url{https://carolinaeuan.shinyapps.io/hcc-vis/}).  }\label{ClustScalpAlpha}
\end{figure}

\begin{figure}
\centering
\subfigure[Before Seizure - Beta band]{\includegraphics[scale=.2]{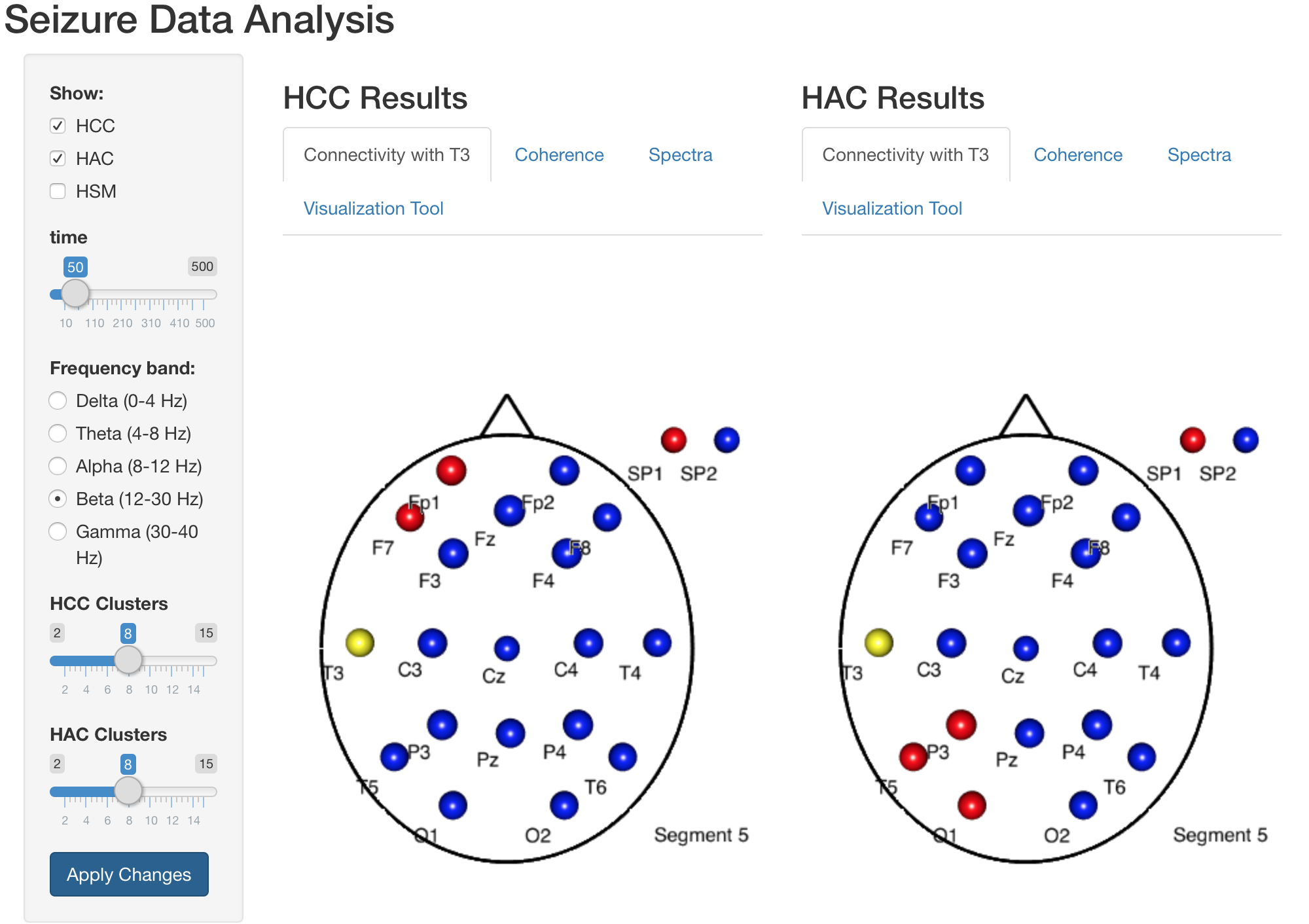}}
\subfigure[Early Seizure - Beta band]{\includegraphics[scale=.2]{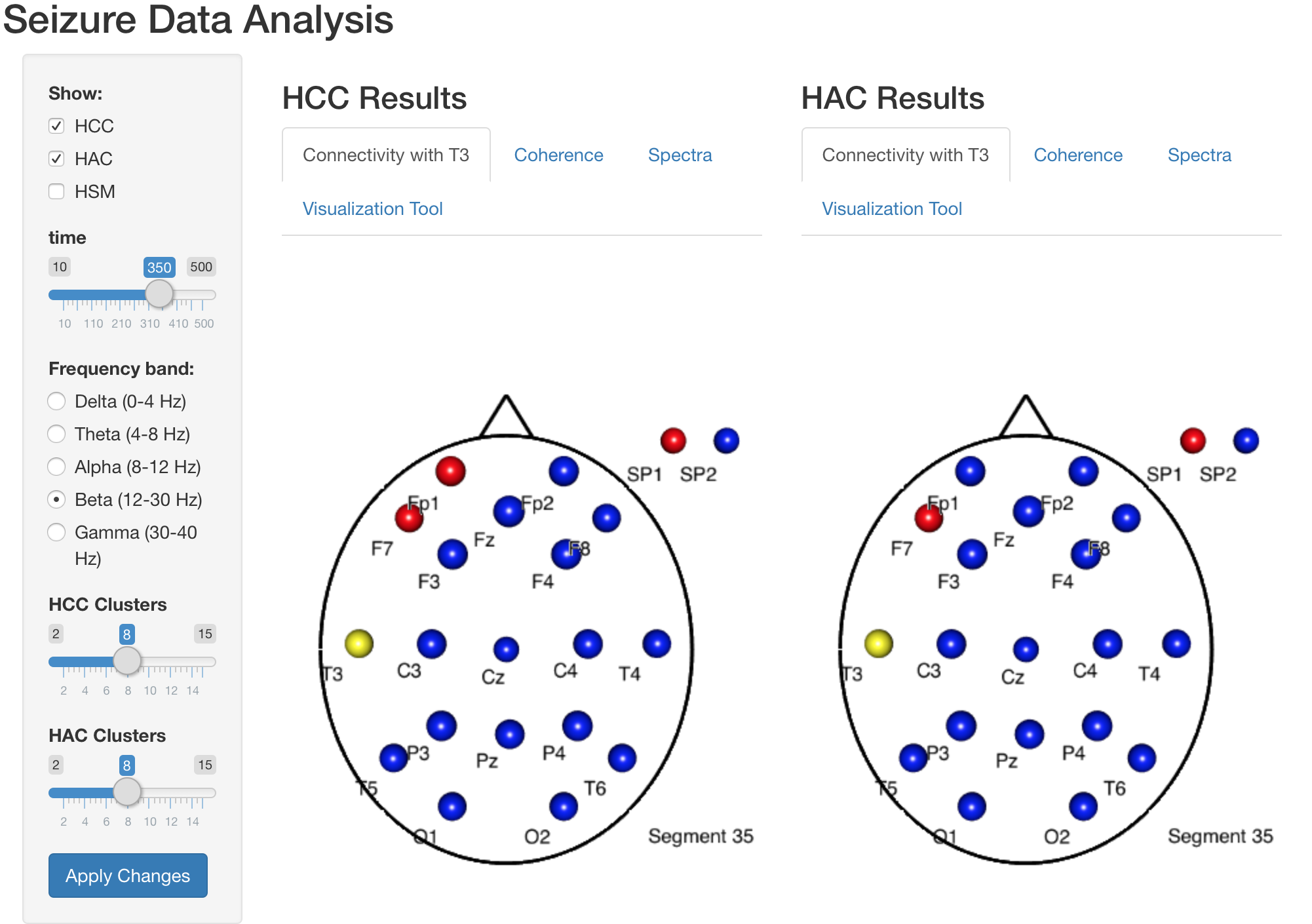}}
\subfigure[Middle Seizure - Beta band]{\includegraphics[scale=.2]{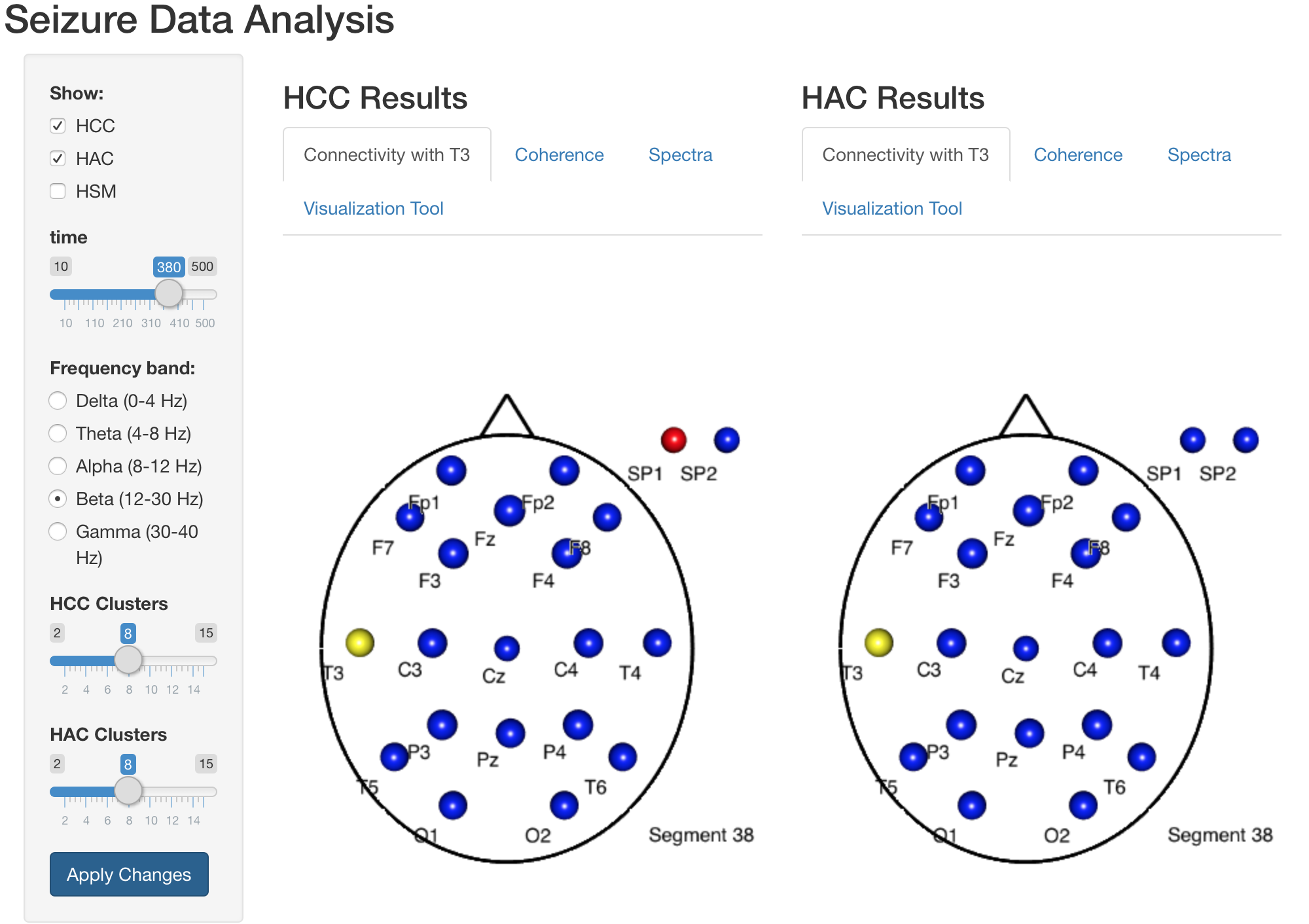}}
\caption{Connectivity with $T3$ based on coherence in the beta band. In each case, 
the HCC results are on the left side and the HAC results are on the right side. These results were produced using the proposed HCC-Vis. }\label{ClustScalpBeta}
\end{figure}

\begin{figure}
\centering
\subfigure[Before Seizure]{\includegraphics[scale=.15]{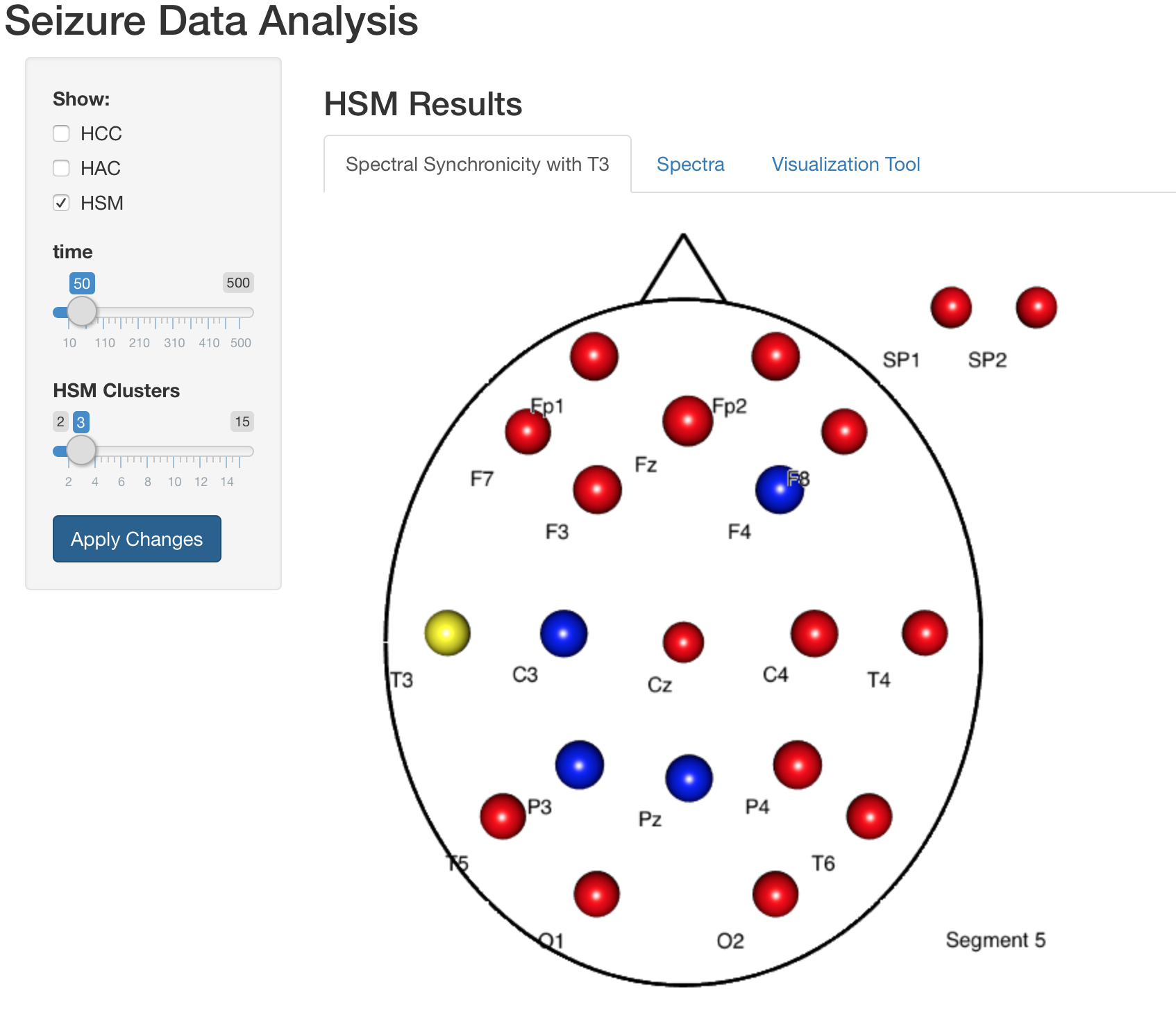}\includegraphics[scale=.15]{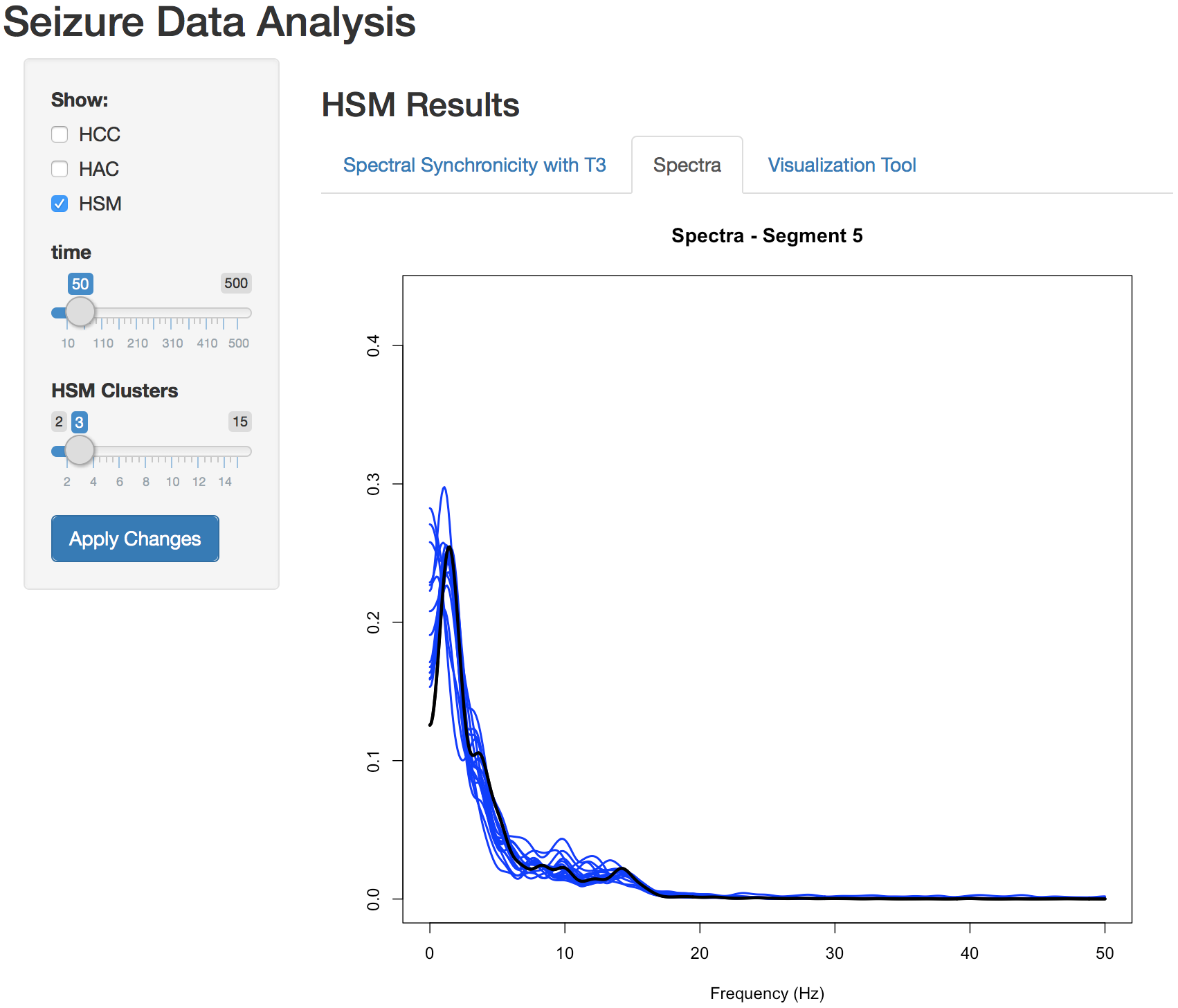}}
\subfigure[Early Seizure]{\includegraphics[scale=.15]{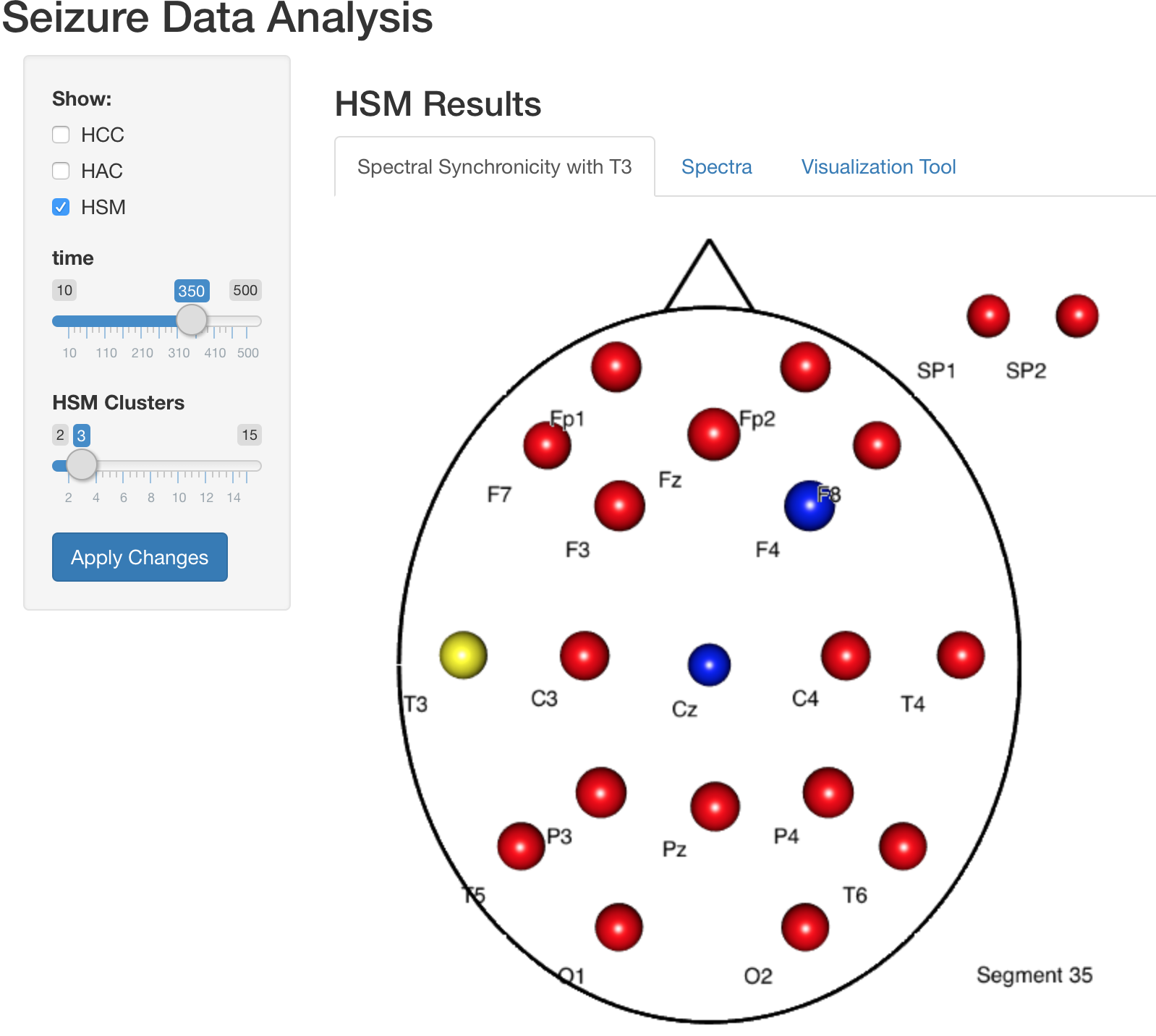}\includegraphics[scale=.15]{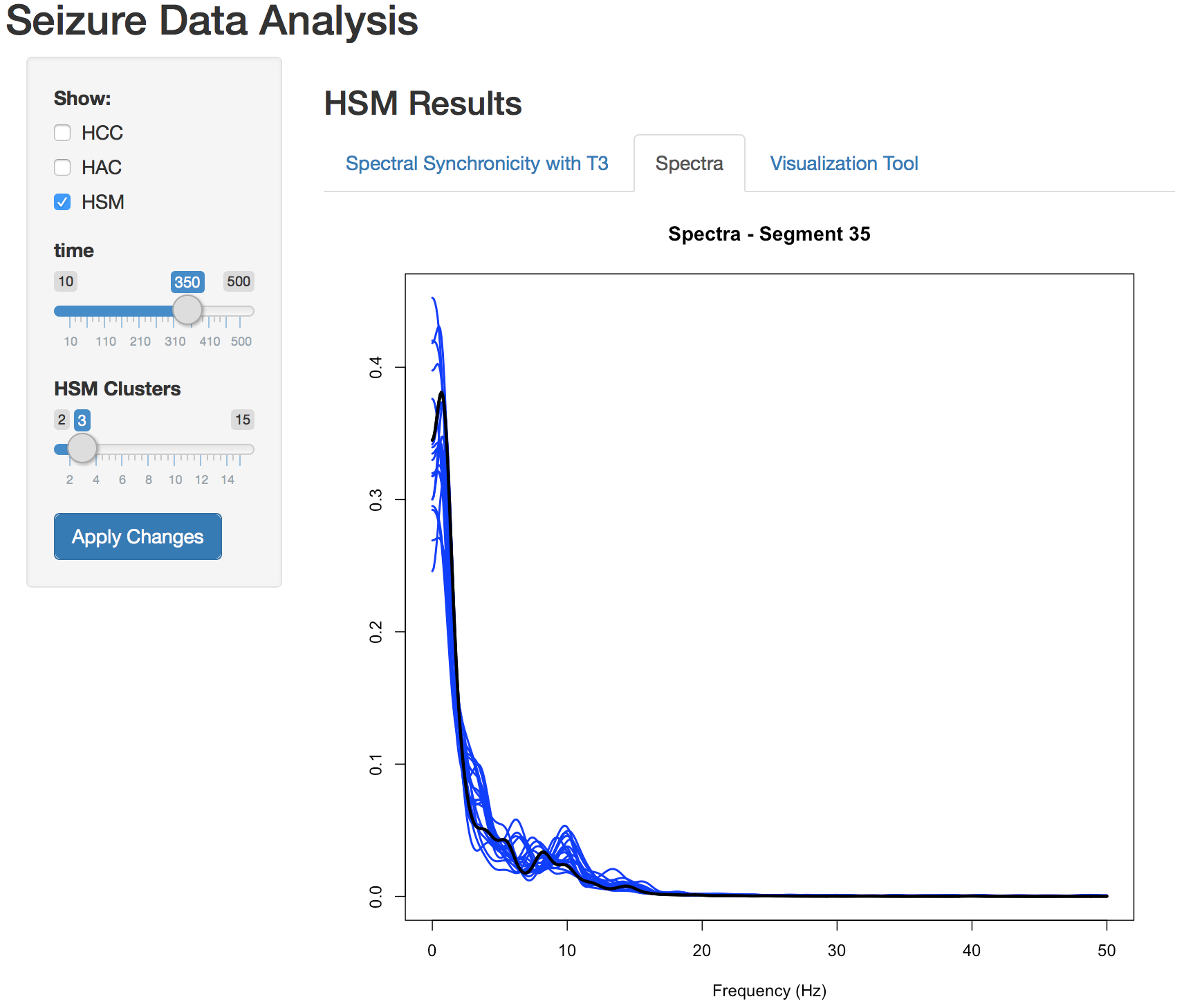}}
\subfigure[Middle Seizure]{\includegraphics[scale=.15]{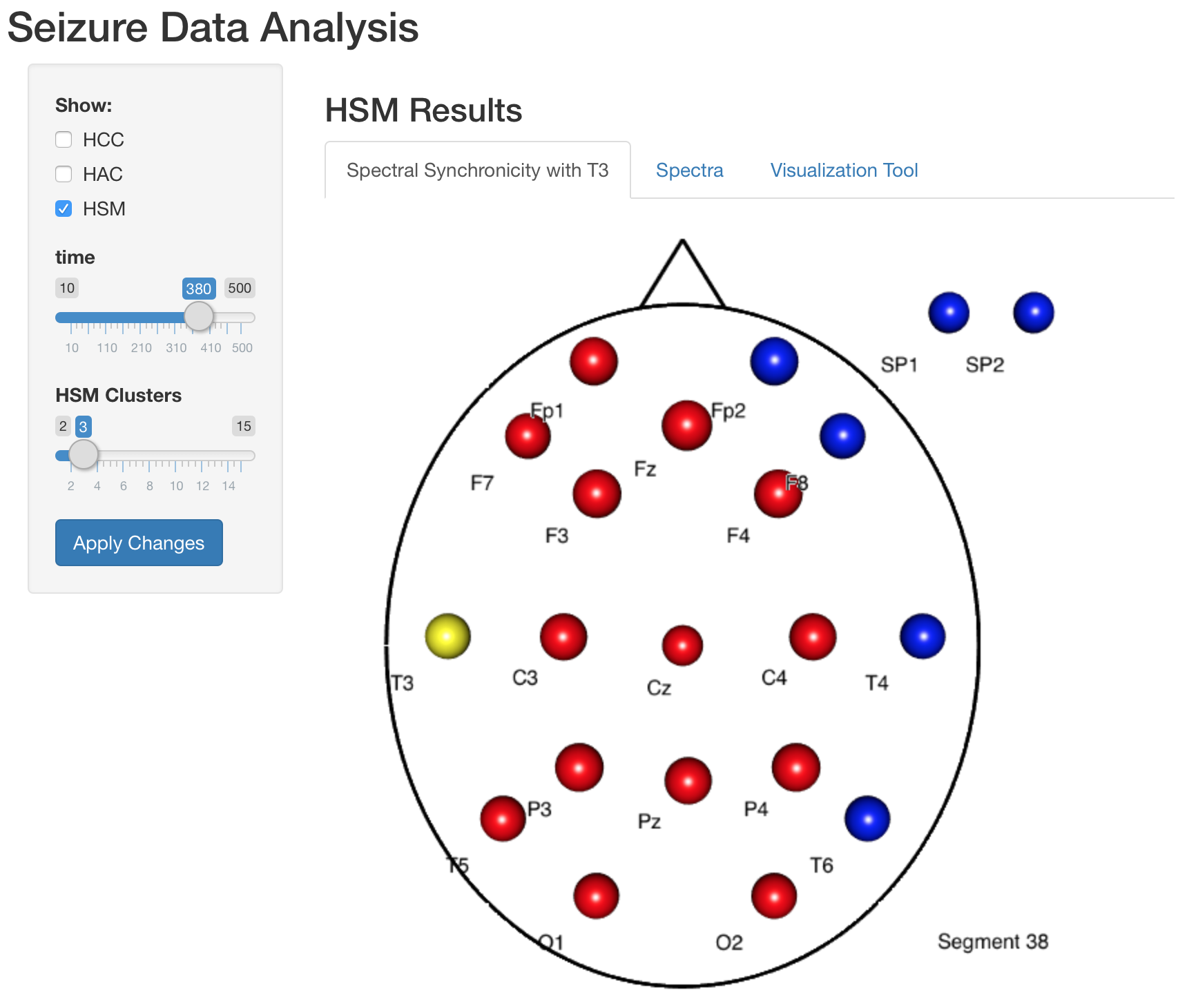}\includegraphics[scale=.15]{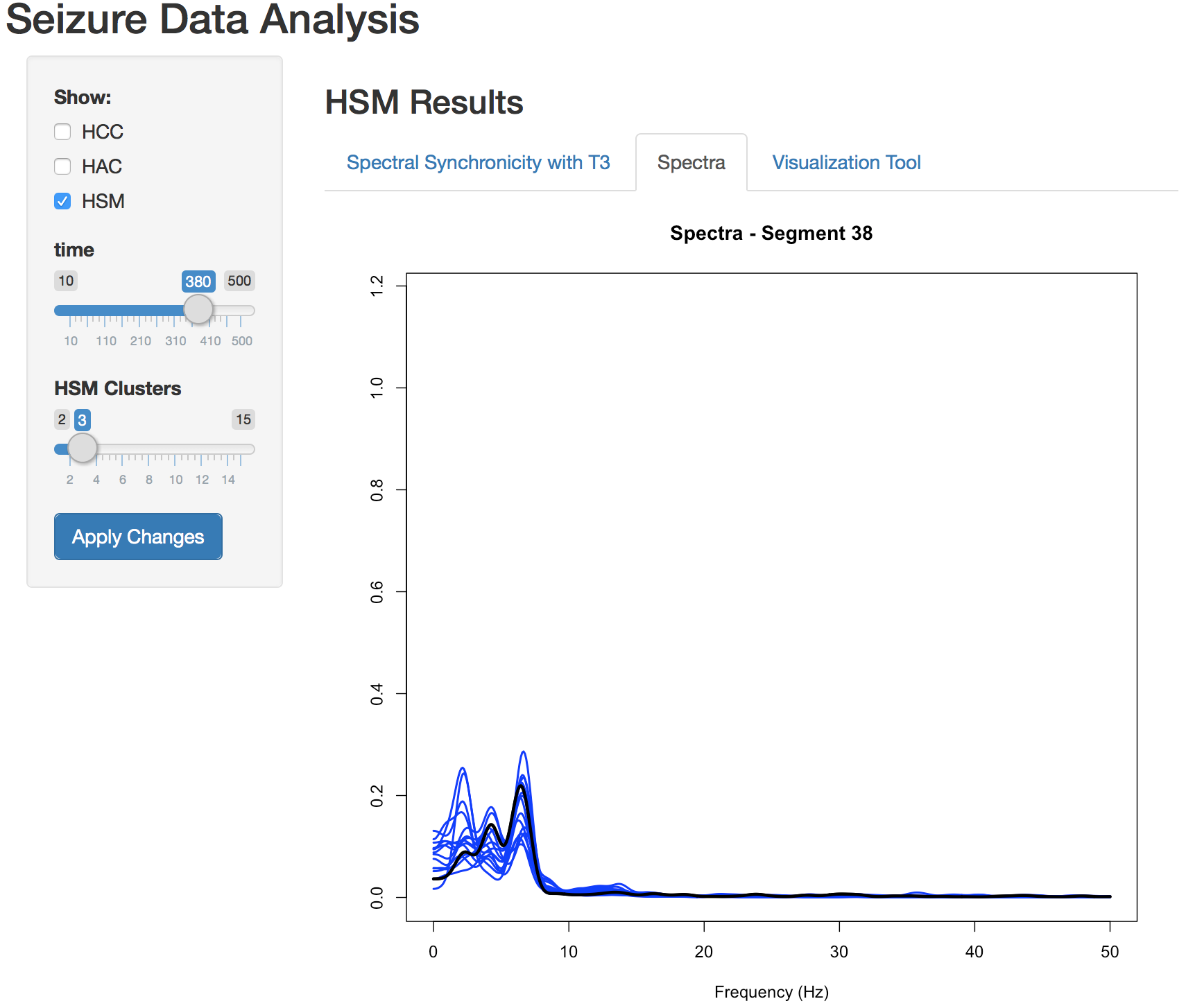}}
\caption{Spectral synchronicity with $T3$ based on HSM clustering using our newly developed HCC-Vis.}\label{ClustScalpHSM}
\end{figure}

Figures \ref{ClustScalpAlpha} and \ref{ClustScalpBeta} show clustering based on coherence of $T3$ and 
other channels, on alpha and beta bands, based on $8$ clusters. Figure  \ref{ClustScalpHSM} show the channels that are spectrally synchronized with $T3$, based on $3$ clusters.
Clustering results from the HCC method on alpha band are different depending on the scenario. 
Connectivity before seizure was between $T3$ and $F7$, $F3$ and $SP1$, which is reasonable since 
they are located closely on the scalp. After seizure started, more channels have a higher 
tendency to belong to the same cluster as $T3$. As a result, 
the left frontal and temporal regions become into a single highly correlated cluster. In the middle seizure, 
the connectivity of $T3$ expands to more channels located in the frontal region and temporal region. 
If we consider HAC, the results in the first scenario are similar to HCC results. However, during the seizure the are less 
channels connected with $T3$. 

The clustering results on the beta band are more similar between the two methods, HCC and HAC. 
The results also suggest that there is strong correlation between channel 
$T3$ and $SP1$ on the beta band. $F7$ also shows connectivity at this amplitude. 
However, this connectivity vanishes after the seizure.   
Clustering results obtained by using the HSM method confirm 
that many of the channels show activity in same 
frequencies. In particular, we identify that $T3$ is not the only channel 
with energy in high frequency bands, alpha and beta.

\begin{figure}
\centering
 \includegraphics[scale=.25]{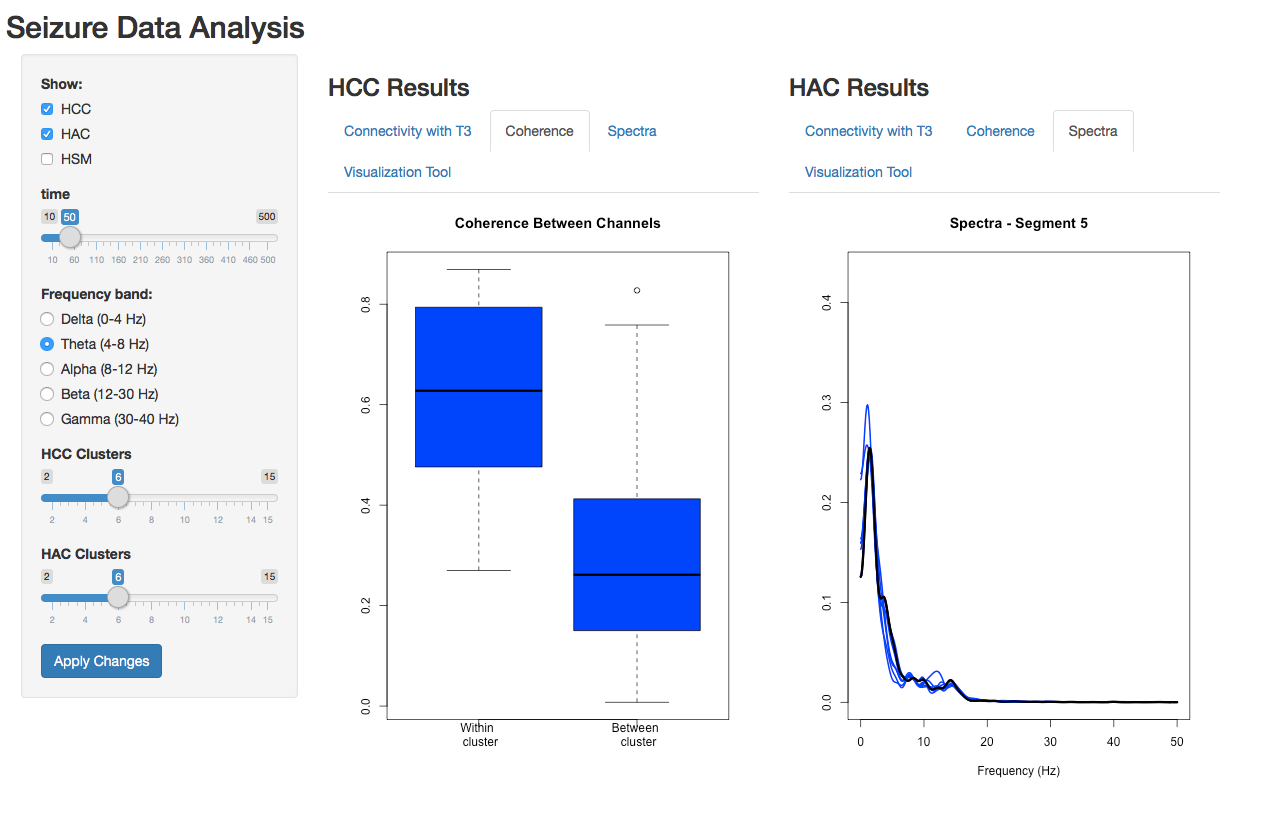}
\caption{HCC-Vis. Additional plots to visualize the connectivity strength and spectral profile of channels connected with T3.}\label{Shi3}
\end{figure}

Different time segments and number of clusters can be explore on the HCC-Vis included in additional material. We also included in the HCC-Vis tabs with the spectra of channels clustered with $T3$ and 
Coherence between members in same cluster as $T3$ for the coherence based clustering methods, HCC and 
HAC. For example, Figure \ref{Shi3} presents these additional plots; the first corresponds to the coherence values within the cluster that T3 belongs, and the second corresponds to the spectral 
density of each member of this cluster; these are displayed when we select the ``Coherence'' and ``Spectra'' tabs, respectively. 
 
To conclude, HCC clusters change between frequency bands and evolve under different scenarios. 
These suggest that the brain network changes between frequency bands and during different
stages of the seizure episode. Also, we identify many channels that showed activity 
in higher frequency bands as $T3$, but they are not necessarily correlated with $T3$.

\section{Discussion and Conclusions}
The main contribution of this paper is the hierarchical coherence clustering (HCC) method 
which uses the notion of a cluster-based coherence rather than coherence between a 
pair of univariate time series. The
simulation studies demostrated the advantages of the HCC method compared to the average and minimum 
coherence methods. In cases where the clusters are independent, the HCC method performs as well as 
the other clustering methods. In the presence of noisy signals, e.g., in EEG data,  
the HCC method performs better than other methods.

We applied the HCC method and the HSM method to EEG data recorded during an epileptic seizure
on the alpha and beta bands. The clustering results show that many channels were active during 
the seizure but not all of them were correlated. Also, by considering the changes in the 
connectivity before and after clustering, we found that the seizure possibly 
affected the connectivity on alpha band. 

Cluster coherence measures the similarity between clusters by taking into account both the within 
cluster structure and between clusters via the eigenvalues. One limitation of this method is the computational time 
required to compute the eigenvalues when the size of the clusters increases. A truncated version of the 
cluster coherence could be used to speed up these computations; in this case, 
a fast algorithm could be used to compute only the $p$ first eigenvalues.

\end{document}